\PassOptionsToPackage{unicode}{hyperref}
\PassOptionsToPackage{hyphens}{url}
\documentclass[
]{article}
\usepackage{lmodern}
\usepackage{amssymb,amsmath}
\usepackage{ifxetex,ifluatex}
\ifnum 0\ifxetex 1\fi\ifluatex 1\fi=0 
  \usepackage[T1]{fontenc}
  \usepackage[utf8]{inputenc}
  \usepackage{textcomp} 
\else 
  \usepackage{unicode-math}
  \defaultfontfeatures{Scale=MatchLowercase}
  \defaultfontfeatures[\rmfamily]{Ligatures=TeX,Scale=1}
\fi
\IfFileExists{upquote.sty}{\usepackage{upquote}}{}
\IfFileExists{microtype.sty}{
  \usepackage[]{microtype}
  \UseMicrotypeSet[protrusion]{basicmath} 
}{}
\makeatletter
\@ifundefined{KOMAClassName}{
  \IfFileExists{parskip.sty}{%
    \usepackage{parskip}
  }{
    \setlength{\parindent}{0pt}
    \setlength{\parskip}{6pt plus 2pt minus 1pt}}
}{
  \KOMAoptions{parskip=half}}
\makeatother
\usepackage{xcolor}
\IfFileExists{xurl.sty}{\usepackage{xurl}}{} 
\IfFileExists{bookmark.sty}{\usepackage{bookmark}}{\usepackage{hyperref}}
\hypersetup{
  pdftitle={Choosing the correlation structure of mixed effect models for experiments with stimuli},
  pdfauthor={Jaromil Frossard \& Olivier Renaud},
  hidelinks,
  pdfcreator={LaTeX via pandoc}}
\usepackage[margin=1in]{geometry}
\usepackage{color}
\usepackage{fancyvrb}

\DefineVerbatimEnvironment{Highlighting}{Verbatim}{commandchars=\\\{\}}
\usepackage{framed}
\definecolor{shadecolor}{RGB}{248,248,248}
\newenvironment{Shaded}{\begin{snugshade}}{\end{snugshade}}

\newcommand{\DataTypeTok}[1]{\textcolor[rgb]{0.13,0.29,0.53}{#1}}
\newcommand{\DecValTok}[1]{\textcolor[rgb]{0.00,0.00,0.81}{#1}}

\newcommand{\KeywordTok}[1]{\textcolor[rgb]{0.13,0.29,0.53}{\textbf{#1}}}
\newcommand{\NormalTok}[1]{#1}
\newcommand{\OperatorTok}[1]{\textcolor[rgb]{0.81,0.36,0.00}{\textbf{#1}}}

\newcommand{\StringTok}[1]{\textcolor[rgb]{0.31,0.60,0.02}{#1}}

\usepackage{graphicx,grffile}
\makeatletter
\def\maxwidth{\ifdim\Gin@nat@width>\linewidth\linewidth\else\Gin@nat@width\fi}
\def\maxheight{\ifdim\Gin@nat@height>\textheight\textheight\else\Gin@nat@height\fi}
\makeatother
\setkeys{Gin}{width=\maxwidth,height=\maxheight,keepaspectratio}
\makeatletter
\def\fps@figure{htbp}
\makeatother
\usepackage{mathrsfs} 
\usepackage{algorithm} 
\usepackage[noend]{algpseudocode} 
\RequirePackage{natbib}

\renewcommand{\DataTypeTok}[1]{{#1}} 
\renewcommand{\baselinestretch}{\baselstr}
\renewcommand{\baselinestretch}{1.2}

\usepackage{booktabs}
\usepackage{longtable}
\usepackage{array}
\usepackage{multirow}
\usepackage{wrapfig}
\usepackage{float}
\usepackage{colortbl}
\usepackage{pdflscape}
\usepackage{tabu}
\usepackage{threeparttable}
\usepackage{threeparttablex}
\usepackage[normalem]{ulem}
\usepackage{makecell}
\usepackage[]{natbib}
\title{Choosing the correlation structure of mixed effect models for
experiments with stimuli}
\author{Jaromil Frossard \& Olivier Renaud}
\date{16-10-2020}

\begin{document}
\maketitle
\begin{abstract}
In many experiments in psychology, participants are measured at many
occasions while reacting to different stimuli. For these experiments,
the mixed effects model with crossed random effects is usually the
appropriate tool to analyse the data because it takes into consideration
the sampling of both participants and stimuli. However, unlike
traditional models such as ANOVA for which many software packages leave
no option in their analyses, these models include several options and
there is no common agreement in the literature on the best choices.
Researchers analysing their data may find difficult to make educated
choices and to assess their consequences. In the present article, we are
focusing on the choice of the correlation structure of the data, because
it is both subtle and influential on the results of the analysis. We
provide an explanation of several correlation structures used in the
literature and their strengths and weaknesses. In addition, we propose a
new one called gANOVA that is the natural extension of the repeated
measures ANOVA model. A large simulation study shows that correlation
structures that are either too simple or too complex fail to deliver
credible results, even for designs with only three (fixed effect)
factors, and that the gANOVA structure is the most reliable in a broad
range of situations. We also show how the design of the experiment
influences the correlation structure of the data. Moreover, we provide R
code to estimate all the correlation structures presented in this
article, as well as functions implemented in an R package to compute our
new proposal.
\end{abstract}

\subsection*{Translational Abstract}

Whether the measures are based on reaction times, evaluations of some
human production, based on electro- or magneto-encephalography (M/EEG)
or task related fMRI data, in many instances in psychology and cognitive
neurosciences, experiments use a given set of stimuli (words, images,
cases, situations, sounds) to elicit the cognitive process of interest.
To draw valid conclusions about this process, one must be able to state
that the observed result is neither specific to the participants that
have been measured nor to the specific stimuli that have been shown, but
that it somehow would be the same if other participants were measured
and other stimuli were used. The fact that the same participant is
evaluated at several occasions and that the same stimulus is evaluated
by several participants creates links or correlations between the
responses. The model that best controls the false positive rate in these
cases is the mixed effects model with crossed random effects. There are
however many different families within this model that are more or less
adequate and there is a certain confusion in the psychological
literature. In this article, we describe them, we associate them with a
name, we link them with \texttt{R} code for simple and complex
experiments and we compare them on their assumptions and their
performances. In particular, we propose a new family called gANOVA and
argue that it has many distinct advantages.

\section*{Introduction}
\label{sec:intro}

\subsection*{Experiments with stimuli}
\label{sec:def}

Researchers in human sciences agree that the participants of an
experiment are merely a sample of possible participants and therefore
that the interest in the data analysis is in the generalisation from the
given sample of participants to a more general population of
participants. Somehow, the analysis has to answer as accurately as
possible the following question: ``Is the observed difference specific
to these participants or is it a general trend of the population?''; or
``Would another set of participants give similar results or not?''.

The above questions are at the heart of the replicability issue and a
statistical analysis that would not take into account that participants
are only some amongst many possible participants would be misleading and
likely not replicable.

Formally, the analysis consists of drawing a statistical inference from
the data, where the term inference emphasizes that one infers from a
sample (of participants) to the population. A model and probability
densities are used to explain the experiment outputs and to draw
inferential statements about the population parameters.

Technically, the statistical model must contain random effects that
account for the participant effects and that permit generalisation. It
is probably a misnomer, but this is often expressed in stating that the
participants are treated as random effects. It does not mean that the
researcher must have selected them purely at random from the entire
population, it just means that different participants might have been
chosen and that the analysis should generalize to all possible
participants. Even if it not labelled as such, this is part of the
assumption behind the error term in many widespread models such as the
simple and multiple regression, the one- and two-sample t-test, and
simple and factorial ANOVA (see e.g.~\citet{howell_statistical_2012}).
For repeated measures ANOVA (rANOVA) and longitudinal data, the
underlying models are more complex, but the rationale is exactly the
same.

In order to measure several instances of the subject response, many
experiments in psychology or neurosciences resort to testing materials
that constitute a certain number of what we will call stimuli. The
research situations include words or sentences in psycholinguistics,
neurolinguistics or any experiment that uses them to evaluate some
cognitive process, emotion-eliciting pictures or sounds in affective
sciences, other type of pictures or video that represent the different
situations that are of interest for the research, representational
objects in cognitive psychology, situations or cases in social
psychology, objects or situations in decision-making neuroscience, and
so on. Many experiments, which utilize electro- or
magneto-encephalography (M/EEG) where event related potentials (ERP) are
extracted, use some sort of stimuli to generate these ``events''.
Similarly, in task related fMRI, the ``tasks'' are often some sort of
stimuli designed to elicit a cognitive or neuronal response.

In an experiment that uses stimuli, the very same questions as above can
be asked: ``Is the observed difference specific to the stimuli used in
this experiment or is it a general trend for all the stimuli?''; or
``Would another set of stimuli give similar results or not?''.

For task-based fMRI, \citet{westfall_fixing_2016} have shown that around
63\% of a random selection of articles extracted from the Neurosynth
database used multiple stimuli. More importantly, they mentioned that
the stimuli were used ``in a context where generalization over stimuli
was clearly indicated.'' Since one wants to be able to generalize the
finding of an experiment to similar experiments but with different
stimuli, the model employed for analysis should contain random effects
that account for stimuli. Said differently, the model must treat stimuli
as random effects
\citep{clark_language-as-fixed-effect_1973, coleman_generalizing_1964, judd_treating_2012, raaijmakers_further_2003, westfall_fixing_2016}.
Mirroring the above argument about participants, a random effect does
not require that stimuli have been selected fully at random from the
entire population of possible stimuli, but it merely reflects the fact
that the researcher could have chosen other stimuli (words, pictures,
objects, situations). The scope of the findings always includes them and
researchers want to state results that generalise to these possible
stimuli. Only a model that treats stimuli as random effects allows such
a statement.

In order to increase replicability, researchers are asked to increase
the number of participants \citep{simmons_false-positive_2011}. But the
above problem will not disappear with a larger number of participants.
On the contrary, it will become even more acute and false positive/ type
I error rate of models that do not treat stimuli as random will increase
(up to 100\%) as the number of participants increases, as shown in
\citet{judd_treating_2012}. To take a meaningless but concerning
example, suppose many laboratories were to select a given number of
landscape images and houses images, show them sequentially to
participants that must press a key as soon as they see a landscape
image. Researchers may record the fixed effects based on experimental
factors (linked to the participants, to the images or the experiment).
Now, suppose that the laboratories add a fixed effect, which we will
call parity, that codes whether the size on disk of the image is an odd
or an even number of bytes.

Using a traditional model where either the mean or the median of trial
reaction times (per participant and per condition) is computed and fed
in an ANOVA or a repeated measure ANOVA with all the fixed effects
including parity, with a sufficiently large number of participants,
almost 100\% of the laboratories would declare parity significant. Even
more worrisome, half of the laboratories would find that the even-sized
images induce significantly faster reaction times than odd-sized images
and conversely for the other half of the laboratories. This has been
shown in many simulation studies
(e.g.~\citet{baayen_mixed-effects_2008},
\citet{judd_treating_2012},\citet{lachaud_tutorial_2011}). The
significance is due to the fact that for a given laboratory the
odd-sized landscape images selected are slightly easier to process than
the even-sized ones, and the converse for another laboratory. Stated
differently, with a model that does not treat stimuli as random, the
material is such that the experiment is always under the alternative
hypothesis concerning parity. And this is true for any characteristic
that can be used to split stimuli (here the images) in categories (by
symmetry the same problem would occur if participants were not treated
as random).

On the contrary, with a model that rightfully treats stimuli as random
effects, the number of laboratories that would declare the parity
significant would be around the nominal level (usually 5\%), which is
what is expected \citep{clark_language-as-fixed-effect_1973}. This is
also confirmed on the above mentioned simulations.

For any experiment containing stimuli, if analysed correctly, the
replicability will hold to the expected extent. If analysed incorrectly,
replicability indices are deluded with significant results in
contradicting directions as in the above example. To achieve high
replicability in a study, the choice of an adequate model and
appropriate data analysis techniques is the first necessary step. The
problem raised above is therefore on top of replicability issues
concerning cherry picking/\(p\)~hacking, researcher's degree-of-freedom,
sample-size or low power issues, pre-registration, or the debate between
the null hypothesis significance testing (NHST) versus Bayesian
approaches. Finally, we note that there exist very specific experimental
situations (or model designs) in which a model which does not explicitly
contain random effects for stimuli is nonetheless statistically
appropriate. We refer to \citet{raaijmakers_how_1999} and
\citet{westfall_fixing_2016} for these particular cases.

\subsection*{Embracing models that account for stimuli}
\label{sec:adoption}

The above findings on the dramatic consequences of omitting the random
effect of stimuli are by no means recent as the first article to
pinpoint them and to propose an appropriate model dates back to the
1960s \citep{coleman_generalizing_1964}. The approach and the model
using random effects for stimuli were fully adopted in psycholinguistics
decades ago due to the very influential paper from
\citet{clark_language-as-fixed-effect_1973}, despite controversy
(\citet{wike_comments_1976}, \citet{forster_more_1976}). Although some
researchers have been using a method called \(F1+F2\) that is not as
adequate as the quasi-\(F\) proposed by Clark, both approaches
explicitly treat the stimuli as random effects (see
\citet{raaijmakers_how_1999} for a survey of the usage of the \(F1+F2\)
and the quasi-\(F\) over time).

The models used today are the crossed random effects mixed effects
models (CRE-MEM) as they do not require fully balanced data and allow
for one or several covariates. These models are part of the family of
mixed effects models (MEM) and sometimes called crossed random effects
models or even mixed effect models. They have been introduced by
\citet{baayen_analyzing_2008}, \citet{lachaud_tutorial_2011},
\citet{locker_use_2007} or \citet{judd_treating_2012} to psychologists,
discussed by \citet{barr_random_2013} and
\citet{bates_parsimonious_2015}, and extremely efficiently implemented
by \citet{bates_fitting_2015} in the R programming language.

The adoption of models that account for stimuli in other fields of
psychology and neuroscience is much less rapid, if present at all. A
wake-up call in the field of social psychology by
\citet{judd_treating_2012} ignited a spark and \citet{bedny_item_2007},
\citet{westfall_fixing_2016} and \citet{burki_accounting_2018} extend
the CRE-MEM models to fMRI and EEG data. Finally, we note that the idea
that some experiments have additional random effects is not specific to
stimuli; see for example the discussion in Psychological Methods on
whether to regard the providers of treatment as a random effect
\citep{crits-christoph_therapists_2003, serlin_should_2003, siemer_assumptions_2003, siemer_power_2003}.

\subsection*{Choice of a model and of its correlation structure}\label{chap:model}

When analysing an experiment, researchers are mostly interested in
testing a few hypotheses. However, this cannot be achieved without
(explicitly or implicitly) building a statistical model, choosing a test
statistic and computing its associated \(p\)~value (or some Bayesian
index). All of the steps in this procedure force the researcher to
choose between many options, whereas for analyses like ANOVA and rANOVA,
many software packages do not let any option for the model. The
consequence of a wrong choice is that the obtained \(p\)~value is
unreliable, will mislead researchers about their findings and lead to
poor replicability. As \citet{box_empirical_1987} warn, ``Essentially,
all models are wrong, but some are useful''. Indeed, even for the
simplest experiment that compares two groups, nobody can guarantee that
the data are Gaussian, with the same variance for the two groups,
independent (and randomly sampled from the population), which are
necessary assumptions of the \(t\)~test or one-way ANOVA. If these
assumptions are not met, one cannot know how misleading the \(p\)~value
obtained by a \(t\)~test will be. There is no ``right model'' for this
simplest experiment, and therefore no sure answer to the research
question. For complex experiments with many experimental factors,
participants and stimuli, there is no ``right model'' either. However,
some models will be more useful in the sense that they will deliver an
answer to the research hypotheses that is more valuable.

Several elements, like the choice of the predictors, test statistic and
distribution may influence the resulting inference, see
\citet{brauer_linear_2018}. In this article we concentrate on the choice
of the correlation structure i.e.~the way to define the random part. For
CRE-MEM, software packages usually let users tune many settings of the
correlation structure and no consensus exists. Researchers are probably
perplexed by these choices and their consequences, and there is even
some confusion in the literature. For example \citet{barr_random_2013}
is cited almost 4000 times, but researchers that cite them do not
necessarily used their prescribed maximal correlation structure (see the
section on families of correlation structures); e.g.~ for the analysis
of an experiment in which morphed male/female faces images are shown to
measure the preferences for gay men, \citet{cassar_no_2020} state that
they use the maximal model as recommended by \citet{barr_random_2013}
but actually set the correlations between random effects to zero and use
only few of the maximal structure variance parameters.

It might be counterintuitive, but as we will show, the choice of the
correlation structure has a large impact on the reliability of
\(p\)~values for hypotheses of interest (i.e.~of factors or fixed
effects). In an article that has raised attention of several media,
\citet{fisher_retracted_2015} showed different makeup settings on face
pictures and linked preferences with hormone levels. After publication,
the authors retracted their article on the sole grounds that they used a
simple correlation structure, usually named random intercept (RI),
instead of a more complex one, which alters seriously the significance
of the main hypothesis (see the discussion in
\citet{retraction_watch_makeup_2016}).

The correlation structure models the similarities between some responses
due to the participants and the stimuli. For instance, two reaction
times from the same participant will be both smaller than average if
this participant is fast, which implies that these responses are
correlated. Similarly, two reaction times from two participants to the
same stimulus might be both small if the stimulus is ``easy'', which
also implies that these reaction times are correlated. A suitable model
should account for all these correlations. If in addition these
correlations depend on the experimental conditions, an even more complex
correlation structure must be included in the model to account for it.
In the literature, CRE-MEMs with a variety of correlation structures are
present, from the simplest ones to very complex ones. In this paper, we
discuss the effect of the choice of the correlation structure on the
\(p\)-values of factors (or fixed effects). We will see that the choice
of the correlation structure is indeed relevant in the rANOVA framework
and was discussed in the literature a few decades ago and we will
compare the findings with the CRE-MEM settings.

In the next sections, we introduce the CRE-MEM model by showing its
similarities and differences with rANOVA. Next, we present the main
CRE-MEM correlation structures used in the literature. We also introduce
a new correlation structure called gANOVA that is the natural
generalization of the rANOVA. Finally, we compare the type I error rates
of all these correlation structures in a simulation study. In the
appendix, the readers will find ready to use \texttt{R} codes for all
correlation structures presented in this article as well as many
extensions designed to help the reader gain understanding of CRE-MEM and
their correlation structures.

In this article, we concentrate on frequentist approach and evaluate the
effect of the model on the \(p\)~values. However, all the argument
developed here are also relevant in a Bayesian framework, as the choice
of the model influences the results equally.

\section*{rANOVA and CRE-MEM}\label{chap:anova}

In order to understand the correlation structure of the models typically
used for analysis, we first investigate the model of rANOVA (for
experiments without stimuli), highlight some of its properties and show
how this model can be extended to CRE-MEM. To illustrate with a research
example, suppose we measure the cardiovascular (CV) response in active
coping on participants that are randomly assigned to one of the two
experimental conditions (clarity of task difficulty: clear vs.~unclear)
and that each participant is separately evaluated in two distinct reward
schemes (reward: attractive vs.~unattractive), as in
\citet{richter_incentive_2006}. This experiment is analysed using a
generic rANOVA with one between-participant factor (clarity of task
difficulty), and one within-participant factor (reward). We note that
the between-participant factor is a feature of the participants because
the participants are not allowed a change of levels during the
experiment. As for many models, each response, is decomposed into
factors (or fixed effects) and random effects. Following the notation of
e.g.~\citet{howell_statistical_2012}, we write the underlying model
using the equation: \begin{align}\label{eq:anova}
y_{ijk} &= \mu + \alpha_j + \psi_k + (\alpha\psi)_{jk} \\
\nonumber
& + \pi_{i} +(\pi\psi)_{ik}  + \epsilon_{ijk},
\end{align} where \(y_{ijk}\) is the response variable, here the
cardiovascular measure of the \(i^{th}\) participant, assigned to
experimental condition \(j\) on the \(k^{th}\) reward scheme. The fixed
portion of the equation is decomposed into the between-participant
effects and within-participant effects. The between-participant effects
are \(\alpha_j\) for \(j \in \{1,\dots, n_j\}\), which corresponds to
the main effect of clarity in our example. The within-participant
effects are \(\psi_k\) for \(k \in \{1,\dots, n_k\}\), which correspond
to the main effect of reward and all interactions that contain it. In
our example, there is one interaction \((\alpha\psi)_{jk}\), which
corresponds to clarity-reward.

The first random element of the equation is composed of the random
intercepts \(\pi_i\) for \(i \in \{1,\dots, n_i\}\) which correspond to
the participant's average cardiovascular responses relative to the
overall mean cardiovascular response. Stated differently, it indicates
how high (or low) in cardiovascular response participant \(i\) is,
compared to the average. The interactions \((\pi\psi)_{ik}\) are the
second random effects included in the model and indicate how each
participant deviates from the average effect of reward on cardiovascular
response. These are sometimes called random slopes since it is
mathematically equivalent if the factor is replaced by a covariate.
Finally, the error term
\(\epsilon_{ijk} \sim \mathcal{N} (0,\sigma_\epsilon ^2)\) captures
everything that cannot be captured by the previous terms. The
correlation structure used in the rANOVA model is implied by the assumed
distribution of the \(\pi_i\)'s and \((\pi\psi)_{ik}\)'s. To have an
exact \(F\) test for the factors in rANOVA, we must assume that the
random effects \(\psi_k\), \((\pi\psi)_{ik}\), follows each a normal
homoscedastic distribution. Therefore, the assumed correlation structure
relies on two random effect (variance) parameters in addition to the
error term variance.

The elements (i.e.~the rows) of the ANOVA table -- as shown by many
statistical software -- correspond exactly to all the terms in
Equation~(\ref{eq:anova}) (except that the highest interaction is
confounded with the error term). This model, including the above
conditions about the normality and homoscedasticity of all random
effects, is behind rANOVA procedures in many statistical software and
therefore the model underlying thousands of published researches that
use rANOVA. Note that other models have been proposed and were discussed
in the 1970s
\citep{box_theorems_1954, huynh_conditions_1970, huynh_estimation_1976, rouanet_comparison_1970, cornfield_average_1956}.
We provide in Appendix~\ref{ap:csranova}, a small summary of the
discussion and arguments that lead to what we know today as a rANOVA.

As explained in the introduction, for experiments with stimuli, a more
complex model is necessary for correct inference. To illustrate the
model, suppose we want to measure priming effects to understand the
conceptual developmental shift. To that end, pictures first of a priming
item and then of the target item were shown to participants, where the
priming falls in one of three categories (priming category:
instrumental, categorical and unrelated, feature of the stimuli) and
children were instructed to name as quickly as possible the target.
Children in 4 age groups between 5 and 9 years old (age group, features
of the participants) are evaluated, as in
\citet{perraudin_contribution_2009}. Suppose finally that the same pairs
stimuli are shown to the same participants in two separate contexts
(context: classroom or playroom, feature of the experimental
manipulation). Since it is the indissociable pair of priming and target
images that causes the cognitive process of interest, it is this pair
(and not the individual image) that constitutes a stimulus. To analyse
this experiment, we use a CRE-MEM which equation is written:

\begin{align}\label{eq:mlm}
y_{imk} =& \mu + \alpha_j + \psi_k + \phi_l + (\alpha\psi)_{jk} +
(\alpha\psi)_{jk}+ (\psi\phi)_{kl}+ (\alpha\psi\phi)_{jkl}\\
\nonumber
&+\pi_{i} +(\pi\psi)_{ik} + (\pi\phi)_{il} + (\pi\psi\phi)_{ikl} \\
\nonumber
&+\omega_{m} +(\omega\psi)_{mk} + (\omega\alpha)_{mj} + (\omega\psi\alpha)_{mkj} \\
\nonumber
&+ {(\pi\omega)}_{im} +(\pi\omega\psi)_{imk} + \epsilon_{imk},
\end{align} with the response \(y_{imk}\) (here the reaction time (RT))
and with a fixed-effect portion written in the first line of
Equation~\ref{eq:mlm} composed of \(\alpha_j\) (features of
participants: here the age group), \(\psi_k\) (features of stimuli: here
the category), \(\phi_l\) (features of experimental manipulation: here
the context), and their interactions, \((\alpha\psi)_{jk}\),
\((\alpha\psi)_{jk}\), \((\psi\phi)_{kl}\) and
\((\alpha\psi\phi)_{jkl}\). The decomposition of the random-effect
portion of the equation has then effects associated to the participants
and their interactions with some factors (second line of
Equation~\ref{eq:mlm}), \(\pi_{i}\), \((\pi\psi)_{ik}\),
\((\pi\phi)_{il}\) and \((\pi\psi\phi)_{ikl}\); effects associated to
the stimuli and their interactions with some factors (third line),
\(\omega_{m}\), \((\omega\psi)_{mk}\), \((\omega\alpha)_{mj}\) and
\((\omega\psi\alpha)_{mkj}\); and effects associated to the
participants-stimuli interactions and to the error term (fourth line),
\({(\pi\omega)}_{im}\), \((\pi\omega\psi)_{imj}\), and
\(\epsilon_{imk}\). Stated differently, as for the previous example,
\(\pi_i\) indicates how fast (or slow) participant \(i\) is in its
responses compared to its (age) group. Similarly, \(\omega_{m}\)
indicates how easy (or hard) stimulus \(m\) is (for its cognitive
assessment), and all subsequent terms modify these effects for specific
conditions. Note that the interaction between sampling units and a
factor is only feasible if the sampling units are evaluated in several
levels of the factor. The correlation structure of this CRE-MEM is
implied by the multivariate distribution of all the random effects
(defined in the three last lines of Equation~\ref{eq:mlm}) and is de
facto more complex than in rANOVA. The different choices and assumptions
for covariance structure are discussed in a following section and are
one of the focus of the present article.

Note that the effects on the fourth line (not including the error) are
seldom if ever mentioned it the literature or included in models. They
correspond to the random effects associated to the interaction
participant-stimuli and model the assumption that some participants will
have e.g.~a greater difference in their responses to different stimuli
than other participants.

\section*{Classification of factors for the CRE-MEM}
\label{chap:clvar}

\renewcommand{\baselinestretch}{1}
\begin{table}[tb]
\centering
\caption{Link between random units and the type of factors. A cross means that a random interaction (i.e. an interaction between the random effect define by the row and the factor (or fixed effect) defined by the column) is estimable or allowed in a CRE-MEM with a fully balanced design.}
\label{tab:ranef}
\begin{tabular}{@{}lcccccc@{}}
\hline
&intercept&$A_P$&$A_S$&$A_{PS}$&$A_M$&$A_O$\\
\hline
Participants& X & & X &  X & X & X\\
Stimuli& X & X &  &  X & X & X\\
Participants:Stimuli& X & & &  & X & X\\
\hline
\end{tabular}
\end{table}
\renewcommand{\baselinestretch}{1.7}

In this article, we will use the term ``factor'' to designate a (fixed
effect) experimental variable that can take a finite number of values,
that will be called levels, e.g.~Sex of participant, with two levels. In
other fields, it might be called a categorical independent variable
(categorical IV) or categorical explanatory variable.

With CRE-MEM, the dichotomy of ``between-participant'' factors and
``within-participant'' factors used in rANOVA is insufficient and we
present here a classification of 5 types of factors for the CRE-MEM. In
the ANOVA framework, between-participant factors (that we will call
\(A_P\)) represent often a feature of the participants, like their sex,
and a participant can be only from one level (or measured in only one
level) of the between-participant factors. Within-participant factors
often represent features of the experimental manipulations~(\(A_M\))
which means that the participants can be measured in multiple levels of
a within-participant factors. This classification is feasible since
there is only one random unit, the participants and no cross random unit
like stimuli. Based on this dichotomy, we know that only
within-participant factors may interact with the random unit to create
random effects.

As viewed in the previous Section, many experiments in psychology are
more complex as they cross participants and stimuli (and therefore must
be analysed with CRE-MEM). In that setting, 3 random units are actually
present: the participants, the stimuli, and their interactions. In order
to know which models are at least feasible (or more precisely, which
random effects can be included in the model), we have to know which
factors may interact with which random units.

For CRE-MEM, as in rANOVA, some factors are either \(A_P\) or \(A_M\) as
they specify a feature of the participants or of the experimental
manipulation. By symmetry with \(A_P\), there might be factors that
specify a feature of the stimuli, called \(A_S\), and factors that
specify a feature of the interaction of participants and stimuli, that
will be called \(A_{PS}\). Finally, some factors indicate a feature of
the specific occurrence, or observation, and will be called \(A_O\). In
more detail, here is the list of potential types of factors:

\begin{enumerate}
\item $A_P$: factors that specify a feature of the participants. The typical example is the sex of the participants. It can also happen that for experimental reasons, participants are (randomly) assigned to a single experimental condition, like a learning method in education or given a specific instruction in social psychology. Experimentally, this condition becomes a feature of the participant and thus the corresponding factor is also classified as $A_P$. In Equation~\ref{eq:mlm}, the (fixed) effects $\alpha_j$ come from a $A_P$ factor and code for the age group in the experiment of \citet{perraudin_contribution_2009}. This type of factor reduces to a between-participant factor in the rANOVA setting.

\item $A_S$: factors that specify a feature of stimuli, in the same way as $A_P$ specify features of participants. The typical example is the valence of an image or the characteristic (frequency, type, \ldots) of a word. In Equation~\ref{eq:mlm}, the effects $\phi_l$ come from a $A_S$ factor and correspond to the priming category.

\item $A_M$: factors that specify a feature of the experimental manipulation. The experimenter usually has the ability to choose the level independently of the participants and of the stimuli, and shows several conditions to the same pair participant-stimulus. Examples are the hemifield of presentation of a target, the lightning or surrounding sound conditions. In Equation~\ref{eq:mlm}, the effects $\psi_k$ come from a $A_M$ factor and correspond to the context. This type of factor reduces to a within-participant factor in the rANOVA.

\item $A_{PS}$: factors that specify a feature of the interaction between a participant and a stimulus, and therefore that cannot vary for a given pair participant-stimulus. Often, this type of factor is the result of constraints on the experimental design: if the same participant cannot see the same stimulus in several conditions, this factor is then specific for each pair of participant and stimulus and is therefore of type $A_{PS}$, for example, if only the high-frequency or only the low-frequency of an image is shown to a given participant. In that case, for a given image, half of the participants will see its high-frequency version and the other half its low-frequency version, and conversely for another image. As an additional example in linguistics, in a novel word experiment design  half the participants learn half learn written words and half learn spoken words, and the association between one word and a level is balanced across participants. 

\item $A_O$: factors that are specific to the observation (or the trial). It may be a physiological or physical measure taken at the precise time of the measures of the response (for a given participant subject to a given stimulus). The position of the trial in the experiment, or the RT to the previous stimulus fall also in this category.
\end{enumerate}

As a rule, for any factor, if the random unit (the participants, the
stimuli or their interactions) is measured in several of its levels,
then a random interaction between this factor and the random unit is
estimable, i.e.~it can be included in the CRE-MEM. As a result, the
interaction with an \(A_p\) factor is only estimable for the stimuli,
the interaction with an \(A_M\) factor is estimable for the
participants, the stimuli and their interaction, etc.
Table~\ref{tab:ranef} gives a summary of the estimable random
interactions for the 5 types of factors. Up to now, we discussed the
cases where the factors are categorical. When dealing with one or more
covariates instead of factors, the classification of predictors and its
consequences on the correlation structure of the data are the same.
However, especially in the presence of interactions, a special care is
needed in the interpretation of the results (e.g.~depending if the
predictors are centred or not). Moreover, this classification and the
approach described above can easily be extended for cases with more than
2 random units (and their interactions).

Finally, note that in rANOVA, the ``wide'' format of the data makes a
clear distinction between the representation of the within-participant
and between-participant factors. In appendix~\ref{ap:typo}, we extend
the representation of the ``wide'' format to data with both participants
and stimuli to include the 5 types of factors. It might be extremely
useful for researchers to identify what random interaction are allowed
in the model presented next.

\section*{Several families of correlation structures}
\label{chap:rs}

\renewcommand{\baselinestretch}{1}
\begin{table}[tb]
\centering
\caption{List of 5 typical experimental designs involving participants and stimuli that will be exemplified in this article (column Use: "Formula" meaning that its \texttt{R} formulas are given in the Appendix~\ref{ap:formula} for all correlation structures, and "Simul" means that it is used in the simulation study. The five different types of factors ($A_P$, $A_S$, $A_M$, $A_{PS}$, and $A_O$) are defined in previous section and the number of levels are given within parentheses.}
\label{tab:model}

\begin{tabular}{lll}
\toprule
Model & Factors & Use\\
\midrule
M1 & Ap(2), As(2), Am(2) & Formula/Simul\\
M2 & Ap(3), As(3), Am(3) & Simul\\
M3 & Ap(3), As(3), Am(3), Am(2) & Formula\\
M4 & Ap(3), As(3), Am(3), Aps(2) & Simul\\
M5 & Ap(3), As(3), Am(3), Am(2), Aps(2), Ao(2) & Formula\\
\bottomrule
\end{tabular}
\end{table}
\renewcommand{\baselinestretch}{1.7}

\renewcommand{\baselinestretch}{1}
\begin{table}[tb]
\centering
\caption{Number of parameters for the correlation structure for the five models described in Table~\ref{tab:model} and for all the discussed correlation structures. A plus sign ($+$) describes a correlation structure that includes the interaction participants:stimuli.}
\label{tab:param}

\begin{tabular}{lrrrrr}
\toprule
 & M1 & M2 & M3 & M4 & M5\\
\midrule
RI & 2 & 2 & 2 & 2 & 2\\
RI-L & 8 & 8 & 16 & 16 & 64\\
MAX & 20 & 90 & 342 & 342 & 5256\\
ZCP & 8 & 18 & 36 & 36 & 144\\
gANOVA & 8 & 8 & 16 & 16 & 64\\
\addlinespace
RI+ & 3 & 3 & 3 & 3 & 3\\
RI-L+ & 9 & 9 & 19 & 17 & 71\\
MAX+ & 21 & 91 & 352 & 343 & 5311\\
ZCP+ & 9 & 19 & 40 & 37 & 154\\
gANOVA+ & 9 & 9 & 19 & 17 & 71\\
\bottomrule
\end{tabular}
\end{table}
\renewcommand{\baselinestretch}{1.7}

With the popularity of mixed effect models and especially thank to the
exceptional capability of the \texttt{lme4} \citep{bates_fitting_2015}
and \texttt{Matrix} packages that benefit from sparse representation of
matrices in the computations, typical experiments in psychology with
several factors can be analysed with cross random effect mixed effect
model (CRE-MEM). However, for the same experiment, there are still many
different possible models, depending on the structure that is assumed
for the random part (e.g.~of equation~(\ref{eq:mlm})). In this section,
we describe the correlation structures that are discussed in the
literature and propose a new one called gANOVA. Several goals are
pursued. First, to list the major models and to give them a name so that
researchers can specify unequivocally the model they used. Second, to
link them with \texttt{R} code of the \texttt{lme4} package
\citep{bates_fitting_2015} for experiments with more than one or two
factors. Third, to explain their assumptions and fourth to compare them
theoretically. In the following section, we will compare them based on
simulations so that some guidelines can be learned.

Note first that all correlation structures discussed in the literature
assume independence between the three groups of random effects
associated with (a)~participants, (b)~stimuli and (c)~their
interactions. For the model of equation~\eqref{eq:mlm}, it implies that
for all following proposals, the random effects on the second line are
independent with the random effects on the third line and the fourth
line. This is a minor assumption if the interaction between participants
and stimuli (c) is included but might be questionable if not.

Second, each of the proposal may include effects coming from the
interaction participants-stimuli (c) even if authors who originally
described these correlation structures did not include them. Those
interaction terms model if some participants are especially good/bad
with a particular stimulus. Below, a ``+'' sign in the name of a
correlation structure indicates its inclusion.

A large number of parameters in the correlation structure imply a more
difficult optimization process and more convergence errors of the
algorithm. And, as for any statistical model, including additional
parameters makes the model ``less wrong'' (in the sense of the goodness
of fit) at the price of reducing its parsimony. In practice, for
CRE-MEM, the usual trade-off between parsimony and goodness of fit is
disturbed by the possible convergence errors of the algorithm. For the
five models presented in Table~\ref{tab:model}, we show in
Table~\ref{tab:param} the number of parameters for each correlation
structure presented next. It clarifies the huge difference between the
proposed correlation structures.

Moreover, the replicability of the findings has become a major worry in
many fields \citep{opensciencecollaboration_estimating_2015}. The
choices carried out when using CRE-MEM should reflect this tendency. For
that purpose, the correlation structure should have the following
desirable properties: the results of an analysis should be reasonably
replicable through experiments, the model should be robust to some
misspecifications, and it must show a high rate of convergence. If the
model is used for testing in a frequentist approach, a good choice will
exhibit a type I error rate close to the nominal level under the null
hypothesis and, at the same time, a high power under the alternative.

\subsection*{The correlation structure with random intercepts (RI)}
\label{chap:ri}

In this simplest case, the correlation structure has only a random
intercept per participant and a random intercept per stimulus. These
intercepts are not correlated, which means that only 1 variance
parameter per random unit is estimated regardless of the number of
factors, hence the value of 2 in the first line and 3 in the sixth line
of Table~\ref{tab:param}. This is the correlation structure that has
been viewed as too simple and caused the retraction of
\citet{fisher_retracted_2015}, as mentioned previously. All the
interaction terms in the second, third and fourth lines of
Equation~\eqref{eq:mlm} are removed, or equivalently, their variances
are set to zero.

For two factors \texttt{A1} and \texttt{A2} and the participant and
stimulus identifier \texttt{PT} and \texttt{ST}, the typical formula of
RI using the \texttt{lme4} package is:

\begin{Shaded}
\begin{Highlighting}[]
\KeywordTok{lmer}\NormalTok{(y }\OperatorTok{~}\StringTok{ }\NormalTok{A1}\OperatorTok{*}\NormalTok{A2 }\OperatorTok{+}\StringTok{ }\NormalTok{(}\DecValTok{1}\OperatorTok{|}\NormalTok{PT) }\OperatorTok{+}\StringTok{ }\NormalTok{(}\DecValTok{1}\OperatorTok{|}\NormalTok{SM), }\DataTypeTok{data =}\NormalTok{ mydata)}
\end{Highlighting}
\end{Shaded}

and for the RI+ structure:

\begin{Shaded}
\begin{Highlighting}[]
\KeywordTok{lmer}\NormalTok{(y }\OperatorTok{~}\StringTok{ }\NormalTok{A1}\OperatorTok{*}\NormalTok{A2 }\OperatorTok{+}\StringTok{ }\NormalTok{(}\DecValTok{1}\OperatorTok{|}\NormalTok{PT) }\OperatorTok{+}\StringTok{ }\NormalTok{(}\DecValTok{1}\OperatorTok{|}\NormalTok{SM) }\OperatorTok{+}\StringTok{ }\NormalTok{(}\DecValTok{1}\OperatorTok{|}\NormalTok{PT}\OperatorTok{:}\NormalTok{SM), }\DataTypeTok{data =}\NormalTok{ mydata)}
\end{Highlighting}
\end{Shaded}

For all families presented in this section, we will suppose that the two
factors \texttt{A1} and \texttt{A2} can interact with participants,
i.e.~are of type \(A_S\), \(A_{PS}\), \(A_M\) or \(A_O\) as presented in
previous section. Several examples for all types of factors, with the
full formulas, are provided in Appendix~\ref{ap:formula}. For space
reason and in order to focus on the part of interest, in the main text
we will summarize the formulas. For the RI and the RI+, it becomes:

\begin{Shaded}
\begin{Highlighting}[]
\KeywordTok{lmer}\NormalTok{(y }\OperatorTok{~}\StringTok{ }\NormalTok{[...] }\OperatorTok{+}\StringTok{ }\NormalTok{(}\DecValTok{1}\OperatorTok{|}\NormalTok{PT)  [...] )}
\end{Highlighting}
\end{Shaded}

In lay language, this correlation structure supposes that some
participants are better than others on each measurement, and that some
stimuli are more difficult than others for all participants alike.
Although this correlation structure is used in the literature, in most
cases, the true correlation structure will most probably be more
complex. Moreover, it does not make sense in fMRI where the overall mean
is arbitrary and often set to zero. Choosing this correlation structure
will reduce the replicability of the results with a gain in parsimony we
do not really need. RI is probably too simple for most applications and
does not provide credible inference \citep{barr_random_2013}.

\subsection*{The correlation structure with random intercepts at each level (RI-L) }\label{chap:ril}

One way to view the rANOVA model (as in Equation~\eqref{eq:anova}) is to
think that it incorporates all interactions between the random effect of
the participant (\(\pi_{i}\)) and the within-participant factor(s) (only
\(\mu\) and \(\alpha_j\) here), to produce all possible random effects
(here \(\pi_{i}\) and \((\pi\psi)_{ik}\)). \citet{bates_fitting_2015}
suggest following the same idea for CRE-MEM, for the random effects of
both participants (\(\pi_{i}\)) and stimuli (\(\omega_{m}\)). This
correlation structure corresponds to random intercepts and interactions
that are IID and spherical
(\(\pi_{i}\sim \mathcal{N}(0,\sigma^2_{\pi})\),
\((\pi\psi)_{ik}\sim \mathcal{N}(0,\sigma^2_{\pi\psi})\),
\((\pi\phi)_{il}\sim\mathcal{N}(0,\sigma^2_{\pi\phi})\), and so on), and
independence between them, for all the elements in the second, third and
fourth lines of Equation~\eqref{eq:mlm}. The typical formula of RI-L
using the \texttt{lme4} package is:

\begin{Shaded}
\begin{Highlighting}[]
\KeywordTok{lmer}\NormalTok{(y }\OperatorTok{~}\StringTok{  }\NormalTok{[...] }\OperatorTok{+}\StringTok{ }\NormalTok{(}\DecValTok{1} \OperatorTok{|}\StringTok{ }\NormalTok{PT) }\OperatorTok{+}\StringTok{ }\NormalTok{(}\DecValTok{1} \OperatorTok{|}\StringTok{ }\NormalTok{PT}\OperatorTok{:}\NormalTok{A1) }\OperatorTok{+}\StringTok{ }\NormalTok{(}\DecValTok{1} \OperatorTok{|}\StringTok{ }\NormalTok{PT}\OperatorTok{:}\NormalTok{A2) }\OperatorTok{+}\StringTok{ }\NormalTok{(}\DecValTok{1} \OperatorTok{|}\StringTok{ }\NormalTok{PT}\OperatorTok{:}\NormalTok{A1}\OperatorTok{:}\NormalTok{A2)  [...] )}
\end{Highlighting}
\end{Shaded}

The RI-L correlation structure keeps a relatively low number of
parameters which does not increase with respect to the number of levels
of the factors (see Table~\ref{tab:param}).

This may seem the natural extension of rANOVA, however with the same
number of parameters, gANOVA includes all the correlation structures
that can be obtained with RI-L and strictly more. Therefore, the
likelihood (and criteria like AIC and BIC) will always be better or
equal when using gANOVA instead of RI-L. More details on the difference
between the two correlation structures are given in the gANOVA part
below. Moreover, we show in appendix that gANOVA has the effect of
almost orthogonalizing the parameters of RI-L, which highly simplifies
the optimization of the likelihood.

\subsection*{The "maximal" correlation structure (MAX)}
\label{chap:max}

The ``maximal'' correlation structure is suggested by
\citet{barr_random_2013} and it is defined by including all possible
random effects associated with the participants on one side and all
possible random effects associated with the stimuli on the other side.
In the model RI-L, these terms do not correlate, but here,
\citet{barr_random_2013} let also a maximal link between random effects,
which means that all random effects can correlate with each other
(within the same random unit or line of equation~(\ref{eq:mlm})). Said
differently, the covariance matrix of the random effects of the same
unit (or line of equation~(\ref{eq:mlm})) is full and unstructured. The
typical formula of MAX using the \texttt{lme4} package is:

\begin{Shaded}
\begin{Highlighting}[]
\KeywordTok{lmer}\NormalTok{(y }\OperatorTok{~}\StringTok{  }\NormalTok{[...]  }\OperatorTok{+}\StringTok{ }\NormalTok{(A1}\OperatorTok{*}\NormalTok{A2 }\OperatorTok{|}\StringTok{ }\NormalTok{PT)  [...] )}
\end{Highlighting}
\end{Shaded}

This correlation structure may seem the appropriate choice without prior
information on the correlation structure. The problem is that it is not
parsimonious enough except for the smallest models. It requires for
example 90 parameters for a model with three factors, see
Table~\ref{tab:param}. Note that the authors did not specified
explicitly how to handle factors with more than two levels. Moreover,
the actual optimization algorithms often do not converge even for small
models (see Table~\ref{tab:conv}). It might therefore be used when the
design has only one factor but is probably not suited for experiments
with two (fixed) factors or more.

\subsection*{The zero-correlation parameter correlation structure (ZCP)}\label{chap:zcp}

The ZCP structure \citep{bates_parsimonious_2015} also includes all the
random effects associated with stimuli and with participants. Unlike the
MAX model, the ZCP model does not include correlations between the
random effects. This means that one variance parameter is estimated for
each effect, but no correlation between the random effects is assumed.
If a factor has three or more levels, there is a twist and the number of
variance parameters that are estimated for each random part will be
equal to the number of degree-of-freedom of the corresponding fixed
factor (or interaction of factors). Said differently, variance
parameters are attached to contrasts of the factor or interaction of
factors and not to the factors themselves.

For two factors \texttt{A1} and \texttt{A2} and the participant
identifier \texttt{PT}, one first transforms the factors \texttt{A1}
(e.g.~with 4 levels) and \texttt{A2} (e.g.~with 3 levels) into coding
variables \texttt{x1a}, \texttt{x1b} and \texttt{x1c}, respectively
\texttt{x2a} and \texttt{x2b} (more information about the necessity to
transform into coding variable is found in Appendix~\ref{ap:formula}).
Then the typical formula using the \texttt{lme4} package is, for ZCP:

\begin{Shaded}
\begin{Highlighting}[]
\KeywordTok{lmer}\NormalTok{(y }\OperatorTok{~}\StringTok{  }\NormalTok{[...] }\OperatorTok{+}\StringTok{ }\NormalTok{((x1a }\OperatorTok{+}\StringTok{ }\NormalTok{x1b }\OperatorTok{+}\StringTok{ }\NormalTok{x1c)}\OperatorTok{*}\NormalTok{(x2a }\OperatorTok{+}\StringTok{ }\NormalTok{x2b) }\OperatorTok{||}\StringTok{ }\NormalTok{PT)  [...] )}
\end{Highlighting}
\end{Shaded}

This correlation structure is relatively parsimonious when all factors
have exactly two levels, but the number of parameters will increase with
respect to the number of levels of the factors (see
Table~\ref{tab:param}). It has the drawback to be dependent on the
choice of the coding of the factors. This correlation structure can be
viewed as a workaround to force lme4 not to add correlations between
random effects, but that this workaround does not give the expected
results with factors that have 3 or more levels. Although, we do not
expect a huge difference in practice, the maximum likelihood and the
inference will depend on the choice of the coding variables (or
contrasts), even when they are forced to be orthonormal. In many
applications, the choice of the coding variable does not correspond to
any hypothesis and is therefore arbitrary. Moreover, this may be an
obstacle for the reproducibility of the data analysis because it will be
challenging to report the coding variables of all the factors and all
their interactions.

\subsection*{The correlation structure of the generalized ANOVA (gANOVA)}\label{chap:ganova}

Compared to the previous structures, we propose a new one called gANOVA
that we believe best generalize rANOVA to CRE-MEM and possess nice
properties. Let first assume a saturated model with all random effects,
the one associated with participants, with stimuli and with their
interaction. As in the experimental design literature, the covariance
structure of random effects is assumed to be minimal, i.e.~each random
effect is independent from the others, and spherical. A correlation
structure with spherical random effects will have random effects that
share the same variance for each level of the same factor. This means
that the number of parameters will not increase with respect to the
number of levels (see the lines gANOVA in Table~\ref{tab:param} for
models M1 vs M2). The model behind gANOVA is the same as RI-L suggested
in \citet{bates_fitting_2015}, as exemplified in
equation~\eqref{eq:mlm}, and the number of parameters is also identical
(compare the lines RI-L and gANOVA in Table~\ref{tab:param}). The
difference with RI-L is that some constraints are assumed on the random
effects. In equation~\eqref{eq:mlm}, those constraints are written:
\(\sum_{k} (\pi\psi)_{ik}=0\ \forall i\),
\(\sum_l (\pi\phi)_{il} =0 ~\forall ~i\),
\(\sum_{k} (\pi\psi\phi)_{ikl} =0~ \forall ~ i,l\), and
\(\sum_{l} (\pi\psi\phi)_{ikl}=0 ~ \forall ~ i,k\). For the algorithm,
those constraints are simply implemented by transforming the factors
into orthonormal coding variables and forcing them to share the same
variance parameter (within each factor).

It is not possible to use \texttt{lme4} to estimate the gANOVA model, as
it needs to satisfy both the constraints of equality of variances for
each level and to use coding variables for the random interactions.
However, a simple modification of the \texttt{lmer} function implemented
in the \texttt{gANOVA} package (
\url{https://github.com/jaromilfrossard/gANOVA}) allows us to perform
this optimization. For two factors \texttt{A1} and \texttt{A2} and the
participant identifier \texttt{PT}, the gANOVA is performed using the
\texttt{gANOVA} package and the formula:

\begin{Shaded}
\begin{Highlighting}[]
\KeywordTok{gANOVA}\NormalTok{(y }\OperatorTok{~}\StringTok{  }\NormalTok{[...]  }\OperatorTok{+}\StringTok{ }\NormalTok{(}\DecValTok{1} \OperatorTok{|}\StringTok{ }\NormalTok{PT }\OperatorTok{|}\StringTok{ }\NormalTok{A1}\OperatorTok{*}\NormalTok{A2) [...] )}
\end{Highlighting}
\end{Shaded}

The first justification for this correlation structure is that it is
much more in line with the tradition of experimental design. Indeed, the
above constraints are exactly as defined by
\citet{cornfield_average_1956} for ANOVA including one or several random
effects (see Appendix~\ref{ap:csranova}). Moreover, one of the first
tools to obtain \(p\)-values for balanced experiments where there is
crossing of random samples of participants and stimuli was the
quasi-\(F\) statistic \citep{winer_statistical_1962}. This statistic is
based on sums of squares. The quasi-\(F\) statistic follows an
approximate \(F\) distribution under some assumptions
\citep{clark_language-as-fixed-effect_1973}. These assumptions are
identical to the one made in rANOVA (independence and homoscedasticity
of the random effects). For balanced data, the model and the implied
correlation structure are the same for quasi-\(F\) and CRE-MEM based on
gANOVA (but the statistic, \(t\)-value and \(p\)-value are computed
differently) and we expect to obtain quite similar results (very close
\(p\)-values). However, gANOVA generalizes naturally to non-balanced
designs since it is a mixed effect model.

Secondly, the possible correlations between all responses assumed by
RI-L are only a (strict) subset of the ones with gANOVA. For some data
sets, both methods lead to the same variance-covariance matrix of the
response (but its decomposition into variances of random effects is
different). Which is similar to say that, in Equation~\ref{eq:mlm}, the
variances of all \(y_{imk}\) and covariances between any two responses
\(y_{imk}\) are the same for gANOVA and RI-L, but its decomposition into
variances of random effects \(\pi_{i},~ \dots, ~(\pi\omega\psi)_{imk}\)
is different, due to the sum-to-zero constraints in gANOVA. In lay
terms, without these constraints, higher order interaction random
effects put restrictions on the variance of the random effects of lower
interaction. Perhaps surprisingly, this implies that the possible
variances and covariances of the responses are \textit{reduced} in RI-L
(compared to gANOVA), and for some data sets the variance-covariance
matrix of the response will be different between the two methods. It
often implies a solution at the boundary of the domain of definition
(one or several variances set to zero) in RI-L during the optimization.
crucially, in those cases, RI-L and gANOVA do not share the same
solution and gANOVA has always a smaller deviance (and AIC, BIC, \ldots)
which suggests a better fit.

The equations below show the relationship between the variance
parameters of both parametrization for a model with one factor:

\begin{align*}
\sigma^2_{RIL;i} & = \sigma_{gANOVA;i}^2 - a \sigma_{gANOVA;F}^2\\
\sigma^2_{RIL;F} & = \sigma_{gANOVA;F}^2 \\
\sigma^2_{RIL;\epsilon} &= \sigma_{gANOVA;\epsilon}^2,
\end{align*}

where \(\sigma_{RIL;i}^2\) and \(\sigma_{gANOVA;i}^2\) are the variances
of the random intercepts for both parametrizations, \(\sigma_{RIL;F}^2\)
and \(\sigma_{gANOVA;F}^2\) are the variances of the random interactions
between participants and the factor, and obviously,
\(\sigma^2_{RIL;\epsilon}\) and \(\sigma_{gANOVA;\epsilon}^2\) are the
variances of the error term. \(a\) is positive constant that depends on
the number of levels of the factor: \(a=1-1/(\textrm{nb. levels})\). See
Appendix \ref{ap:c-uc} for the full derivation of this example.

Several comments can be made. First, one has to be aware that the
interpretation is different between the two covariance structures.
Second, if \(\sigma_{gANOVA;i}^2 - a \sigma_{gANOVA;F}^2\) is positive,
RI-L and gANOVA will produce the same variance-covariance of the
responses and the same deviance. The fit is exactly the same. However,
there will cases where this term is negative. In that case, RI-L cannot
attain the optimum and is forced to set a variance to zero (and to
adjust the two other ones), leading to a poorer fit compared to the
solution of gANOVA. This leads to the conclusion that gANOVA is strictly
better than RI-L. This effect is shown in the simulations of next
section.

Concerning now the comparison between gANOVA and ZCP, they are the same
model for designs that have factors with exactly two levels. But it is
not the case when at least one factor has 3 levels or more. With gANOVA,
the interactions of such factors with random effects are assumed to be
spherical which imposes the same variance parameters for each coding
variables of the factors. On the contrary, ZCP will have a new variance
parameter for each new coding variable and adding these new parameters
has two drawbacks. First, the number of parameters to estimate increases
which implies more fluctuating estimations (see Table~\ref{tab:param}).
Moreover, these new parameters are usually not dictated by theoretical
ground but more by the convenience of an existing \texttt{R} formula.
Secondly, the correlation structure, the maximum likelihood and the
inference depend on this arbitrary choice of the coding variable, even
when they are forced to be orthonormal. There are infinitely many groups
of coding variables that may be used for a single dataset and for each
of which ZCP will give a different \(p\)-value. This arbitrariness in a
model that is precisely design-driven is not desirable. On the other
hand, the sphericity assumption in gANOVA keeps one variance parameter
for all coding variables of a given factor (or interaction). This will
reduce the number of parameters and the arbitrariness of the coding of
factors as the results will be independent to the choice of the coding
of the factors.

Finally, the constraints used in gANOVA (almost) orthogonalize the
random effects (they would be orthogonal if considered as fixed effects)
such that the parameters have a small mutual influence in comparison
with RI-L. Figure~\ref{fig:curve} shows an example of the likelihood
within the space of the parameters. For the same data (one sampling
unit, one \(I_M\) factor and replications) and model, we see that the
two ridges defining the two profile likelihoods cross almost at
90\(^\circ\) at the optimum in the gANOVA case but is far more inclined
for RI-L. This suggests less dependency between the parameters and a
better optimization process for gANOVA. In higher dimension, it is known
that all optimization suffers from the curse of dimensionality, and the
better independence of gANOVA parameters is clearly an asset.

\subsection*{The correlation structure based on PCA (CS-PCA)}
\label{chap:rs-pca}

\citet{bates_parsimonious_2015} proposed a heuristic to find the
appropriate correlation structure. This correlation structure has a
data-driven approach and therefore changes even between two experiments
sharing the same design. To compare it to the previous methods, we
summarize the proposition of \citet{bates_parsimonious_2015} by a fully
defined algorithm in Algorithm~\ref{alg:randpca}. The general idea
behind this method is to use principal component analysis (PCA) on the
estimated maximal correlation structure (from the MAX model) to estimate
the real dimensionality of the random effects and therefore diminish the
number of parameters; by assuming that the true correlation structure is
of lower dimension than the one defined by the MAX model, one restricts
to a subspace in which it is hoped that most of the variability of the
random effects lives. For readers interested in full details, by
deleting random effects of the model (suppressing the higher-level
interactions first), one then matches the correlation structure to the
estimated dimensionality. Based on the new maximal dimensionality,
\citet{bates_parsimonious_2015} proposed to reduce the number of
parameters according to a test or goodness of fit; first one decides
whether to drop the covariances between random effects and one selects
the correlation structure based on a test, then one decides whether to
drop random effects one by one beginning with the higher interaction
levels. One stops this procedure when it does not improve the model
anymore. The selected correlation structure is then compared to a last
one by adding or subtracting the covariance between random effects.

This algorithm, fully displayed in appendix \ref{chap:pcaalgo}, will
choose a correlation structure that is a subset of the correlation
structure defined by MAX. However, it is possible to imagine new
algorithms that choose a correlation structure that is a subset of the
ZCP, RI-L or gANOVA correlation structures. Moreover, being based on a
MAX correlation structure, CS-PCA will have computational problems in
complex designs.

Note that even if the goal of the algorithm is to match the data, some
design-driven mechanisms persist, like reducing the higher interaction
levels first or, keeping the dimension of the random effects based on
the design disregarding the directions of the eigen vectors of the PCA.

It is not possible to produce meaningful theoretical comparison with the
other correlation structures discussed above. We will compare it through
simulations in the following section. However, the correlation structure
CS-PCA, being mainly data-driven, may seem to come from a different
family than the design-driven correlation structure RI-L, ZCP, MAX or
gANOVA. However, all correlation structures can be summarized as a
function or algorithm of the design and of the data. The main difference
is that the procedures RI-L, ZCP, MAX and gANOVA will mostly use
information about the design to select the correlation structure and
CS-PCA will in addition use information from the data. As recalled
several times, they will all be false, and the goal is to select the
most useful one.

\section*{Simulation study}\label{chap:simul}

\renewcommand{\baselinestretch}{1}
\begin{table}[tb]
\centering
\caption{Percentage of convergence error for all simulations ($N_{sim} = 4000$) under the null hypothesis. Results are split by rows according to the simulation settings based on (1)~the sample size for stimuli, (2)~the true correlation between random effects, (3)~the presence/absence of random effects associated with the participants:stimuli interaction and (4)~the size of the design. The columns represent the type of estimation: all 7 correlation structures are assumed with~(+) and without~(-) the interaction participants:stimuli. All simulations are with 18 participants. The dash~"-" indicates settings without simulations. MAX and to a lesser extent CS-PCA present problems of convergence.}
\label{tab:conv}

\begin{tabular}{llllllllllllllllll}
\toprule
\multicolumn{1}{c}{ } & \multicolumn{1}{c}{ } & \multicolumn{1}{c}{ } & \multicolumn{1}{c}{ } & \multicolumn{2}{c}{RI} & \multicolumn{2}{c}{RI-L} & \multicolumn{2}{c}{MAX} & \multicolumn{2}{c}{ZCP-sum} & \multicolumn{2}{c}{ZCP-poly} & \multicolumn{2}{c}{gANOVA} & \multicolumn{2}{c}{CS-PCA} \\
\cmidrule(l{3pt}r{3pt}){5-6} \cmidrule(l{3pt}r{3pt}){7-8} \cmidrule(l{3pt}r{3pt}){9-10} \cmidrule(l{3pt}r{3pt}){11-12} \cmidrule(l{3pt}r{3pt}){13-14} \cmidrule(l{3pt}r{3pt}){15-16} \cmidrule(l{3pt}r{3pt}){17-18}
  &   &   &   & - & + & - & + & - & + & - & + & - & + & - & + & - & +\\
\rowcolor{gray!6}
\midrule
 &  &  & M1 & 0.0 & 0.0 & 0.0 & 0.0 & 0.1 & 0.1 & 0.1 & 0.0 & 0.0 & 0.0 & 0.0 & 0.0 & 0.0 & 0.0\\

\rowcolor{gray!6}
 &  &  & M2 & 0.0 & 0.0 & 0.0 & 0.0 & 12.7 & 15.5 & 0.0 & 0.0 & 0.0 & 0.0 & 0.0 & 0.0 & 0.3 & 0.7\\

\rowcolor{gray!6}
 &  & \multirow{-3}{*}{\raggedright\arraybackslash no PT:SM} & M4 & 0.0 & 0.0 & 0.0 & 0.0 & - & - & 0.0 & 0.0 & 0.0 & 0.0 & 0.0 & 0.0 & - & -\\

 &  &  & M1 & 0.0 & 0.0 & 0.0 & 0.0 & 0.1 & 0.1 & 0.0 & 0.0 & 0.0 & 0.0 & 0.0 & 0.0 & 0.0 & 0.0\\

 &  &  & M2 & 0.0 & 0.0 & 0.0 & 0.0 & 13.4 & 9.4 & 0.0 & 0.0 & 0.0 & 0.0 & 0.0 & 0.0 & 0.5 & 0.4\\

 & \multirow{-6}{*}{\raggedright\arraybackslash \rotatebox{90}{spheric.}} & \multirow{-3}{*}{\raggedright\arraybackslash PT:SM} & M4 & 0.0 & 0.0 & 0.0 & 0.0 & - & - & 0.0 & 0.0 & 0.0 & 0.0 & 0.0 & 0.0 & - & -\\

\rowcolor{gray!6}
 &  &  & M1 & 0.0 & 0.0 & 0.0 & 0.0 & 4.4 & 8.8 & 0.0 & 0.0 & 0.0 & 0.0 & 0.0 & 0.0 & 0.0 & 0.0\\

\rowcolor{gray!6}
 &  &  & M2 & 0.0 & 0.0 & 0.0 & 0.0 & 29.7 & 35.0 & 0.0 & 0.0 & 0.0 & 0.0 & 0.0 & 0.0 & 6.1 & 8.2\\

\rowcolor{gray!6}
 &  & \multirow{-3}{*}{\raggedright\arraybackslash no PT:SM} & M4 & 0.0 & 0.0 & 0.0 & 0.0 & - & - & 0.0 & 0.0 & 0.0 & 0.0 & 0.0 & 0.0 & - & -\\

 &  &  & M1 & 0.0 & 0.0 & 0.0 & 0.0 & 8.8 & 9.6 & 0.0 & 0.0 & 0.0 & 0.0 & 0.0 & 0.0 & 0.0 & 0.0\\

 &  &  & M2 & 0.0 & 0.0 & 0.0 & 0.0 & 39.1 & 34.0 & 0.0 & 0.1 & 0.0 & 0.0 & 0.0 & 0.0 & 10.0 & 9.3\\

\multirow{-12}{*}{\raggedright\arraybackslash 18} & \multirow{-6}{*}{\raggedright\arraybackslash \rotatebox{90}{corr.}} & \multirow{-3}{*}{\raggedright\arraybackslash PT:SM} & M4 & 0.0 & 0.0 & 0.0 & 0.0 & - & - & 0.0 & 0.0 & 0.0 & 0.0 & 0.0 & 0.0 & - & -\\
\cmidrule{1-18}
\rowcolor{gray!6}
 &  &  & M1 & 0.0 & 0.0 & 0.0 & 0.0 & 0.0 & 0.0 & 0.0 & 0.0 & 0.0 & 0.0 & 0.0 & 0.0 & 0.0 & 0.0\\

\rowcolor{gray!6}
 &  & \multirow{-2}{*}{\raggedright\arraybackslash no PT:SM} & M2 & 0.0 & 0.0 & 0.0 & 0.0 & 15.8 & 19.2 & 0.1 & 0.0 & 0.0 & 0.0 & 0.0 & 0.0 & 0.9 & 2.9\\

 &  &  & M1 & 0.0 & 0.0 & 0.0 & 0.0 & 0.0 & 0.0 & 0.0 & 0.0 & 0.0 & 0.0 & 0.0 & 0.0 & 0.0 & 0.0\\

 & \multirow{-4}{*}{\raggedright\arraybackslash \rotatebox{90}{spheric.}} & \multirow{-2}{*}{\raggedright\arraybackslash PT:SM} & M2 & 0.0 & 0.0 & 0.0 & 0.1 & 14.9 & 15.4 & 0.1 & 0.0 & 0.0 & 0.0 & 0.0 & 0.0 & 2.4 & 2.5\\

\rowcolor{gray!6}
 &  &  & M1 & 0.0 & 0.0 & 0.0 & 0.0 & 3.7 & 11.8 & 0.0 & 0.0 & 0.0 & 0.0 & 0.0 & 0.0 & 0.0 & 0.0\\

\rowcolor{gray!6}
 &  & \multirow{-2}{*}{\raggedright\arraybackslash no PT:SM} & M2 & 0.0 & 0.0 & 0.0 & 0.0 & 30.0 & 41.5 & 0.1 & 0.0 & 0.0 & 0.0 & 0.0 & 0.0 & 7.2 & 14.2\\

 &  &  & M1 & 0.0 & 0.0 & 0.0 & 0.0 & 10.4 & 10.2 & 0.0 & 0.0 & 0.0 & 0.0 & 0.0 & 0.0 & 0.0 & 0.0\\

\multirow{-8}{*}{\raggedright\arraybackslash 36} & \multirow{-4}{*}{\raggedright\arraybackslash \rotatebox{90}{corr.}} & \multirow{-2}{*}{\raggedright\arraybackslash PT:SM} & M2 & 0.0 & 0.0 & 0.0 & 0.0 & 41.0 & 34.6 & 0.0 & 0.1 & 0.0 & 0.0 & 0.0 & 0.0 & 16.4 & 14.8\\
\bottomrule
\end{tabular}
\end{table}
\renewcommand{\baselinestretch}{1.7}

\renewcommand{\baselinestretch}{1}
\renewcommand{\arraystretch}{.73}
\begin{table}[!htb]
\centering
\caption{Type I error rate for three factors of the common model (M2 of in Table \ref{tab:param}). Correct methods should be close the nominal level $\alpha=.050$. The first column indicates the true correlation between random effects (homoscedastic or correlated). The second one indicates if the true model is generated with the interaction participants-stimuli. The third column indicates if model is estimated assuming the interaction participants:stimuli~(+) or not~(-). The RI and CS-PCA correlation structures show huge deviations from the nominal level. Confidence intervals are computed using \cite{agresti_approximate_1998}. Bold font corresponds to nominal level (5\%) within the confidence interval, red font corresponds to confidence interval above the nominal level and italic font corresponds to confidence interval below the nominal level. }
\label{tab:type1:common}

\begin{tabular}{llllllllll}
\toprule
  &   &   & RI & RI-L & MAX & ZCP-sum & ZCP-poly & gANOVA & CS-PCA\\
\midrule
\addlinespace[0.3em]
\multicolumn{10}{l}{\textbf{As}}\\
\hspace{1em} &  &  & \bgroup\fontsize{11}{13}\selectfont \textcolor{red}{.136}\egroup{} & \bgroup\fontsize{11}{13}\selectfont \textbf{.051}\egroup{} & \bgroup\fontsize{11}{13}\selectfont \textbf{.051}\egroup{} & \bgroup\fontsize{11}{13}\selectfont \textbf{.049}\egroup{} & \bgroup\fontsize{11}{13}\selectfont \textbf{.052}\egroup{} & \bgroup\fontsize{11}{13}\selectfont \textbf{.051}\egroup{} & \bgroup\fontsize{11}{13}\selectfont \textbf{.052}\egroup{}\\

\hspace{1em} &  & \multirow{-2}{*}{\raggedright\arraybackslash \bgroup\fontsize{11}{13}\selectfont -\egroup{}} & \bgroup\fontsize{7}{9}\selectfont \textcolor{red}{[.126;.148]}\egroup{} & \bgroup\fontsize{7}{9}\selectfont \textbf{[.044;.058]}\egroup{} & \bgroup\fontsize{7}{9}\selectfont \textbf{[.044;.059]}\egroup{} & \bgroup\fontsize{7}{9}\selectfont \textbf{[.043;.056]}\egroup{} & \bgroup\fontsize{7}{9}\selectfont \textbf{[.045;.059]}\egroup{} & \bgroup\fontsize{7}{9}\selectfont \textbf{[.044;.058]}\egroup{} & \bgroup\fontsize{7}{9}\selectfont \textbf{[.045;.059]}\egroup{}\\

\hspace{1em} &  &  & \bgroup\fontsize{11}{13}\selectfont \textcolor{red}{.136}\egroup{} & \bgroup\fontsize{11}{13}\selectfont \textbf{.053}\egroup{} & \bgroup\fontsize{11}{13}\selectfont \textbf{.053}\egroup{} & \bgroup\fontsize{11}{13}\selectfont \textbf{.053}\egroup{} & \bgroup\fontsize{11}{13}\selectfont \textbf{.054}\egroup{} & \bgroup\fontsize{11}{13}\selectfont \textbf{.053}\egroup{} & \bgroup\fontsize{11}{13}\selectfont \textbf{.055}\egroup{}\\

\hspace{1em} & \multirow{-4}{*}{\raggedright\arraybackslash \bgroup\fontsize{11}{13}\selectfont no PT:SM\egroup{}} & \multirow{-2}{*}{\raggedright\arraybackslash \bgroup\fontsize{11}{13}\selectfont +\egroup{}} & \bgroup\fontsize{7}{9}\selectfont \textcolor{red}{[.126;.148]}\egroup{} & \bgroup\fontsize{7}{9}\selectfont \textbf{[.047;.061]}\egroup{} & \bgroup\fontsize{7}{9}\selectfont \textbf{[.046;.061]}\egroup{} & \bgroup\fontsize{7}{9}\selectfont \textbf{[.046;.060]}\egroup{} & \bgroup\fontsize{7}{9}\selectfont \textbf{[.048;.062]}\egroup{} & \bgroup\fontsize{7}{9}\selectfont \textbf{[.047;.061]}\egroup{} & \bgroup\fontsize{7}{9}\selectfont \textbf{[.048;.062]}\egroup{}\\

\hspace{1em}\hspace{1em} &  &  & \bgroup\fontsize{11}{13}\selectfont \textcolor{red}{.139}\egroup{} & \bgroup\fontsize{11}{13}\selectfont \textbf{.054}\egroup{} & \bgroup\fontsize{11}{13}\selectfont \textcolor{red}{.059}\egroup{} & \bgroup\fontsize{11}{13}\selectfont \textbf{.053}\egroup{} & \bgroup\fontsize{11}{13}\selectfont \textcolor{red}{.058}\egroup{} & \bgroup\fontsize{11}{13}\selectfont \textbf{.054}\egroup{} & \bgroup\fontsize{11}{13}\selectfont \textcolor{red}{.058}\egroup{}\\

\hspace{1em} &  & \multirow{-2}{*}{\raggedright\arraybackslash \bgroup\fontsize{11}{13}\selectfont -\egroup{}} & \bgroup\fontsize{7}{9}\selectfont \textcolor{red}{[.129;.150]}\egroup{} & \bgroup\fontsize{7}{9}\selectfont \textbf{[.047;.061]}\egroup{} & \bgroup\fontsize{7}{9}\selectfont \textcolor{red}{[.052;.068]}\egroup{} & \bgroup\fontsize{7}{9}\selectfont \textbf{[.046;.060]}\egroup{} & \bgroup\fontsize{7}{9}\selectfont \textcolor{red}{[.051;.065]}\egroup{} & \bgroup\fontsize{7}{9}\selectfont \textbf{[.047;.061]}\egroup{} & \bgroup\fontsize{7}{9}\selectfont \textcolor{red}{[.051;.066]}\egroup{}\\

 &  &  & \bgroup\fontsize{11}{13}\selectfont \textcolor{red}{.139}\egroup{} & \bgroup\fontsize{11}{13}\selectfont \textbf{.054}\egroup{} & \bgroup\fontsize{11}{13}\selectfont \textcolor{red}{.059}\egroup{} & \bgroup\fontsize{11}{13}\selectfont \textbf{.053}\egroup{} & \bgroup\fontsize{11}{13}\selectfont \textcolor{red}{.058}\egroup{} & \bgroup\fontsize{11}{13}\selectfont \textbf{.054}\egroup{} & \bgroup\fontsize{11}{13}\selectfont \textcolor{red}{.058}\egroup{}\\

\hspace{1em}\multirow{-8}{*}{\raggedright\arraybackslash \rotatebox{90}{\bgroup\fontsize{11}{13}\selectfont spheric.\egroup{}}} & \multirow{-4}{*}{\raggedright\arraybackslash \bgroup\fontsize{11}{13}\selectfont PT:SM\egroup{}} & \multirow{-2}{*}{\raggedright\arraybackslash \bgroup\fontsize{11}{13}\selectfont +\egroup{}} & \bgroup\fontsize{7}{9}\selectfont \textcolor{red}{[.129;.150]}\egroup{} & \bgroup\fontsize{7}{9}\selectfont \textbf{[.047;.061]}\egroup{} & \bgroup\fontsize{7}{9}\selectfont \textcolor{red}{[.051;.067]}\egroup{} & \bgroup\fontsize{7}{9}\selectfont \textbf{[.047;.060]}\egroup{} & \bgroup\fontsize{7}{9}\selectfont \textcolor{red}{[.051;.065]}\egroup{} & \bgroup\fontsize{7}{9}\selectfont \textbf{[.047;.061]}\egroup{} & \bgroup\fontsize{7}{9}\selectfont \textcolor{red}{[.051;.066]}\egroup{}\\
\cmidrule{1-10}
\hspace{1em} &  &  & \bgroup\fontsize{11}{13}\selectfont \textcolor{red}{.134}\egroup{} & \bgroup\fontsize{11}{13}\selectfont \textbf{.050}\egroup{} & \bgroup\fontsize{11}{13}\selectfont \textbf{.056}\egroup{} & \bgroup\fontsize{11}{13}\selectfont \textbf{.046}\egroup{} & \bgroup\fontsize{11}{13}\selectfont \textbf{.050}\egroup{} & \bgroup\fontsize{11}{13}\selectfont \textbf{.050}\egroup{} & \bgroup\fontsize{11}{13}\selectfont \textbf{.050}\egroup{}\\

\hspace{1em} &  & \multirow{-2}{*}{\raggedright\arraybackslash \bgroup\fontsize{11}{13}\selectfont -\egroup{}} & \bgroup\fontsize{7}{9}\selectfont \textcolor{red}{[.124;.145]}\egroup{} & \bgroup\fontsize{7}{9}\selectfont \textbf{[.044;.058]}\egroup{} & \bgroup\fontsize{7}{9}\selectfont \textbf{[.048;.065]}\egroup{} & \bgroup\fontsize{7}{9}\selectfont \textbf{[.040;.053]}\egroup{} & \bgroup\fontsize{7}{9}\selectfont \textbf{[.043;.057]}\egroup{} & \bgroup\fontsize{7}{9}\selectfont \textbf{[.044;.058]}\egroup{} & \bgroup\fontsize{7}{9}\selectfont \textbf{[.043;.057]}\egroup{}\\

\hspace{1em} &  &  & \bgroup\fontsize{11}{13}\selectfont \textcolor{red}{.134}\egroup{} & \bgroup\fontsize{11}{13}\selectfont \textbf{.052}\egroup{} & \bgroup\fontsize{11}{13}\selectfont \textbf{.050}\egroup{} & \bgroup\fontsize{11}{13}\selectfont \textbf{.048}\egroup{} & \bgroup\fontsize{11}{13}\selectfont \textbf{.052}\egroup{} & \bgroup\fontsize{11}{13}\selectfont \textbf{.052}\egroup{} & \bgroup\fontsize{11}{13}\selectfont \textbf{.052}\egroup{}\\

\hspace{1em} & \multirow{-4}{*}{\raggedright\arraybackslash \bgroup\fontsize{11}{13}\selectfont no PT:SM\egroup{}} & \multirow{-2}{*}{\raggedright\arraybackslash \bgroup\fontsize{11}{13}\selectfont +\egroup{}} & \bgroup\fontsize{7}{9}\selectfont \textcolor{red}{[.124;.145]}\egroup{} & \bgroup\fontsize{7}{9}\selectfont \textbf{[.046;.060]}\egroup{} & \bgroup\fontsize{7}{9}\selectfont \textbf{[.042;.060]}\egroup{} & \bgroup\fontsize{7}{9}\selectfont \textbf{[.042;.055]}\egroup{} & \bgroup\fontsize{7}{9}\selectfont \textbf{[.046;.060]}\egroup{} & \bgroup\fontsize{7}{9}\selectfont \textbf{[.046;.060]}\egroup{} & \bgroup\fontsize{7}{9}\selectfont \textbf{[.045;.060]}\egroup{}\\

\hspace{1em} &  &  & \bgroup\fontsize{11}{13}\selectfont \textcolor{red}{.140}\egroup{} & \bgroup\fontsize{11}{13}\selectfont \textbf{.051}\egroup{} & \bgroup\fontsize{11}{13}\selectfont \textbf{.055}\egroup{} & \bgroup\fontsize{11}{13}\selectfont \textbf{.044}\egroup{} & \bgroup\fontsize{11}{13}\selectfont \textbf{.053}\egroup{} & \bgroup\fontsize{11}{13}\selectfont \textbf{.051}\egroup{} & \bgroup\fontsize{11}{13}\selectfont \textbf{.056}\egroup{}\\

\hspace{1em} &  & \multirow{-2}{*}{\raggedright\arraybackslash \bgroup\fontsize{11}{13}\selectfont -\egroup{}} & \bgroup\fontsize{7}{9}\selectfont \textcolor{red}{[.129;.151]}\egroup{} & \bgroup\fontsize{7}{9}\selectfont \textbf{[.044;.058]}\egroup{} & \bgroup\fontsize{7}{9}\selectfont \textbf{[.047;.065]}\egroup{} & \bgroup\fontsize{7}{9}\selectfont \textbf{[.039;.051]}\egroup{} & \bgroup\fontsize{7}{9}\selectfont \textbf{[.046;.060]}\egroup{} & \bgroup\fontsize{7}{9}\selectfont \textbf{[.044;.058]}\egroup{} & \bgroup\fontsize{7}{9}\selectfont \textbf{[.049;.065]}\egroup{}\\

\hspace{1em} &  &  & \bgroup\fontsize{11}{13}\selectfont \textcolor{red}{.140}\egroup{} & \bgroup\fontsize{11}{13}\selectfont \textbf{.051}\egroup{} & \bgroup\fontsize{11}{13}\selectfont \textbf{.051}\egroup{} & \bgroup\fontsize{11}{13}\selectfont \textbf{.045}\egroup{} & \bgroup\fontsize{11}{13}\selectfont \textbf{.053}\egroup{} & \bgroup\fontsize{11}{13}\selectfont \textbf{.051}\egroup{} & \bgroup\fontsize{11}{13}\selectfont \textcolor{red}{.057}\egroup{}\\

\hspace{1em}\multirow{-8}{*}{\raggedright\arraybackslash \rotatebox{90}{\bgroup\fontsize{11}{13}\selectfont corr.\egroup{}}} & \multirow{-4}{*}{\raggedright\arraybackslash \bgroup\fontsize{11}{13}\selectfont PT:SM\egroup{}} & \multirow{-2}{*}{\raggedright\arraybackslash \bgroup\fontsize{11}{13}\selectfont +\egroup{}} & \bgroup\fontsize{7}{9}\selectfont \textcolor{red}{[.129;.151]}\egroup{} & \bgroup\fontsize{7}{9}\selectfont \textbf{[.044;.058]}\egroup{} & \bgroup\fontsize{7}{9}\selectfont \textbf{[.043;.060]}\egroup{} & \bgroup\fontsize{7}{9}\selectfont \textbf{[.039;.052]}\egroup{} & \bgroup\fontsize{7}{9}\selectfont \textbf{[.046;.060]}\egroup{} & \bgroup\fontsize{7}{9}\selectfont \textbf{[.044;.058]}\egroup{} & \bgroup\fontsize{7}{9}\selectfont \textcolor{red}{[.050;.066]}\egroup{}\\
\cmidrule{1-10}
\addlinespace[0.3em]
\multicolumn{10}{l}{\textbf{Ap:Am}}\\
\hspace{1em} &  &  & \bgroup\fontsize{11}{13}\selectfont \textcolor{red}{.723}\egroup{} & \bgroup\fontsize{11}{13}\selectfont \textcolor{red}{.059}\egroup{} & \bgroup\fontsize{11}{13}\selectfont \textcolor{red}{.065}\egroup{} & \bgroup\fontsize{11}{13}\selectfont \textcolor{red}{.077}\egroup{} & \bgroup\fontsize{11}{13}\selectfont \textcolor{red}{.069}\egroup{} & \bgroup\fontsize{11}{13}\selectfont \textcolor{red}{.058}\egroup{} & \bgroup\fontsize{11}{13}\selectfont \textcolor{red}{.077}\egroup{}\\

\hspace{1em} &  & \multirow{-2}{*}{\raggedright\arraybackslash \bgroup\fontsize{11}{13}\selectfont -\egroup{}} & \bgroup\fontsize{7}{9}\selectfont \textcolor{red}{[.709;.737]}\egroup{} & \bgroup\fontsize{7}{9}\selectfont \textcolor{red}{[.052;.066]}\egroup{} & \bgroup\fontsize{7}{9}\selectfont \textcolor{red}{[.057;.074]}\egroup{} & \bgroup\fontsize{7}{9}\selectfont \textcolor{red}{[.069;.086]}\egroup{} & \bgroup\fontsize{7}{9}\selectfont \textcolor{red}{[.061;.077]}\egroup{} & \bgroup\fontsize{7}{9}\selectfont \textcolor{red}{[.051;.066]}\egroup{} & \bgroup\fontsize{7}{9}\selectfont \textcolor{red}{[.069;.085]}\egroup{}\\

\hspace{1em} &  &  & \bgroup\fontsize{11}{13}\selectfont \textcolor{red}{.809}\egroup{} & \bgroup\fontsize{11}{13}\selectfont \textbf{.055}\egroup{} & \bgroup\fontsize{11}{13}\selectfont \textcolor{red}{.064}\egroup{} & \bgroup\fontsize{11}{13}\selectfont \textcolor{red}{.075}\egroup{} & \bgroup\fontsize{11}{13}\selectfont \textcolor{red}{.063}\egroup{} & \bgroup\fontsize{11}{13}\selectfont \textbf{.054}\egroup{} & \bgroup\fontsize{11}{13}\selectfont \textcolor{red}{.068}\egroup{}\\

\hspace{1em} & \multirow{-4}{*}{\raggedright\arraybackslash \bgroup\fontsize{11}{13}\selectfont no PT:SM\egroup{}} & \multirow{-2}{*}{\raggedright\arraybackslash \bgroup\fontsize{11}{13}\selectfont +\egroup{}} & \bgroup\fontsize{7}{9}\selectfont \textcolor{red}{[.797;.821]}\egroup{} & \bgroup\fontsize{7}{9}\selectfont \textbf{[.048;.062]}\egroup{} & \bgroup\fontsize{7}{9}\selectfont \textcolor{red}{[.056;.073]}\egroup{} & \bgroup\fontsize{7}{9}\selectfont \textcolor{red}{[.067;.083]}\egroup{} & \bgroup\fontsize{7}{9}\selectfont \textcolor{red}{[.056;.071]}\egroup{} & \bgroup\fontsize{7}{9}\selectfont \textbf{[.048;.062]}\egroup{} & \bgroup\fontsize{7}{9}\selectfont \textcolor{red}{[.060;.076]}\egroup{}\\

\hspace{1em} &  &  & \bgroup\fontsize{11}{13}\selectfont \textcolor{red}{.782}\egroup{} & \bgroup\fontsize{11}{13}\selectfont \textbf{.054}\egroup{} & \bgroup\fontsize{11}{13}\selectfont \textcolor{red}{.059}\egroup{} & \bgroup\fontsize{11}{13}\selectfont \textcolor{red}{.072}\egroup{} & \bgroup\fontsize{11}{13}\selectfont \textcolor{red}{.064}\egroup{} & \bgroup\fontsize{11}{13}\selectfont \textbf{.054}\egroup{} & \bgroup\fontsize{11}{13}\selectfont \textcolor{red}{.066}\egroup{}\\

\hspace{1em} &  & \multirow{-2}{*}{\raggedright\arraybackslash \bgroup\fontsize{11}{13}\selectfont -\egroup{}} & \bgroup\fontsize{7}{9}\selectfont \textcolor{red}{[.769;.795]}\egroup{} & \bgroup\fontsize{7}{9}\selectfont \textbf{[.048;.062]}\egroup{} & \bgroup\fontsize{7}{9}\selectfont \textcolor{red}{[.052;.068]}\egroup{} & \bgroup\fontsize{7}{9}\selectfont \textcolor{red}{[.064;.081]}\egroup{} & \bgroup\fontsize{7}{9}\selectfont \textcolor{red}{[.057;.072]}\egroup{} & \bgroup\fontsize{7}{9}\selectfont \textbf{[.048;.062]}\egroup{} & \bgroup\fontsize{7}{9}\selectfont \textcolor{red}{[.059;.074]}\egroup{}\\

\hspace{1em} &  &  & \bgroup\fontsize{11}{13}\selectfont \textcolor{red}{.824}\egroup{} & \bgroup\fontsize{11}{13}\selectfont \textbf{.054}\egroup{} & \bgroup\fontsize{11}{13}\selectfont \textcolor{red}{.059}\egroup{} & \bgroup\fontsize{11}{13}\selectfont \textcolor{red}{.072}\egroup{} & \bgroup\fontsize{11}{13}\selectfont \textcolor{red}{.064}\egroup{} & \bgroup\fontsize{11}{13}\selectfont \textbf{.054}\egroup{} & \bgroup\fontsize{11}{13}\selectfont \textcolor{red}{.067}\egroup{}\\

\hspace{1em}\multirow{-8}{*}{\raggedright\arraybackslash \rotatebox{90}{\bgroup\fontsize{11}{13}\selectfont spheric.\egroup{}}} & \multirow{-4}{*}{\raggedright\arraybackslash \bgroup\fontsize{11}{13}\selectfont PT:SM\egroup{}} & \multirow{-2}{*}{\raggedright\arraybackslash \bgroup\fontsize{11}{13}\selectfont +\egroup{}} & \bgroup\fontsize{7}{9}\selectfont \textcolor{red}{[.812;.836]}\egroup{} & \bgroup\fontsize{7}{9}\selectfont \textbf{[.048;.062]}\egroup{} & \bgroup\fontsize{7}{9}\selectfont \textcolor{red}{[.052;.068]}\egroup{} & \bgroup\fontsize{7}{9}\selectfont \textcolor{red}{[.064;.080]}\egroup{} & \bgroup\fontsize{7}{9}\selectfont \textcolor{red}{[.057;.072]}\egroup{} & \bgroup\fontsize{7}{9}\selectfont \textbf{[.048;.062]}\egroup{} & \bgroup\fontsize{7}{9}\selectfont \textcolor{red}{[.060;.076]}\egroup{}\\
\cmidrule{1-10}
\hspace{1em} &  &  & \bgroup\fontsize{11}{13}\selectfont \textcolor{red}{.698}\egroup{} & \bgroup\fontsize{11}{13}\selectfont \textcolor{red}{.059}\egroup{} & \bgroup\fontsize{11}{13}\selectfont \textbf{.055}\egroup{} & \bgroup\fontsize{11}{13}\selectfont \textcolor{red}{.075}\egroup{} & \bgroup\fontsize{11}{13}\selectfont \textcolor{red}{.067}\egroup{} & \bgroup\fontsize{11}{13}\selectfont \textcolor{red}{.059}\egroup{} & \bgroup\fontsize{11}{13}\selectfont \textcolor{red}{.077}\egroup{}\\

\hspace{1em} &  & \multirow{-2}{*}{\raggedright\arraybackslash \bgroup\fontsize{11}{13}\selectfont -\egroup{}} & \bgroup\fontsize{7}{9}\selectfont \textcolor{red}{[.684;.712]}\egroup{} & \bgroup\fontsize{7}{9}\selectfont \textcolor{red}{[.052;.067]}\egroup{} & \bgroup\fontsize{7}{9}\selectfont \textbf{[.047;.064]}\egroup{} & \bgroup\fontsize{7}{9}\selectfont \textcolor{red}{[.068;.084]}\egroup{} & \bgroup\fontsize{7}{9}\selectfont \textcolor{red}{[.060;.075]}\egroup{} & \bgroup\fontsize{7}{9}\selectfont \textcolor{red}{[.052;.067]}\egroup{} & \bgroup\fontsize{7}{9}\selectfont \textcolor{red}{[.068;.086]}\egroup{}\\

\hspace{1em} &  &  & \bgroup\fontsize{11}{13}\selectfont \textcolor{red}{.794}\egroup{} & \bgroup\fontsize{11}{13}\selectfont \textbf{.053}\egroup{} & \bgroup\fontsize{11}{13}\selectfont \textbf{.053}\egroup{} & \bgroup\fontsize{11}{13}\selectfont \textcolor{red}{.072}\egroup{} & \bgroup\fontsize{11}{13}\selectfont \textcolor{red}{.063}\egroup{} & \bgroup\fontsize{11}{13}\selectfont \textbf{.053}\egroup{} & \bgroup\fontsize{11}{13}\selectfont \textcolor{red}{.075}\egroup{}\\

\hspace{1em} & \multirow{-4}{*}{\raggedright\arraybackslash \bgroup\fontsize{11}{13}\selectfont no PT:SM\egroup{}} & \multirow{-2}{*}{\raggedright\arraybackslash \bgroup\fontsize{11}{13}\selectfont +\egroup{}} & \bgroup\fontsize{7}{9}\selectfont \textcolor{red}{[.781;.806]}\egroup{} & \bgroup\fontsize{7}{9}\selectfont \textbf{[.047;.061]}\egroup{} & \bgroup\fontsize{7}{9}\selectfont \textbf{[.044;.062]}\egroup{} & \bgroup\fontsize{7}{9}\selectfont \textcolor{red}{[.064;.080]}\egroup{} & \bgroup\fontsize{7}{9}\selectfont \textcolor{red}{[.056;.071]}\egroup{} & \bgroup\fontsize{7}{9}\selectfont \textbf{[.047;.061]}\egroup{} & \bgroup\fontsize{7}{9}\selectfont \textcolor{red}{[.067;.084]}\egroup{}\\

\hspace{1em} &  &  & \bgroup\fontsize{11}{13}\selectfont \textcolor{red}{.770}\egroup{} & \bgroup\fontsize{11}{13}\selectfont \textbf{.054}\egroup{} & \bgroup\fontsize{11}{13}\selectfont \textbf{.058}\egroup{} & \bgroup\fontsize{11}{13}\selectfont \textcolor{red}{.070}\egroup{} & \bgroup\fontsize{11}{13}\selectfont \textcolor{red}{.064}\egroup{} & \bgroup\fontsize{11}{13}\selectfont \textbf{.054}\egroup{} & \bgroup\fontsize{11}{13}\selectfont \textcolor{red}{.073}\egroup{}\\

\hspace{1em} &  & \multirow{-2}{*}{\raggedright\arraybackslash \bgroup\fontsize{11}{13}\selectfont -\egroup{}} & \bgroup\fontsize{7}{9}\selectfont \textcolor{red}{[.757;.783]}\egroup{} & \bgroup\fontsize{7}{9}\selectfont \textbf{[.048;.062]}\egroup{} & \bgroup\fontsize{7}{9}\selectfont \textbf{[.049;.068]}\egroup{} & \bgroup\fontsize{7}{9}\selectfont \textcolor{red}{[.062;.078]}\egroup{} & \bgroup\fontsize{7}{9}\selectfont \textcolor{red}{[.057;.072]}\egroup{} & \bgroup\fontsize{7}{9}\selectfont \textbf{[.048;.062]}\egroup{} & \bgroup\fontsize{7}{9}\selectfont \textcolor{red}{[.065;.083]}\egroup{}\\

\hspace{1em} &  &  & \bgroup\fontsize{11}{13}\selectfont \textcolor{red}{.809}\egroup{} & \bgroup\fontsize{11}{13}\selectfont \textbf{.054}\egroup{} & \bgroup\fontsize{11}{13}\selectfont \textbf{.058}\egroup{} & \bgroup\fontsize{11}{13}\selectfont \textcolor{red}{.069}\egroup{} & \bgroup\fontsize{11}{13}\selectfont \textcolor{red}{.064}\egroup{} & \bgroup\fontsize{11}{13}\selectfont \textbf{.054}\egroup{} & \bgroup\fontsize{11}{13}\selectfont \textcolor{red}{.073}\egroup{}\\

\hspace{1em}\multirow{-8}{*}{\raggedright\arraybackslash \rotatebox{90}{\bgroup\fontsize{11}{13}\selectfont corr.\egroup{}}} & \multirow{-4}{*}{\raggedright\arraybackslash \bgroup\fontsize{11}{13}\selectfont PT:SM\egroup{}} & \multirow{-2}{*}{\raggedright\arraybackslash \bgroup\fontsize{11}{13}\selectfont +\egroup{}} & \bgroup\fontsize{7}{9}\selectfont \textcolor{red}{[.797;.821]}\egroup{} & \bgroup\fontsize{7}{9}\selectfont \textbf{[.047;.061]}\egroup{} & \bgroup\fontsize{7}{9}\selectfont \textbf{[.049;.067]}\egroup{} & \bgroup\fontsize{7}{9}\selectfont \textcolor{red}{[.062;.078]}\egroup{} & \bgroup\fontsize{7}{9}\selectfont \textcolor{red}{[.056;.072]}\egroup{} & \bgroup\fontsize{7}{9}\selectfont \textbf{[.047;.061]}\egroup{} & \bgroup\fontsize{7}{9}\selectfont \textcolor{red}{[.065;.083]}\egroup{}\\
\cmidrule{1-10}
\addlinespace[0.3em]
\multicolumn{10}{l}{\textbf{Ap:As:Am}}\\
\hspace{1em} &  &  & \bgroup\fontsize{11}{13}\selectfont \textcolor{red}{.250}\egroup{} & \bgroup\fontsize{11}{13}\selectfont \textcolor{red}{.075}\egroup{} & \bgroup\fontsize{11}{13}\selectfont \em{.036}\egroup{} & \bgroup\fontsize{11}{13}\selectfont \textcolor{red}{.103}\egroup{} & \bgroup\fontsize{11}{13}\selectfont \textcolor{red}{.087}\egroup{} & \bgroup\fontsize{11}{13}\selectfont \textcolor{red}{.075}\egroup{} & \bgroup\fontsize{11}{13}\selectfont \textcolor{red}{.409}\egroup{}\\

\hspace{1em} &  & \multirow{-2}{*}{\raggedright\arraybackslash \bgroup\fontsize{11}{13}\selectfont -\egroup{}} & \bgroup\fontsize{7}{9}\selectfont \textcolor{red}{[.237;.264]}\egroup{} & \bgroup\fontsize{7}{9}\selectfont \textcolor{red}{[.067;.083]}\egroup{} & \bgroup\fontsize{7}{9}\selectfont \em{[.030;.043]}\egroup{} & \bgroup\fontsize{7}{9}\selectfont \textcolor{red}{[.094;.112]}\egroup{} & \bgroup\fontsize{7}{9}\selectfont \textcolor{red}{[.079;.096]}\egroup{} & \bgroup\fontsize{7}{9}\selectfont \textcolor{red}{[.067;.083]}\egroup{} & \bgroup\fontsize{7}{9}\selectfont \textcolor{red}{[.394;.425]}\egroup{}\\

\hspace{1em} &  &  & \bgroup\fontsize{11}{13}\selectfont \textcolor{red}{.446}\egroup{} & \bgroup\fontsize{11}{13}\selectfont \textbf{.051}\egroup{} & \bgroup\fontsize{11}{13}\selectfont \em{.040}\egroup{} & \bgroup\fontsize{11}{13}\selectfont \textcolor{red}{.092}\egroup{} & \bgroup\fontsize{11}{13}\selectfont \textcolor{red}{.070}\egroup{} & \bgroup\fontsize{11}{13}\selectfont \textbf{.051}\egroup{} & \bgroup\fontsize{11}{13}\selectfont \textcolor{red}{.249}\egroup{}\\

\hspace{1em} & \multirow{-4}{*}{\raggedright\arraybackslash \bgroup\fontsize{11}{13}\selectfont no PT:SM\egroup{}} & \multirow{-2}{*}{\raggedright\arraybackslash \bgroup\fontsize{11}{13}\selectfont +\egroup{}} & \bgroup\fontsize{7}{9}\selectfont \textcolor{red}{[.431;.462]}\egroup{} & \bgroup\fontsize{7}{9}\selectfont \textbf{[.045;.059]}\egroup{} & \bgroup\fontsize{7}{9}\selectfont \em{[.034;.047]}\egroup{} & \bgroup\fontsize{7}{9}\selectfont \textcolor{red}{[.083;.101]}\egroup{} & \bgroup\fontsize{7}{9}\selectfont \textcolor{red}{[.063;.079]}\egroup{} & \bgroup\fontsize{7}{9}\selectfont \textbf{[.045;.059]}\egroup{} & \bgroup\fontsize{7}{9}\selectfont \textcolor{red}{[.235;.263]}\egroup{}\\

\hspace{1em} &  &  & \bgroup\fontsize{11}{13}\selectfont \textcolor{red}{.360}\egroup{} & \bgroup\fontsize{11}{13}\selectfont \textbf{.049}\egroup{} & \bgroup\fontsize{11}{13}\selectfont \em{.039}\egroup{} & \bgroup\fontsize{11}{13}\selectfont \textcolor{red}{.085}\egroup{} & \bgroup\fontsize{11}{13}\selectfont \textcolor{red}{.060}\egroup{} & \bgroup\fontsize{11}{13}\selectfont \textbf{.049}\egroup{} & \bgroup\fontsize{11}{13}\selectfont \textcolor{red}{.195}\egroup{}\\

\hspace{1em} &  & \multirow{-2}{*}{\raggedright\arraybackslash \bgroup\fontsize{11}{13}\selectfont -\egroup{}} & \bgroup\fontsize{7}{9}\selectfont \textcolor{red}{[.345;.375]}\egroup{} & \bgroup\fontsize{7}{9}\selectfont \textbf{[.043;.056]}\egroup{} & \bgroup\fontsize{7}{9}\selectfont \em{[.033;.046]}\egroup{} & \bgroup\fontsize{7}{9}\selectfont \textcolor{red}{[.077;.094]}\egroup{} & \bgroup\fontsize{7}{9}\selectfont \textcolor{red}{[.053;.068]}\egroup{} & \bgroup\fontsize{7}{9}\selectfont \textbf{[.043;.056]}\egroup{} & \bgroup\fontsize{7}{9}\selectfont \textcolor{red}{[.183;.208]}\egroup{}\\

\hspace{1em} &  &  & \bgroup\fontsize{11}{13}\selectfont \textcolor{red}{.463}\egroup{} & \bgroup\fontsize{11}{13}\selectfont \textbf{.049}\egroup{} & \bgroup\fontsize{11}{13}\selectfont \em{.040}\egroup{} & \bgroup\fontsize{11}{13}\selectfont \textcolor{red}{.085}\egroup{} & \bgroup\fontsize{11}{13}\selectfont \textcolor{red}{.059}\egroup{} & \bgroup\fontsize{11}{13}\selectfont \textbf{.049}\egroup{} & \bgroup\fontsize{11}{13}\selectfont \textcolor{red}{.189}\egroup{}\\

\hspace{1em}\multirow{-8}{*}{\raggedright\arraybackslash \rotatebox{90}{\bgroup\fontsize{11}{13}\selectfont spheric.\egroup{}}} & \multirow{-4}{*}{\raggedright\arraybackslash \bgroup\fontsize{11}{13}\selectfont PT:SM\egroup{}} & \multirow{-2}{*}{\raggedright\arraybackslash \bgroup\fontsize{11}{13}\selectfont +\egroup{}} & \bgroup\fontsize{7}{9}\selectfont \textcolor{red}{[.448;.479]}\egroup{} & \bgroup\fontsize{7}{9}\selectfont \textbf{[.042;.056]}\egroup{} & \bgroup\fontsize{7}{9}\selectfont \em{[.034;.047]}\egroup{} & \bgroup\fontsize{7}{9}\selectfont \textcolor{red}{[.076;.094]}\egroup{} & \bgroup\fontsize{7}{9}\selectfont \textcolor{red}{[.052;.067]}\egroup{} & \bgroup\fontsize{7}{9}\selectfont \textbf{[.042;.056]}\egroup{} & \bgroup\fontsize{7}{9}\selectfont \textcolor{red}{[.177;.202]}\egroup{}\\
\cmidrule{1-10}
\hspace{1em} &  &  & \bgroup\fontsize{11}{13}\selectfont \textcolor{red}{.245}\egroup{} & \bgroup\fontsize{11}{13}\selectfont \textcolor{red}{.083}\egroup{} & \bgroup\fontsize{11}{13}\selectfont \em{.026}\egroup{} & \bgroup\fontsize{11}{13}\selectfont \textcolor{red}{.113}\egroup{} & \bgroup\fontsize{11}{13}\selectfont \textcolor{red}{.086}\egroup{} & \bgroup\fontsize{11}{13}\selectfont \textcolor{red}{.083}\egroup{} & \bgroup\fontsize{11}{13}\selectfont \textcolor{red}{.457}\egroup{}\\

\hspace{1em} &  & \multirow{-2}{*}{\raggedright\arraybackslash \bgroup\fontsize{11}{13}\selectfont -\egroup{}} & \bgroup\fontsize{7}{9}\selectfont \textcolor{red}{[.232;.259]}\egroup{} & \bgroup\fontsize{7}{9}\selectfont \textcolor{red}{[.075;.092]}\egroup{} & \bgroup\fontsize{7}{9}\selectfont \em{[.020;.032]}\egroup{} & \bgroup\fontsize{7}{9}\selectfont \textcolor{red}{[.104;.124]}\egroup{} & \bgroup\fontsize{7}{9}\selectfont \textcolor{red}{[.078;.096]}\egroup{} & \bgroup\fontsize{7}{9}\selectfont \textcolor{red}{[.075;.092]}\egroup{} & \bgroup\fontsize{7}{9}\selectfont \textcolor{red}{[.441;.473]}\egroup{}\\

\hspace{1em} &  &  & \bgroup\fontsize{11}{13}\selectfont \textcolor{red}{.420}\egroup{} & \bgroup\fontsize{11}{13}\selectfont \textcolor{red}{.060}\egroup{} & \bgroup\fontsize{11}{13}\selectfont \em{.028}\egroup{} & \bgroup\fontsize{11}{13}\selectfont \textcolor{red}{.103}\egroup{} & \bgroup\fontsize{11}{13}\selectfont \textcolor{red}{.069}\egroup{} & \bgroup\fontsize{11}{13}\selectfont \textcolor{red}{.060}\egroup{} & \bgroup\fontsize{11}{13}\selectfont \textcolor{red}{.494}\egroup{}\\

\hspace{1em} & \multirow{-4}{*}{\raggedright\arraybackslash \bgroup\fontsize{11}{13}\selectfont no PT:SM\egroup{}} & \multirow{-2}{*}{\raggedright\arraybackslash \bgroup\fontsize{11}{13}\selectfont +\egroup{}} & \bgroup\fontsize{7}{9}\selectfont \textcolor{red}{[.405;.436]}\egroup{} & \bgroup\fontsize{7}{9}\selectfont \textcolor{red}{[.053;.068]}\egroup{} & \bgroup\fontsize{7}{9}\selectfont \em{[.022;.035]}\egroup{} & \bgroup\fontsize{7}{9}\selectfont \textcolor{red}{[.094;.113]}\egroup{} & \bgroup\fontsize{7}{9}\selectfont \textcolor{red}{[.062;.077]}\egroup{} & \bgroup\fontsize{7}{9}\selectfont \textcolor{red}{[.053;.068]}\egroup{} & \bgroup\fontsize{7}{9}\selectfont \textcolor{red}{[.478;.511]}\egroup{}\\

\hspace{1em} &  &  & \bgroup\fontsize{11}{13}\selectfont \textcolor{red}{.351}\egroup{} & \bgroup\fontsize{11}{13}\selectfont \textcolor{red}{.058}\egroup{} & \bgroup\fontsize{11}{13}\selectfont \em{.028}\egroup{} & \bgroup\fontsize{11}{13}\selectfont \textcolor{red}{.106}\egroup{} & \bgroup\fontsize{11}{13}\selectfont \textcolor{red}{.068}\egroup{} & \bgroup\fontsize{11}{13}\selectfont \textcolor{red}{.058}\egroup{} & \bgroup\fontsize{11}{13}\selectfont \textcolor{red}{.502}\egroup{}\\

\hspace{1em} &  & \multirow{-2}{*}{\raggedright\arraybackslash \bgroup\fontsize{11}{13}\selectfont -\egroup{}} & \bgroup\fontsize{7}{9}\selectfont \textcolor{red}{[.337;.366]}\egroup{} & \bgroup\fontsize{7}{9}\selectfont \textcolor{red}{[.051;.065]}\egroup{} & \bgroup\fontsize{7}{9}\selectfont \em{[.022;.036]}\egroup{} & \bgroup\fontsize{7}{9}\selectfont \textcolor{red}{[.097;.116]}\egroup{} & \bgroup\fontsize{7}{9}\selectfont \textcolor{red}{[.061;.077]}\egroup{} & \bgroup\fontsize{7}{9}\selectfont \textcolor{red}{[.051;.065]}\egroup{} & \bgroup\fontsize{7}{9}\selectfont \textcolor{red}{[.485;.519]}\egroup{}\\

\hspace{1em} &  &  & \bgroup\fontsize{11}{13}\selectfont \textcolor{red}{.448}\egroup{} & \bgroup\fontsize{11}{13}\selectfont \textcolor{red}{.057}\egroup{} & \bgroup\fontsize{11}{13}\selectfont \em{.027}\egroup{} & \bgroup\fontsize{11}{13}\selectfont \textcolor{red}{.106}\egroup{} & \bgroup\fontsize{11}{13}\selectfont \textcolor{red}{.067}\egroup{} & \bgroup\fontsize{11}{13}\selectfont \textcolor{red}{.056}\egroup{} & \bgroup\fontsize{11}{13}\selectfont \textcolor{red}{.493}\egroup{}\\

\hspace{1em}\multirow{-8}{*}{\raggedright\arraybackslash \rotatebox{90}{\bgroup\fontsize{11}{13}\selectfont corr.\egroup{}}} & \multirow{-4}{*}{\raggedright\arraybackslash \bgroup\fontsize{11}{13}\selectfont PT:SM\egroup{}} & \multirow{-2}{*}{\raggedright\arraybackslash \bgroup\fontsize{11}{13}\selectfont +\egroup{}} & \bgroup\fontsize{7}{9}\selectfont \textcolor{red}{[.433;.463]}\egroup{} & \bgroup\fontsize{7}{9}\selectfont \textcolor{red}{[.050;.064]}\egroup{} & \bgroup\fontsize{7}{9}\selectfont \em{[.022;.034]}\egroup{} & \bgroup\fontsize{7}{9}\selectfont \textcolor{red}{[.097;.116]}\egroup{} & \bgroup\fontsize{7}{9}\selectfont \textcolor{red}{[.059;.075]}\egroup{} & \bgroup\fontsize{7}{9}\selectfont \textcolor{red}{[.050;.064]}\egroup{} & \bgroup\fontsize{7}{9}\selectfont \textcolor{red}{[.477;.510]}\egroup{}\\
\bottomrule
\end{tabular}
\end{table}

\renewcommand{\arraystretch}{1}
\renewcommand{\baselinestretch}{1.7}

\renewcommand{\baselinestretch}{1}
\renewcommand{\arraystretch}{.73}
\begin{table}[!htb]
\centering
\caption{Type I error rate  for three factors of the common model (M1 of in Table \ref{tab:param}) with 36 stimuli. Correct methods should be close the nominal level $\alpha=.050$. The data are generated without random intercepts. In this setting, the type I error rates of RI-L  deviate strongly from the nominal level, whereas gANOVA rates stay close to it. Bold font corresponds to nominal level (5\%) within the confidence interval, red font corresponds to confidence interval above the nominal level and italic font corresponds to confidence interval below the nominal level.}
\label{tab:ganova_ril_main}

\begin{tabular}{lllllll}
\toprule
  &   &   & RI-L & RI-L+ & gANOVA & gANOVA+\\
\midrule
 &  & \bgroup\fontsize{11}{13}\selectfont no PT:SM\egroup{} & \bgroup\fontsize{11}{13}\selectfont \em{.005}\egroup{} \bgroup\fontsize{7}{9}\selectfont \em{[.003;.008]}\egroup{} & \bgroup\fontsize{11}{13}\selectfont \em{.005}\egroup{} \bgroup\fontsize{7}{9}\selectfont \em{[.003;.008]}\egroup{} & \bgroup\fontsize{11}{13}\selectfont \textbf{.046}\egroup{} \bgroup\fontsize{7}{9}\selectfont \textbf{[.040;.054]}\egroup{} & \bgroup\fontsize{11}{13}\selectfont \textbf{.046}\egroup{} \bgroup\fontsize{7}{9}\selectfont \textbf{[.040;.054]}\egroup{}\\

 & \multirow{-2}{*}{\raggedright\arraybackslash \bgroup\fontsize{11}{13}\selectfont corr.\egroup{}} & \bgroup\fontsize{11}{13}\selectfont PT:SM\egroup{} & \bgroup\fontsize{11}{13}\selectfont \em{.005}\egroup{} \bgroup\fontsize{7}{9}\selectfont \em{[.004;.008]}\egroup{} & \bgroup\fontsize{11}{13}\selectfont \em{.005}\egroup{} \bgroup\fontsize{7}{9}\selectfont \em{[.004;.008]}\egroup{} & \bgroup\fontsize{11}{13}\selectfont \textbf{.049}\egroup{} \bgroup\fontsize{7}{9}\selectfont \textbf{[.043;.056]}\egroup{} & \bgroup\fontsize{11}{13}\selectfont \textbf{.049}\egroup{} \bgroup\fontsize{7}{9}\selectfont \textbf{[.043;.056]}\egroup{}\\

 &  & \bgroup\fontsize{11}{13}\selectfont no PT:SM\egroup{} & \bgroup\fontsize{11}{13}\selectfont \em{.005}\egroup{} \bgroup\fontsize{7}{9}\selectfont \em{[.004;.008]}\egroup{} & \bgroup\fontsize{11}{13}\selectfont \em{.005}\egroup{} \bgroup\fontsize{7}{9}\selectfont \em{[.004;.008]}\egroup{} & \bgroup\fontsize{11}{13}\selectfont \textbf{.051}\egroup{} \bgroup\fontsize{7}{9}\selectfont \textbf{[.045;.058]}\egroup{} & \bgroup\fontsize{11}{13}\selectfont \textbf{.051}\egroup{} \bgroup\fontsize{7}{9}\selectfont \textbf{[.045;.058]}\egroup{}\\

\multirow{-4}{*}{\raggedright\arraybackslash \bgroup\fontsize{11}{13}\selectfont Ap\egroup{}} & \multirow{-2}{*}{\raggedright\arraybackslash \bgroup\fontsize{11}{13}\selectfont spheric.\egroup{}} & \bgroup\fontsize{11}{13}\selectfont PT:SM\egroup{} & \bgroup\fontsize{11}{13}\selectfont \em{.005}\egroup{} \bgroup\fontsize{7}{9}\selectfont \em{[.003;.008]}\egroup{} & \bgroup\fontsize{11}{13}\selectfont \em{.005}\egroup{} \bgroup\fontsize{7}{9}\selectfont \em{[.003;.008]}\egroup{} & \bgroup\fontsize{11}{13}\selectfont \em{.039}\egroup{} \bgroup\fontsize{7}{9}\selectfont \em{[.033;.045]}\egroup{} & \bgroup\fontsize{11}{13}\selectfont \em{.039}\egroup{} \bgroup\fontsize{7}{9}\selectfont \em{[.033;.045]}\egroup{}\\
\cmidrule{1-7}
 &  & \bgroup\fontsize{11}{13}\selectfont no PT:SM\egroup{} & \bgroup\fontsize{11}{13}\selectfont \textcolor{red}{.119}\egroup{} \bgroup\fontsize{7}{9}\selectfont \textcolor{red}{[.109;.129]}\egroup{} & \bgroup\fontsize{11}{13}\selectfont \textcolor{red}{.119}\egroup{} \bgroup\fontsize{7}{9}\selectfont \textcolor{red}{[.109;.129]}\egroup{} & \bgroup\fontsize{11}{13}\selectfont \textbf{.047}\egroup{} \bgroup\fontsize{7}{9}\selectfont \textbf{[.041;.054]}\egroup{} & \bgroup\fontsize{11}{13}\selectfont \textbf{.047}\egroup{} \bgroup\fontsize{7}{9}\selectfont \textbf{[.041;.054]}\egroup{}\\

 & \multirow{-2}{*}{\raggedright\arraybackslash \bgroup\fontsize{11}{13}\selectfont corr.\egroup{}} & \bgroup\fontsize{11}{13}\selectfont PT:SM\egroup{} & \bgroup\fontsize{11}{13}\selectfont \textcolor{red}{.117}\egroup{} \bgroup\fontsize{7}{9}\selectfont \textcolor{red}{[.108;.128]}\egroup{} & \bgroup\fontsize{11}{13}\selectfont \textcolor{red}{.117}\egroup{} \bgroup\fontsize{7}{9}\selectfont \textcolor{red}{[.108;.128]}\egroup{} & \bgroup\fontsize{11}{13}\selectfont \textbf{.050}\egroup{} \bgroup\fontsize{7}{9}\selectfont \textbf{[.043;.057]}\egroup{} & \bgroup\fontsize{11}{13}\selectfont \textbf{.050}\egroup{} \bgroup\fontsize{7}{9}\selectfont \textbf{[.043;.057]}\egroup{}\\

 &  & \bgroup\fontsize{11}{13}\selectfont no PT:SM\egroup{} & \bgroup\fontsize{11}{13}\selectfont \textcolor{red}{.117}\egroup{} \bgroup\fontsize{7}{9}\selectfont \textcolor{red}{[.107;.127]}\egroup{} & \bgroup\fontsize{11}{13}\selectfont \textcolor{red}{.117}\egroup{} \bgroup\fontsize{7}{9}\selectfont \textcolor{red}{[.107;.127]}\egroup{} & \bgroup\fontsize{11}{13}\selectfont \textbf{.046}\egroup{} \bgroup\fontsize{7}{9}\selectfont \textbf{[.040;.054]}\egroup{} & \bgroup\fontsize{11}{13}\selectfont \textbf{.046}\egroup{} \bgroup\fontsize{7}{9}\selectfont \textbf{[.040;.054]}\egroup{}\\

\multirow{-4}{*}{\raggedright\arraybackslash \bgroup\fontsize{11}{13}\selectfont Am\egroup{}} & \multirow{-2}{*}{\raggedright\arraybackslash \bgroup\fontsize{11}{13}\selectfont spheric.\egroup{}} & \bgroup\fontsize{11}{13}\selectfont PT:SM\egroup{} & \bgroup\fontsize{11}{13}\selectfont \textcolor{red}{.107}\egroup{} \bgroup\fontsize{7}{9}\selectfont \textcolor{red}{[.098;.117]}\egroup{} & \bgroup\fontsize{11}{13}\selectfont \textcolor{red}{.107}\egroup{} \bgroup\fontsize{7}{9}\selectfont \textcolor{red}{[.098;.117]}\egroup{} & \bgroup\fontsize{11}{13}\selectfont \textbf{.051}\egroup{} \bgroup\fontsize{7}{9}\selectfont \textbf{[.044;.058]}\egroup{} & \bgroup\fontsize{11}{13}\selectfont \textbf{.051}\egroup{} \bgroup\fontsize{7}{9}\selectfont \textbf{[.044;.058]}\egroup{}\\
\cmidrule{1-7}
 &  & \bgroup\fontsize{11}{13}\selectfont no PT:SM\egroup{} & \bgroup\fontsize{11}{13}\selectfont \textcolor{red}{.114}\egroup{} \bgroup\fontsize{7}{9}\selectfont \textcolor{red}{[.105;.125]}\egroup{} & \bgroup\fontsize{11}{13}\selectfont \textcolor{red}{.114}\egroup{} \bgroup\fontsize{7}{9}\selectfont \textcolor{red}{[.105;.125]}\egroup{} & \bgroup\fontsize{11}{13}\selectfont \textbf{.052}\egroup{} \bgroup\fontsize{7}{9}\selectfont \textbf{[.045;.059]}\egroup{} & \bgroup\fontsize{11}{13}\selectfont \textbf{.052}\egroup{} \bgroup\fontsize{7}{9}\selectfont \textbf{[.045;.059]}\egroup{}\\

 & \multirow{-2}{*}{\raggedright\arraybackslash \bgroup\fontsize{11}{13}\selectfont corr.\egroup{}} & \bgroup\fontsize{11}{13}\selectfont PT:SM\egroup{} & \bgroup\fontsize{11}{13}\selectfont \textcolor{red}{.114}\egroup{} \bgroup\fontsize{7}{9}\selectfont \textcolor{red}{[.104;.124]}\egroup{} & \bgroup\fontsize{11}{13}\selectfont \textcolor{red}{.113}\egroup{} \bgroup\fontsize{7}{9}\selectfont \textcolor{red}{[.104;.123]}\egroup{} & \bgroup\fontsize{11}{13}\selectfont \textbf{.051}\egroup{} \bgroup\fontsize{7}{9}\selectfont \textbf{[.044;.058]}\egroup{} & \bgroup\fontsize{11}{13}\selectfont \textbf{.051}\egroup{} \bgroup\fontsize{7}{9}\selectfont \textbf{[.044;.058]}\egroup{}\\

 &  & \bgroup\fontsize{11}{13}\selectfont no PT:SM\egroup{} & \bgroup\fontsize{11}{13}\selectfont \textcolor{red}{.113}\egroup{} \bgroup\fontsize{7}{9}\selectfont \textcolor{red}{[.103;.123]}\egroup{} & \bgroup\fontsize{11}{13}\selectfont \textcolor{red}{.113}\egroup{} \bgroup\fontsize{7}{9}\selectfont \textcolor{red}{[.103;.123]}\egroup{} & \bgroup\fontsize{11}{13}\selectfont \textbf{.047}\egroup{} \bgroup\fontsize{7}{9}\selectfont \textbf{[.041;.054]}\egroup{} & \bgroup\fontsize{11}{13}\selectfont \textbf{.047}\egroup{} \bgroup\fontsize{7}{9}\selectfont \textbf{[.041;.054]}\egroup{}\\

\multirow{-4}{*}{\raggedright\arraybackslash \bgroup\fontsize{11}{13}\selectfont Ap:As\egroup{}} & \multirow{-2}{*}{\raggedright\arraybackslash \bgroup\fontsize{11}{13}\selectfont spheric.\egroup{}} & \bgroup\fontsize{11}{13}\selectfont PT:SM\egroup{} & \bgroup\fontsize{11}{13}\selectfont \textcolor{red}{.111}\egroup{} \bgroup\fontsize{7}{9}\selectfont \textcolor{red}{[.101;.121]}\egroup{} & \bgroup\fontsize{11}{13}\selectfont \textcolor{red}{.110}\egroup{} \bgroup\fontsize{7}{9}\selectfont \textcolor{red}{[.101;.121]}\egroup{} & \bgroup\fontsize{11}{13}\selectfont \textbf{.050}\egroup{} \bgroup\fontsize{7}{9}\selectfont \textbf{[.044;.058]}\egroup{} & \bgroup\fontsize{11}{13}\selectfont \textbf{.050}\egroup{} \bgroup\fontsize{7}{9}\selectfont \textbf{[.044;.058]}\egroup{}\\
\bottomrule
\end{tabular}
\end{table}
\renewcommand{\arraystretch}{1}
\renewcommand{\baselinestretch}{1.7}

\renewcommand{\baselinestretch}{1}
\begin{figure}[tb]
\includegraphics[width=160mm]{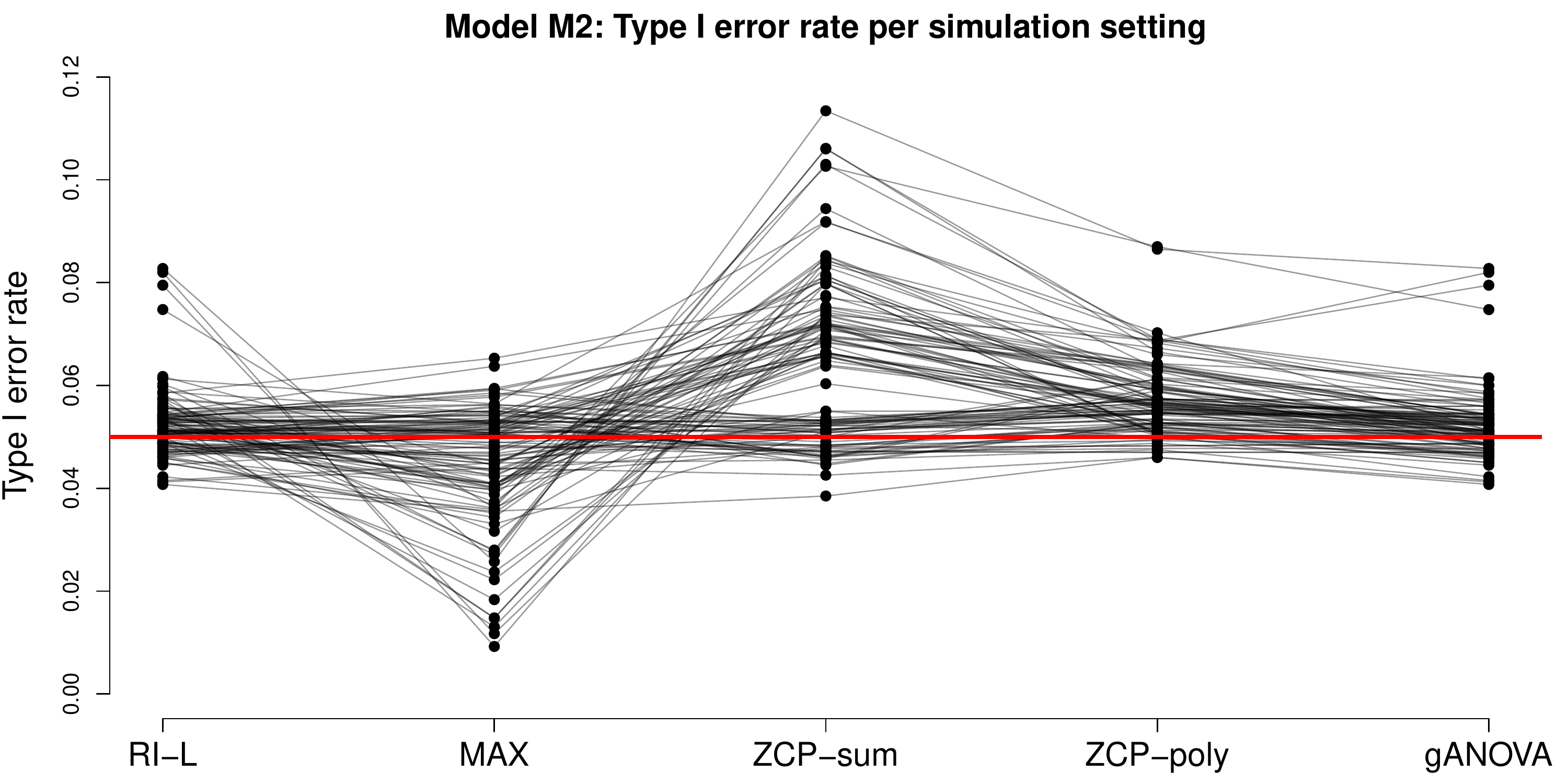}
\caption{Display of type I error rates for all simulation settings for model M2 (2 sample sizes~$\times$~2 correlations of random effects~$\times$~2 interactions in simulation~$\times$~2 interactions in estimation $\times$ 7 effects~=~112 settings). RI-L and gANOVA produce results closer to the nominal level $\alpha=.050$ represented by a red line.}
\label{fig:type1:common}
\end{figure}
\renewcommand{\baselinestretch}{1.7}

\renewcommand{\baselinestretch}{1}
\begin{figure}[tb]
\includegraphics[width=160mm]{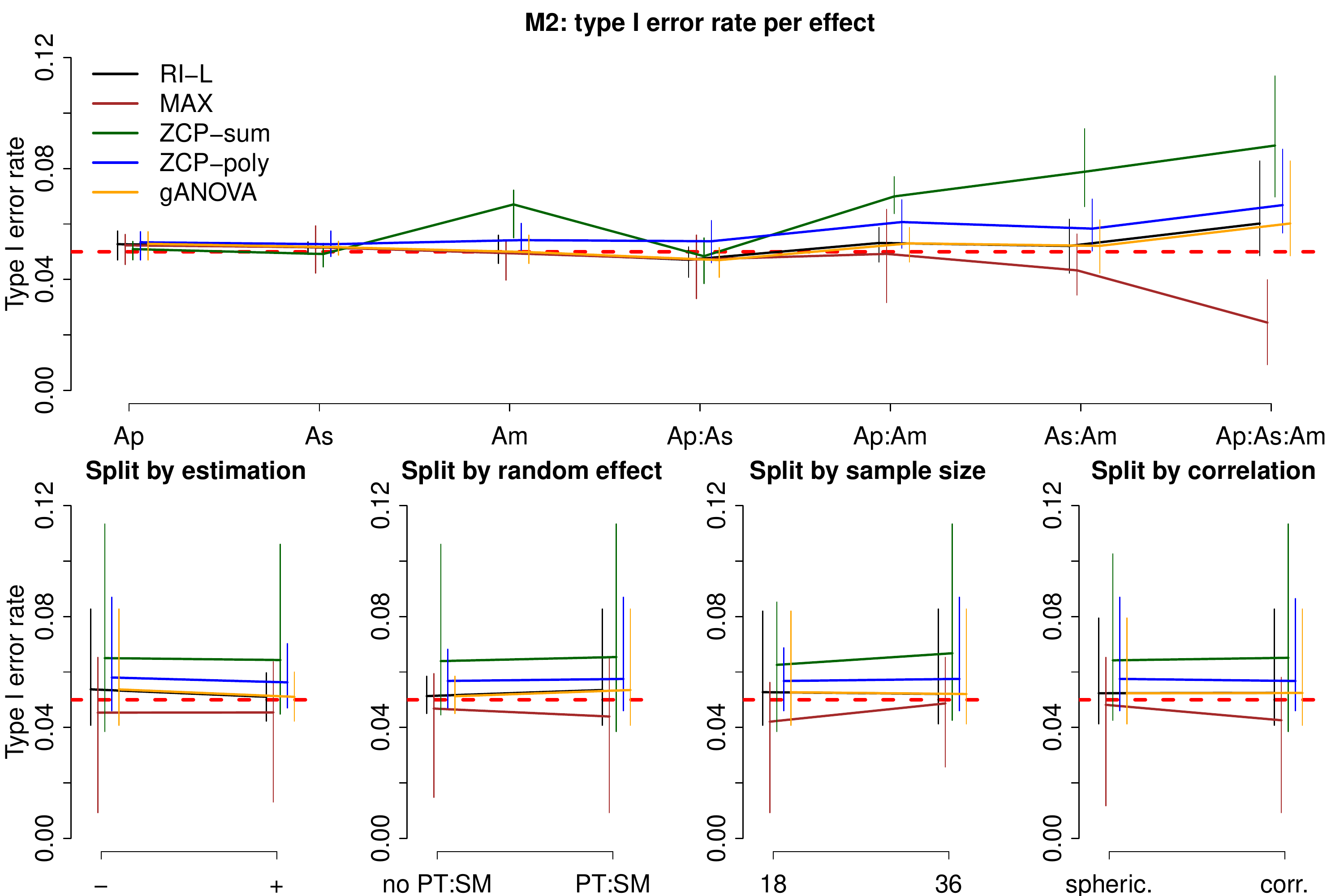}
\caption{Display of type I error rates for the model M2 aggregated according to the simulations settings. The vertical lines indicate the range of all simulations within the condition. RI-L and gANOVA are the closest to the nominal level $\alpha=.050$ represented by a red dashed line. The factor $A_M$ and its interaction produce higher deviation from the nominal level across all correlation structures. No other simulation setting tends to have an effect on the type I error rate.}
\label{fig:type1:common:split}
\end{figure}
\renewcommand{\baselinestretch}{1.7}

\renewcommand{\baselinestretch}{1}
\begin{figure}[tb]
\includegraphics[width=160mm]{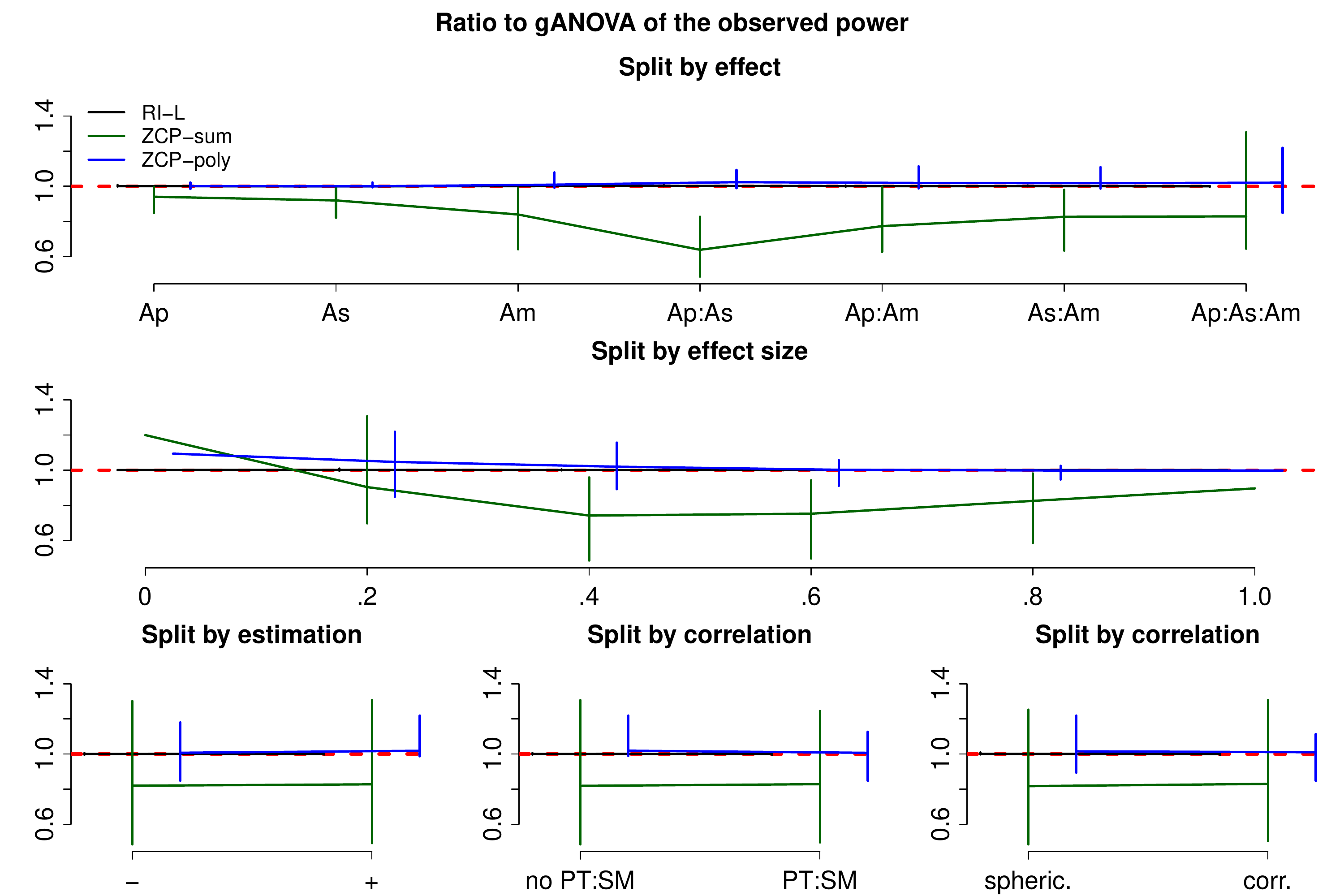}
\caption{Ratio of uncorrected average observed powers of RI-L, ZCP-sum and ZCP-poly compared to the one of gANOVA. A method has larger power than gANOVA if the ratio is bigger than one. The vertical lines indicate the range of all simulations within the condition. ZCP-poly is liberal which explains that for lower effects size, gANOVA has a slightly lower power than ZCP-poly, but the deviation reduces for higher effect sizes.}
\label{fig:observed:power}
\end{figure}
\renewcommand{\baselinestretch}{1.7}

\renewcommand{\baselinestretch}{1}
\begin{figure}[tb]
\includegraphics[width=160mm]{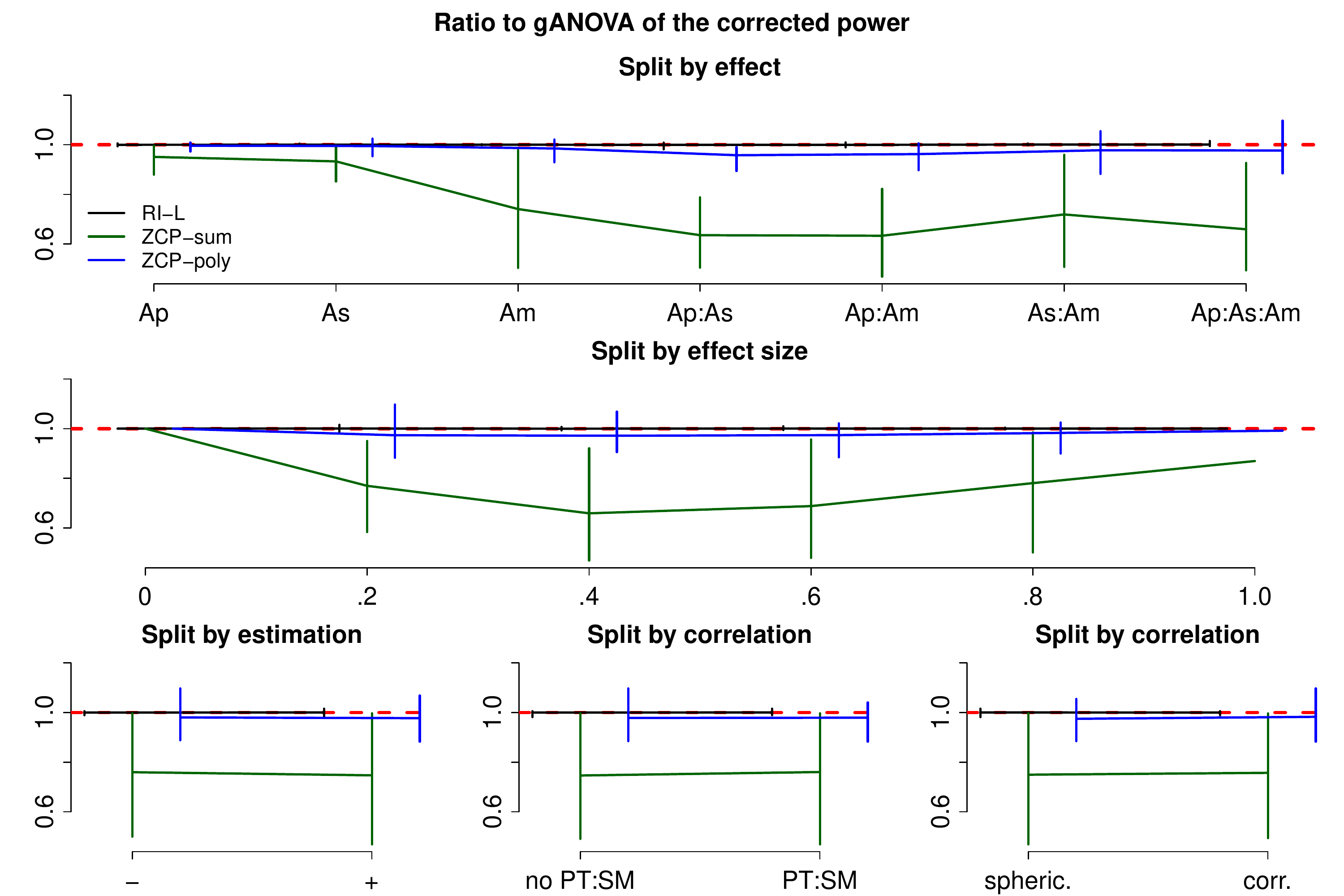}
\caption{Ratio of corrected average observed powers of RI-L, ZCP-sum and ZCP-poly compared to the one of gANOVA. A method has larger power than gANOVA if the ratio is bigger than one. The vertical lines indicate the range of all simulations within the condition. In all simulation settings, no correlation structure performs better than gANOVA.}
\label{fig:corrected:power}
\end{figure}
\renewcommand{\baselinestretch}{1.7}

This simulation study is designed to compare the above correlation
structures when performing tests on the (fixed) factors, which is
usually the principal interest of researchers. We focus on the type I
error rate and the convergence rate of the methods. The type I error
rate is the average number of rejected null hypotheses per simulation
settings. It should be close to the nominal level, that is set here to
~\(\alpha = 5\%\); a lower value indicates a conservative method and a
higher value indicates a liberal one. Moreover, to evaluate the power of
the tests, we recorded, under the alternative hypothesis, the average
number of true positives (the empirical power). A higher number of true
positives indicates a more powerful method.

\subsection*{Simulating the datasets}

For the principal simulation, we choose several settings in order to
match likely experimental designs. 4000 samples were simulated in order
to have small confidence interval of our metrics. The settings vary
according to 3 different designs, 2 different sample sizes, 2 different
patterns of correlation of random effects, and the fact that random
effects for the interaction participants:stimuli are included or not.

The 3 experimental designs are: a small design with only 2 levels per
factor (M1 in Tables~\ref{tab:model} and~\ref{tab:param}), a rather
common design with 3 levels per factor (M2 in Tables~\ref{tab:model}
and~\ref{tab:param}) and a larger design with more factors (M4 in
Tables~\ref{tab:model} and~\ref{tab:param}). The designs M1 and M2 have
factors of type \(A_P\),\(A_S\), \(A_M\) and M4 has an additional factor
of type \(A_{PS}\). Two patterns of correlation between random effects
are used for the generation of the data. In the first case, the random
effects are spherical (spheric.) and in the second case, the random
effects are fully correlated (corr.); the fully correlated covariance
matrix is such that random effects spanned a space of half of the
dimension of the random effects (but not in the canonical directions).
In order to give more importance to the main effects, the standard
deviations of random effects are halved when increasing an order (or
degree) of interaction. Moreover, all standard deviations of the random
effects associated to stimulus and interaction participants:stimuli are
shrunken by \(0.9\), respectively \(0.8\). Each design is simulated with
random effects associated to the interaction participants:stimuli
(PT:SM) and without (no PT:SM). Moreover, the small (M1) and common (M2)
designs are simulated with 2 different sample sizes: 18 participants and
18 stimuli, and 18 participants and 36 stimuli. The large model (M4) was
only simulated using 18 participants and 18 stimuli to reduce
computation time.

To evaluate further the methods, two additional simulation extensions
have been carried out. Because decreasing the standard deviations as the
interaction level increases favours RI-L correlation structure, we
change the standard deviations of random effects to highlight the
difference of gANOVA and RI-L on the type I error rate. In that case, we
simulate data without random intercepts while all others standard
deviations of random effects are kept the same. The second extension
concerns a much larger sample size with 180 participants.

Finally, we produce a power analysis by increasing the true effects. For
each parameter, we first estimated a maximum value such that the
empirical power exceeds 90\%. Then, each factor is tested by multiplying
this value by \(.2\), \(.4\), \(.6\), \(.8\) and \(1\). For factors with
more than 3 levels, we let all factor parameters increase
simultaneously. Moreover, to reduce the computation time, we increased
all factors simultaneously (all main effects and all interactions).

\subsection*{Fitting of the data}

The randomly generated data are fitted using all the presented
correlation structures: the random intercepts (RI), the random
intercepts at each level (RI-L), the maximal (MAX), the zero-correlation
parameter (ZCP), the generalized ANOVA (gANOVA) and correlation
structure based on PCA (CS-PCA). ZCP is computed once with the default
(non-orthonormal) ``sum'' coding (ZCP-sum) and once with a
``polynomial'' (and orthonormal) coding (ZCP-poly). Each model is
estimated with (+) and without (-) including the random effects
associated to the interaction participant-stimulus. Moreover, the
significance is evaluated using the type III test with Satterthwaite's
approximation of the degrees of freedom using the \texttt{lmerTest}
package \citep{kuznetsova_lmertest_2017} and the restricted maximum
likelihood (REML) estimation \citep{bates_fitting_2015}. The larger
model (M4) was only estimated using RI, ZCP and gANOVA to reduce
computation time.

To reduce the convergence error, each model is first optimized using the
default \texttt{BOBYQA} optimizer \citep{powell_bobyqa_2009}, then the
Nelder-Mead optimizer \citep{nelder_simplex_1965}, then from the
\texttt{optimx} package \citep{nash_unifying_2011} the \texttt{nlminb}
optimizer and finally the \texttt{L-BFGS-B} optimizer. We stop the
procedure when a solution is found without convergence error. If all
optimizers fail, we declare a failure of convergence for that sample.

\subsection*{Evaluation of simulation}

Table~\ref{tab:conv} shows the percentage of samples with convergence
error based on 4000 simulated samples for all simulation settings
(designs M1, M2 and M4 in Table~\ref{tab:param}). We deduce that MAX is
not scalable to even moderately sized designs because with only 3 levels
per factor we recorded up to 40\% of convergence error. Moreover, even
with a very careful implementation of the CS-PCA by
Algorithm~\ref{alg:randpca} did not reach a low number of convergence
error. For the other correlation structures, we achieve a high
convergence rate. This indicates that using several optimizers seems a
good practice to reduce convergence error.

Table~\ref{tab:type1:common} shows estimated type I error rates with
their confidence intervals \citep{agresti_approximate_1998} for the
common model (M2) globally on all simulation settings. The too liberal
type I error rates are shown in red and the too conservative ones in
italic. The rates are computed using only the samples without
convergence error. This might have an effect on the results for MAX. One
sees that RI and CS-PCA are globally too liberal as their type I error
rates show huge deviations from the nominal level. The second
observation is that including the interaction participants:stimuli does
not influence the number of convergence error (see ``+'' versus ``-'' in
Table~\ref{tab:conv}) nor does it increase the type I error rate.
Interestingly, the ZCP-poly and ZCP-sum exhibit differences in their
type I error rate. It implies that the choice of the coding variable of
the random effects will influence the results of the tests, which is not
a desirable property. The orthonormal coding (polynomial) has a better
control on the type I error rate.

The type I error rate seems reasonably close to the nominal level for
gANOVA, ZCP-poly, and RI-L. To show the results more graphically,
Figure~\ref{fig:type1:common} plots the type I error rates of
Table~\ref{tab:type1:common} for the five best correlation structures:
RI-L, MAX, ZCP-sum, ZCP-poly, and gANOVA. Best methods are those with
points close to the red line. The superiority of ZCP-poly compared to
the ZCP-sum is noticeable. Moreover, gANOVA and RI-L seem superior to
ZCP-poly. By aggregating the results with respect to the simulation
settings (Figure~\ref{fig:type1:common:split} bottom plots), we see that
neither the type of effects, the sample size, the third interaction in
the data generation or estimation show influence on the type I error
rate. And, on average, gANOVA and RI-L perform better than ZCP-poly. By
contrast, the type of factor influences the type I error rate
(Figure~\ref{fig:type1:common:split} top plot). We see that the factor
\(A_M\) (or the interactions with this type of factor) induces type I
error rates that deviate more from the nominal level.

In the supplementary material, the results for all factors and the three
models M1, M2 and M4 are displayed, as well as for the much larger
sample size of 180 participants. The above findings are similar for all
simulations.

The difference between gANOVA and RI-L is visible in
Table~\ref{tab:ganova_ril_main} which shows simulations with data from
model M2 with a null standard deviation for the random intercepts. For
these datasets, gANOVA stay close to the nominal level but RI-L shows
large deviations (liberal or conservative). In addition to the
theoretical remarks on the difference between RI-L and gANOVA given in
the previous section and Appendix~\ref{ap:c-uc}, these simulations
confirm that gANOVA is strictly better than RI-L for reporting tests of
(fixed) factors.

Moreover, the parsimony of gANOVA seems to endow it an advantage for the
power of the tests. The details for all power results are provided in
the supplementary material. Figure~\ref{fig:observed:power} summarizes
the findings by computing the ratio of uncorrected average observed
powers of RI-L, ZCP-sum and ZCP-poly compared to the one of gANOVA. A
method has larger (better) power than gANOVA if the ratio is bigger than
one and smaller power if inferior to 1. One sees that ZCP-poly and
gANOVA perform clearly better than ZCP-sum and that RI-L is close to
gANOVA due to the choice of the variance parameters of the random
effects. As seen in Figure~\ref{fig:type1:common}, ZCP-poly has a higher
type I error rate for most of the simulation settings under the null
hypothesis. However, by increasing the effect size, this deviation
decreases. This means that the better control of the type I error rate
of gANOVA is not achieved at the expense of the power of the test.
Moreover, when the power is corrected by using a critical value set at
the nominal level (Figure~\ref{fig:corrected:power}), gANOVA performs
better than ZCP-poly. No other simulation settings have a big influence
on the average difference of power between the methods.

\section*{Conclusion}\label{chap:concl}

Using CRE-MEM in psychology is a growing practice and there is a
diversity of correlation structures that are available and that are
used. These correlation structures directly depend on the design of
experiment and we develop a classification of factors in order to help
users in the planning of the experiment and its analysis. All
correlation structures do not share the same advantages and there is no
predefined tools to help researchers select the appropriate correlation
structure given the experiment. Depending on the goal of the analysis
some properties are more important to control than others.

In the case of hypothesis testing, we have shown that the gANOVA
correlation structure has many desirable properties. Simulations show
that it controls the type I error rate without a loss in power even with
misspecifications of the model. It also provides a high rate of
convergence and is scalable to complex datasets. Moreover, it is in line
with the experimental design tradition represented by the ANOVA/rANOVA
framework, which is helpful for the interpretation of the results for
researchers familiar with the ANOVA.

\section*{Acknowledgement}

We are particularly grateful for the assistance given by Audrey Bürki
that greatly improved this manuscript. She provided many comments coming
from her extended reading of this paper and her expertise with CRE-MEM,
and to Tom Maullin-Sapey whose suggestions have been very helpful in
improving the manuscript, although any errors are our own.

\appendix

\section{Algorithm for PCA method of section \ref{chap:rs-pca}}
\label{chap:pcaalgo}

\begin{algorithm}[H]
\caption{Correlation structure based on PCA}\label{alg:randpca}
\begin{algorithmic}[1]
      \State Choose a model selection procedure $\mathscr{P}$.
      \State Estimate the model based on MAX $\mathscr{S}_{max}$.
      \For{\texttt{participants} and \texttt{stimuli}}
      \State Perform PCA to find the dimensionality $r_{\mathscr{S}_{max}}$ of the random effects.
      \State Drop random effects with higher interaction levels to match $r_{\mathscr{S}_{max}}$.
      \EndFor
      \State Define the new random structure $\mathscr{S}_{PCA}^+$.
      \State Drop covariance between random effects and define this random structure $\mathscr{S}_{PCA}^-$.
      \State Choose between $\mathscr{S}_{PCA}^+$ and $\mathscr{S}_{PCA}^-$ using $\mathscr{P}$. The chosen random structure is called $\mathscr{S}_{reduced}$.
      \While{$\mathscr{P}$ suggests the smaller correlation structure}
      \State $\mathscr{S}_{reduced}^0$ is defined by dropping from $\mathscr{S}_{reduced}$ the random effect of the higher interaction levels.
      \State Choose between $\mathscr{S}_{reduced}$ and $\mathscr{S}_{reduced}^0$ using $\mathscr{P}$.
      \State Update $\mathscr{S}_{reduced}$ by the previous choice.
      \EndWhile
      \State Given the choice made in 8, add or drop covariance to $\mathscr{S}_{reduced}$ to create $\mathscr{S}_{reduced}^1$.
      \State Choose between $\mathscr{S}_{reduced}$ and $\mathscr{S}_{reduced}^1$ using $\mathscr{P}$.
\end{algorithmic}
\end{algorithm}

\section{Correlation structure of the rANOVA} \label{ap:csranova}

In the earlier times when the repeated measures ANOVA model was first
discussed, an abundant literature on the choice of the model and
especially on the choice of the correlation structure was written. At
that time, the necessity to compute everything by hand gave constraints
on the analysis and all test statistics were based on sum of squares.
Under specific assumptions, these statistics possess an exact \(F\)
distribution. Let us first mention that \citet{huynh_conditions_1970}
showed that the sphericity (or circularity) of the covariance structure
was a necessary and sufficient condition for an exact test.
\citet{box_theorems_1954} and \citet{huynh_estimation_1976} proposed
modifications in the degrees of freedom -- called
\(\epsilon\)~correction -- when this condition is not fulfilled. Note
that even earlier, a discussion was engaged whether to include in the
correlation structure the interactions between the participant and the
fixed effects ( \((\pi\psi)_{ik}\) in Equation~\ref{eq:anova}).
\citet{rouanet_comparison_1970}, for example, compare the models with
and without them. Ultimately, the model that includes all random
interactions became the reference. Statistical software like SPSS or
Statistica use it as if it was the only possible random structure for
rANOVA.

On a more technical side, for the fixed effects, some constraints have
to be chosen since the model is otherwise overparametrized and no
estimation or test can be obtained \citep{cardinal_anova_2013}. In order
to keep the interpretation of main effects as in the ANOVA tradition, we
use the sigma-restricted parametrizations, which corresponds in
Equation~\ref{eq:anova} to the following constraints:
\(\sum_{j}\alpha_j=0\), \(\sum_{j} (\alpha\psi)_{jk}=0\ \forall k\) and
\(\sum_{k} (\alpha\psi)_{jk}=0\ \forall j\), and so on. Concerning
random interactions, which are technically interactions between a fixed
effect and a random unit, it is worth asking if this type of constraints
should also be also applied. In a very influential paper,
\citet{cornfield_average_1956} , with a construction called
``pigeonhole'' that includes both fixed and random factors as special
cases, show that the distributional behaviour for factorial designs lead
to constraints only for one margin:
\(\sum_{k} (\pi\psi)_{ik}=0\ \forall i\) , see
e.g.~\citet{montgomery_design_2017}. If these constraints are not
included e.g.~when simulating data, some variances will be inflated.

The discussion in the previous paragraphs shows that there was a debate
over several decades on the best model for rANOVA, both for the fixed
part and for the correlation structure. It is therefore not surprising
that concerning the more recently proposed CRE-MEM, a similar debate is
ongoing. It is noteworthy that it concerns exactly the same questions on
the best correlation structure and the correlation structures for mixed
effect models that are presented in this article take root in the
above-mentioned literature on rANOVA.

\section{The database representation of the 5 types of factors}\label{ap:typo}

\renewcommand{\baselinestretch}{1}
\begin{figure}[tb]
\includegraphics[width=130mm]{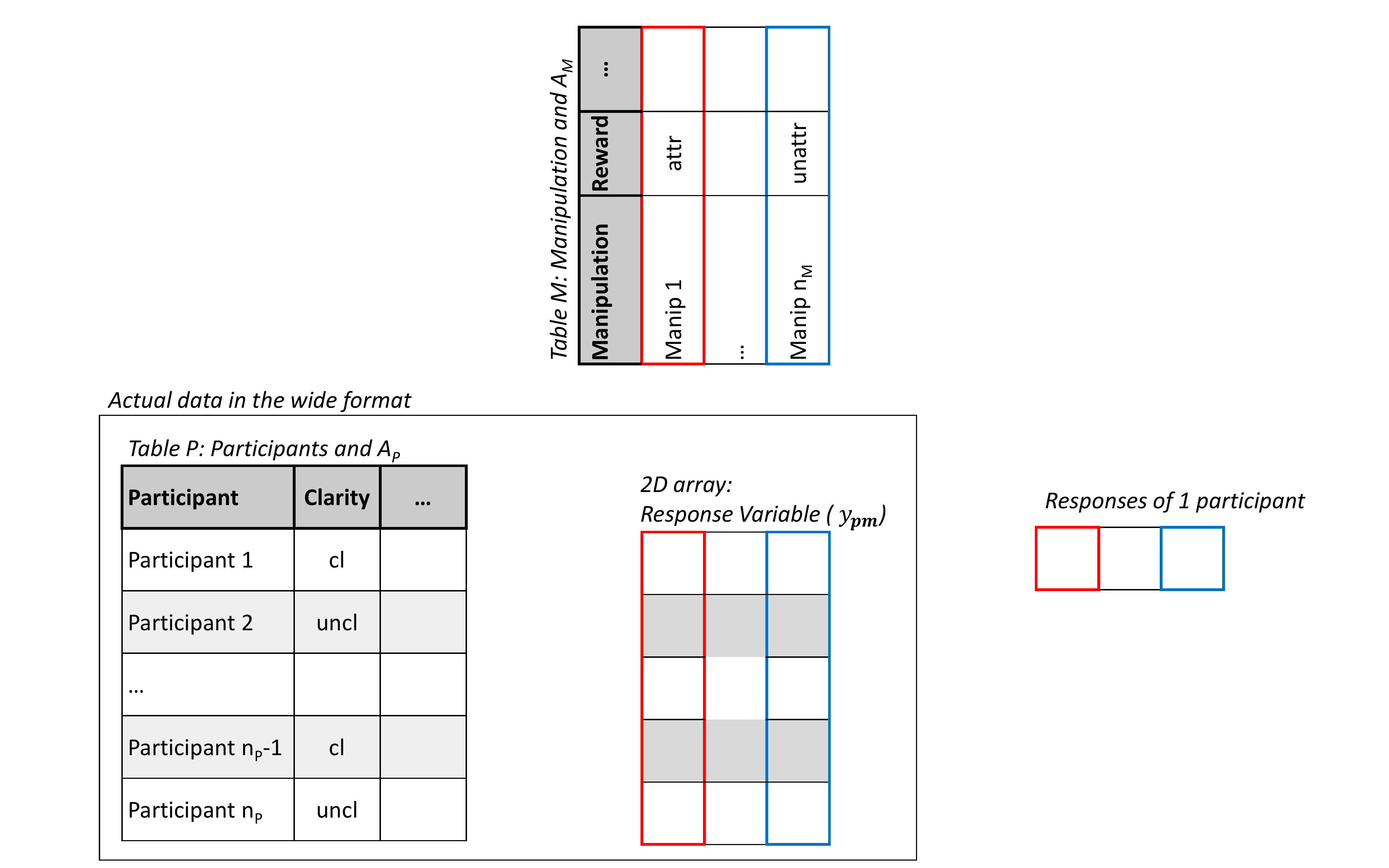}
\caption{Representation of the factors and the responses of a rANOVA in the "wide" format. The responses are stored in a 2D array that stem from the crossing of 2 tables. Table~$P$ stores the participants and their features $A_P$ (or between-participant factors) and Table~$M$ stores the features of the experimental manipulations $A_M$ (or within-participant factors). }
\label{fig:varranova}
\end{figure}
\renewcommand{\baselinestretch}{1.7}

\renewcommand{\baselinestretch}{1}
\begin{figure}[tb]
\includegraphics[width=130mm]{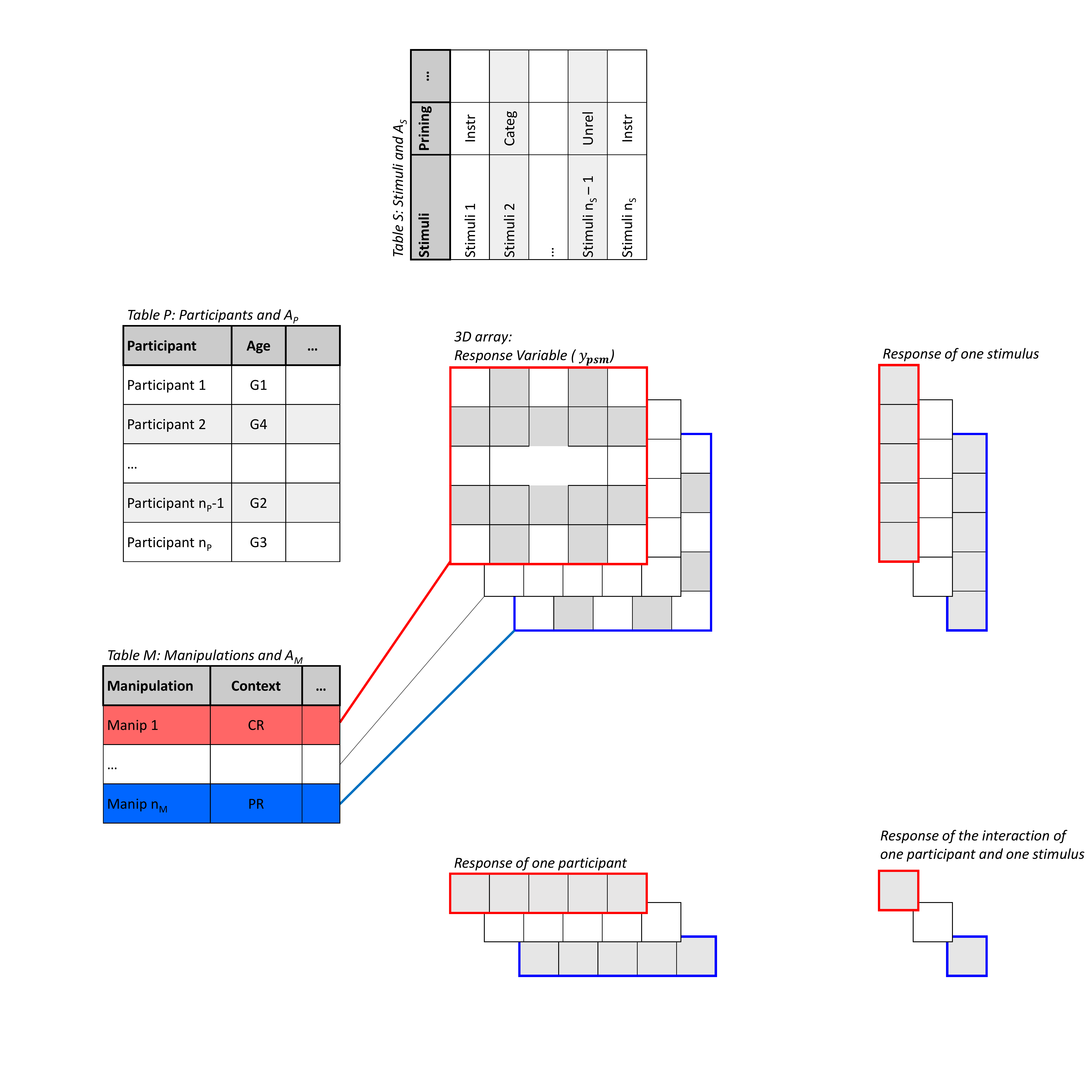}
\caption{Representation of the factors and the responses of CRE-MEM. The responses are stored in a 3D array that is the result of the crossing of Table~$P$, Table~$S$ and Table~$M$. Participants are associated with the Table~$P$ and the levels of factor $A_P$ will be identical for each "floor". Stimuli are associated with the Table~$S$ and factors $A_S$ will takes the same value for each "vertical slide". And the experimental manipulations are associated with the Table~$M$ and each "slice" of the 3D array.}
\label{fig:lmm}
\end{figure}
\renewcommand{\baselinestretch}{1.7}

As a complement to the section on classification of factors, we propose
here a generalisation of the ``wide'' format for the 5 types of factors.
In the rANOVA framework, software like SPSS or Statistica use data in
the ``wide'' format. This format makes explicit the difference between
within-participant factors (\(A_P\)) and between-participant factors
(\(A_M\)), see Figure~\ref{fig:varranova}. Each line of the data
represents one participant and the ``data'' columns are split into
between-participant factors, and columns which values are the response
recorded in one level of the within-participant factors. So, the
responses are stored in a two-dimensional (2D) array which entries, by
rows, correspond to the participants and by column to the
within-participant levels. This 2D array is the results of the crossing
of one table storing the participant and \(A_P\) factors, and one table
storing the experimental manipulations and the \(A_M\) factors.

The ``wide'' format clearly shows the fundamental difference between
within-participant and between-participant factors. This representation
can be extended to CRE-MEM. In that setting, we cross 3 tables (instead
of 2): one for the participants and the factors \(A_P\) (Table~\(P\)),
one for the stimuli and the factors \(A_S\) (Table~\(S\)) and one for
the experimental manipulations and the factors \(A_M\) (Table~\(M\)).
The crossing of those 3 tables creates a 3D array (like a building, of
dimension \([ n_P,~n_S,~n_M]\)) in which the responses can be stored.
This construction is represented in the Figure~\ref{fig:lmm}. In this
example, each participant is represented by one row of the Table~\(P\)
and each stimulus is represented by one row (here rotated by
\(90^{\circ}\)) of the Table~\(S\). Then, the responses of one
participant are a stored in one ``floor'' (of dimension
\([ 1,~n_S,~n_M]\)), the responses based on a given stimulus are stored
in one ``vertical slide'' from forefront to back (of dimension
\([ n_P,~1,~n_M]\)) and the responses in one experimental manipulation
is one ``slice'' (of dimension \([ n_P,~n_S,~1]\)) of the 3D array. The
responses associated to one pair participant-stimulus is consequently a
``pile'' (of dimension \([ 1,~1,~n_M]\)) of the 3D array. Note that to
simplify the representation the factors \(A_{PS}\) and \(A_{O}\) are not
represented in Figure~\ref{fig:lmm}. However, they could be associated
both to their own table (Table~\(PS\) and Table~\(O\)) and each entry of
the Table~\(PS\) would be associated with one ``pile'' of the 3D array.
Finally, each entry of the Table~\(O\) would be associated with one cell
of the 3D array.

This representation is a tool to understand which interactions between
fixed effects and random units are allowed in a model. Each ``floor''
(the responses by participants) crosses all levels of the \(A_S\) and
\(A_M\) factors, which implies that participants are measured in all
levels of \(A_S\) and \(A_M\). Moreover, each ``floor'' is composed of
multiple ``piles'' and multiple ``cells'' which means that a participant
will be measured in multiple levels of the factors \(A_{PS}\) and
\(A_{O}\). All the random interactions participant:\(A_S\),
participant:\(A_M\), participant:\(A_{PS}\), participant:\(A_{O}\) are
then allowed for CRE-MEM. The same rational is applied for stimuli: each
stimulus is represented in by one ``vertical slide'' and will be
measured in all levels of the \(A_P\) and \(A_M\), then the random
interaction stimuli:\(A_P\), stimuli:\(A_M\), stimuli:\(A_{PS}\),
stimuli:\(A_{O}\) are feasible. To understand which random interaction
associated to the interaction participants:stimuli can be included in
the model we apply the same strategy for the ``pile'' and each
interaction crosses all levels of the factors \(A_M\). And each
``pile''" is composed of multiple cells (which are associated to factors
\(A_O\)). Which means that the random interactions between
participants:stimuli:\(A_M\) and participants:stimuli:\(A_O\) are
feasible. These findings are summarized in Table~\ref{tab:ranef}.

\section{Matrix formulation of the CRE-MEM} \label{ap:notation}

\subsection{General Notation for mixed models}\label{ap:mlm}

Following \citet{bates_fitting_2015}, we define the mixed linear model:

\begin{equation}\label{eq:mlm_model}
y = X \beta +Z\gamma +\epsilon
\end{equation}

where \(y\) is the response, the fixed part of the design is \(X\) and
the random part is \(Z\). The fixed parameters are \(\beta\), the random
effects are \(\gamma \sim (0, \Sigma)\) and the error terms are
\(\epsilon \sim (0,\sigma^2I)\). For a CRE-MEM, we split the random
effects into \(G\) independent components
\(\gamma = \left[\gamma_1^\top | \dots |\gamma_g^\top | \dots |\gamma_G^\top \right]^\top\)
and their associated design matrices
\(Z = \left[Z_1 | \dots |Z_g | \dots |Z_G \right]\) which led to the
decomposition of the covariance matrix of the random effects into
\(\Sigma = \textrm{diag}(\Sigma_1, \dots, \Sigma_g, \dots, \Sigma_G)\).
The covariance matrix of the response variable \(y\) can then be written
as
\(\Omega = Z\Sigma Z^\top+ I\sigma^2 = Z_1\Sigma_1 Z_1^\top+\dots+Z_g\Sigma_g Z_g^\top+\dots+Z_G\Sigma_G Z_G^\top+ I\sigma^2\).

\subsection{Generalized ANOVA and RI-L for the CRE-MEM}

The gANOVA is written following the Equation~\ref{eq:mlm_model} with
specific constraints on the random effects. Independence between the
three groups of random components is specified using
\(\gamma = \left[\gamma_P^\top |\gamma_S^\top |\gamma_{P:S}^\top \right]^\top\)
and by defining a covariance matrix of observations split into the parts
relative to participants, stimuli and their interactions:

\begin{equation}\label{eq:decomp_var}
\Omega = Z\Sigma Z^\top+ I\sigma^2 = 
Z_P\Sigma_P Z_P^\top + 
Z_S\Sigma_S Z_S^\top +
Z_{P:S}\Sigma_{P:S} Z_{P:S}^\top + I\sigma^2,
\end{equation}

where
\(Z_P = \left(X_{parti}^{\top} * [\boldsymbol{1} | X_S| X_{PS} | X_M | X_O]^\top \right)^\top\),
\(Z_S = \left(X_{stimulus}^{\top} *[\boldsymbol{1} | X_P | X_{PS}| X_M | X_O]^\top \right)^\top\),
\(Z_{P:S} = \left( (X_{parti}^\top * X_{stimulus}^\top) * [\boldsymbol{1} | X_M | X_O]^\top \right)^\top\)
and the Khatri-Rao product \citep{khatri_solutions_1968} is written
using \(*\). \(X_{parti}\) and \(X_{stimulus}\) are matrices dummy
coding for the participants and stimuli, and \(\Sigma_P\), \(\Sigma_S\),
\(\Sigma_{P:S}\) are covariance matrices.

If the dataset has no missing value and after the appropriate
permutation (\(P_P\),\(P_S\) and \(P_{PS}\)), the covariance matrices
are written as block diagonal matrices of the individual covariance
structure: \(\Sigma_P = P_P(I_P \otimes \Sigma_{P}^0)P_P^\top\),
\(\Sigma_S = P_S(I_S \otimes \Sigma_{S}^0)P_S^\top\) and
\(\Sigma_{P:S} = P_{P:S}(I_{P:S} \otimes \Sigma_{P:S}^0)P_{P:S}^\top\).
gANOVA assumes that the matrices \(\Sigma_{P}^0\), \(\Sigma_{S}^0\) and
\(\Sigma_{PS}^0\) are diagonal matrices with the same value for the
random effects (i.e.~contrasts) associated to the same factor.

For gANOVA, the \(X_{\cdot}\) matrices \(X_S\), \(X_P\), \(X_{PS}\),
\(X_M\) and \(X_O\) are written using orthonormal contrasts \(C_\cdot\)
(e.g.: \texttt{contr.poly}) and overparametrized dummy-coded design
matrices \(X_\cdot^0\) such that \(X_\cdot = X_\cdot^0 C^-_\cdot\) where
\(C^-_\cdot\) is the generalized inverse of \(C_\cdot\). If \(C_\cdot\)
is orthonormal gives the properties \(C_\cdot C_\cdot^- = I\) and
\(C_\cdot^-C_\cdot^{-\top} = I -a{\bf 11}^\top\) where \(a\) is a
positive value that depends on the dimension of \(C_\cdot\).

Concerning the RI-L model, the only difference with gANOVA is that there
are no constraints on the random effects. Its matrix formulation is
therefore the same as gANOVA except that the contrasts \(C_\cdot\) are
not used (or are replaced by an identity matrix \(I\)). Note that the
\(X_\cdot^0\) matrices are used to construct the fixed part of the
design and are usually associated with contrasts. Depending on the
hypothesis, these contrasts may be represented by non-orthonormal
matrices (e.g.: using \texttt{contr.sum}).

\section{Comparison of the gANOVA and RI-L model} \label{ap:c-uc}

\renewcommand{\baselinestretch}{1}
\begin{figure}[!htb]
\includegraphics[width=160mm]{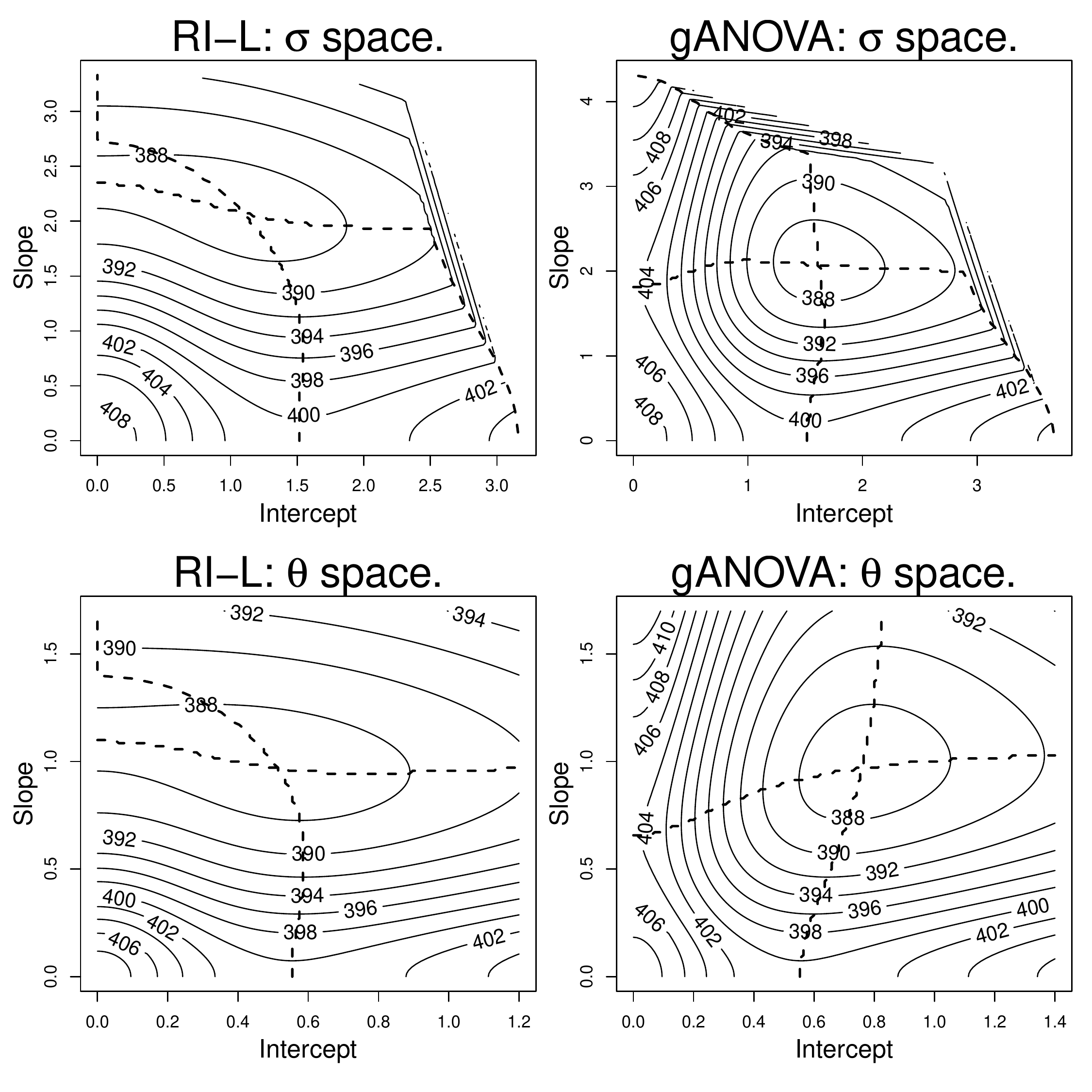}
\caption{Likelihood of RI-L and gANOVA for a model with one sampling units and 1 factor $A_M$ with replications and assuming random intercepts and random slopes. The two top figures represent the likelihood in the $\sigma$ parameters space (standard deviation of random effects), and two bottom one represents the $\theta$ parameters space (space of optimization parameters; see Equation~(4) in \citet{bates_fitting_2015} for an exact definition of the $\theta$'s.). The dotted lines are the  two ridges defining the profile likelihoods. We see in this example that the gANOVA tends to orthogonalize the profile likelihoods as they cross almost at a 90$^\circ$ angle, which suggest less dependency of the two parameters and an easier optimization process.}
\label{fig:curve}
\end{figure}
\renewcommand{\baselinestretch}{1.7}

In this appendix, we provide evidence that gANOVA has a strictly broader
decomposition of the correlation structure than RI-L. For the sake of
the argument, we focus the comparison to a model with only one sampling
unit (the participants), balanced design and replications (one factor
\(A_M\)), which means that there will be 3 variances parameters: one for
random intercepts, one for the random interactions and one for the
residuals. The replications will make the variance of the random
intercepts estimable. The gANOVA and RI-L correlation structure differ
by the constraints on the random effects and those constraints are
represented by contrast matrices. We will call constraint~(c) the design
of gANOVA and unconstraint~(uc) the one of RI-L.

In that setting, the RI-L model is written:

\begin{equation*}
y = X\beta + Z_{uc}\gamma_{uc} + \epsilon,
\end{equation*}

where \(\gamma_{uc} \sim ( 0 , I_{N_P} \otimes \Sigma_{uc})\),
\(\epsilon = (0,I\sigma_{uc;\epsilon}^2)\) for \(N_P\) the number of
participants. \(\Sigma_{uc}\) is a diagonal covariance matrix of
dimension \(N_M + 1\):
\(\Sigma_{uc} = \textrm{diag}(\sigma^2_{uc;i}, I_{N_M} \sigma^2_{uc;T})\).

The design matrix of the random effects is written
\(Z_{uc} = \left(Z^\top* \begin{bmatrix}\bf{1} & X_{uc}\end{bmatrix} ^\top\right)^\top\)
and \(*\) denotes the column-wise Khatri-Rao product
\citep{khatri_solutions_1968}. Assuming a balanced design, we write:

\begin{align*}
 Z_{uc} &= \left(({\bf 1}_{N_P} \otimes I_{N_M +1} )^\top* \begin{bmatrix}\bf{1} & {\bf 1}_{N_P} \otimes X_{uc;p}\end{bmatrix} ^\top\right)^\top \\
 &= \left(I_{N_P} \otimes \begin{bmatrix}{\bf 1}_{N_M} & X_{uc;p}\end{bmatrix}\right),
\end{align*}

where \(X_{uc;p}\) is a \(N_MN_R\times N_M\) matrix representing the
overparametrized design of one participant for a \(A_M\) factor of
\(N_M\) levels and assuming \(N_R\) replications in each cell. The
covariance matrix of the response \(y\) is:

\begin{align*}
Z_{uc} (I_{N_P} \otimes \Sigma_{uc}) Z_{uc}^\top + I\sigma_{uc;\epsilon}^2&= \left(I_{N_P}\otimes \begin{bmatrix}{\bf 1}_{N_M} & X_{uc;p}\end{bmatrix} \right)   (I_{N_P} \otimes \Sigma_{uc}) \left(I_{N_P}\otimes \begin{bmatrix}{\bf 1}_{N_M} & X_{uc;p}\end{bmatrix} ^\top\right)+ I\sigma_{uc;\epsilon}^2\\
& = \left(I_{N_P}\otimes \begin{bmatrix}{\bf 1}_{N_M}& X_{uc;p}\end{bmatrix} \right)  \left(I_{N_P}\otimes \Sigma_{uc} \begin{bmatrix}{\bf 1}_{N_M} & X_{uc;p}\end{bmatrix} ^\top\right)+ I\sigma_{uc;\epsilon}^2\\
& = I_{N_P}\otimes (\begin{bmatrix}{\bf 1}_{N_M} & X_{uc;p}\end{bmatrix}\Sigma_{uc} \begin{bmatrix}{\bf 1}_{N_M} & X_{uc;p}\end{bmatrix} ^\top)+ I\sigma_{uc;\epsilon}^2\\
& = I_{N_P}\otimes (\begin{bmatrix}{\bf 1}_{N_M} & X_{uc;p}\end{bmatrix}\Sigma_{uc} \begin{bmatrix}{\bf 1}_{N_M} & X_{uc;p}\end{bmatrix} ^\top+ I_{N_M}\sigma_{uc;\epsilon}^2).
\end{align*}

The covariance matrix of the response is a block diagonal matrix with
block-elements of the form:

\begin{align*}
\begin{bmatrix}{\bf 1}_{N_M} & X_{uc;p}\end{bmatrix}\Sigma_{uc} \begin{bmatrix}{\bf 1}_{N_M} & X_{uc;p}\end{bmatrix} ^\top+ I_{N_M}\sigma_{uc;\epsilon}^2
= {\bf 1}_{N_M} {\bf 1}_{N_M} ^\top  \sigma_{uc;i}^2 + X_{uc;p} X_{uc;p}^\top \sigma_{uc;F} + I_{N_M}\sigma_{uc;\epsilon}^2.
\end{align*}

Similarly, gANOVA is written:

\begin{equation*}
y = X\beta + Z_{c}\gamma_{c} + \epsilon,
\end{equation*}

where \(\gamma_{c} \sim ( 0 , I_{N_P} \otimes \Sigma_{c})\),
\(\epsilon = (0,I\sigma_{c;\epsilon}^2)\). \(\Sigma_{c}\) is a diagonal
matrix covariance matrix of dimension \(N_M\) :
\(\Sigma_{c} = \textrm{diag}(\sigma^2_{c;i}, I_{N_M-1} \sigma^2_{c;T})\).

The design matrix of the random effects is written using the orthonormal
contrast \(C\):
\(Z_{c} = \left(Z^\top* \begin{bmatrix}\bf{1} & X_{uc}C^{-}\end{bmatrix} ^\top\right)^\top\)
and assuming that it is balanced, it becomes:
\(Z_{c} = \left(I_{N_P}\otimes \begin{bmatrix}\bf{1} & X_{uc;p}C^-\end{bmatrix}\right)\).
The covariance matrix of the response is then:

\begin{align*}
Z_{c} (I_{N_P} \otimes \Sigma_{c}) Z_{c}^\top + I\sigma_{c;\epsilon}^2 &= I_{N_P}\otimes \left(\begin{bmatrix}{\bf 1}_{N_M} & X_{c;p}\end{bmatrix}\Sigma_{c} \begin{bmatrix}{\bf 1}_{N_M} & X_{c;p}\end{bmatrix} ^\top+ I_{N_M}\sigma_{c;\epsilon}^2 \right)\\
& = I_{N_P}\otimes \left(\begin{bmatrix}{\bf 1}_{N_M} & X_{uc;p}C^-\end{bmatrix}\Sigma_{c} \begin{bmatrix}{\bf 1}_{N_M} & X_{uc;p}C^-\end{bmatrix} ^\top+ I_{N_M}\sigma_{c;\epsilon}^2 \right).
\end{align*}

Using the properties of the orthonormal contrasts, the block-elements of
the covariance matrix simplify:

\begin{align*}
\begin{bmatrix}{\bf 1}_{N_M} & X_{uc;p}C^-\end{bmatrix}\Sigma_{c} \begin{bmatrix}{\bf 1}_{N_M}
& X_{uc;p}C^-\end{bmatrix} ^\top+ I_{N_M}\sigma_{c;\epsilon}^2 & = {\bf 1}_{N_M} {\bf 1}_{N_M} ^\top ~ \sigma_{c;i}^2 + X_{uc;p}C^-C^{-\top} X_{uc;p}^\top \sigma_{c;F} + I\sigma_{c;\epsilon}^2
\\
& = {\bf 1}_{N_M} {\bf 1}_{N_M} ^\top \sigma_{c;i}^2 + X_{uc;p}({ I -\bf 11}^\top a) X_{uc;p}^\top \sigma_{c;F} + I\sigma_{c;\epsilon}^2
\\
& = {\bf 1}_{N_M} {\bf 1}_{N_M} ^\top (\sigma_{c;i}^2  -a \sigma_{c;F}^2 ) +  X_{uc;p} X_{uc;p}^\top \sigma_{c;F} + I\sigma_{c;\epsilon}^2,
\end{align*}

where \(a\) is a positive fraction defined by the number of levels of
\(A_M\): \(a = 1-1/N_M\). The covariance matrices of the 2 models are
equal if and only if:

\begin{align*}
\sigma^2_{uc;i} &=\sigma_{c;i}^2 - a \sigma_{c;F}^2\\ 
\sigma^2_{uc;F} &= \sigma_{c;F}^2 \\
\sigma^2_{uc;\epsilon} &= \sigma_{c;\epsilon}^2.
\end{align*}

These equalities show us that the 2 models are equal for some values of
the variances of random effects. But, the first equality tells us that
the 2 models are not equal when \(\sigma_{c;i}^2 < a \sigma_{c;F}^2\).
Which means that adding the constraint of the contrasts \(C\) increases
the range of acceptable covariance matrices of the observations. With
more factors in the model, higher level interactions put similar
conditions on lower level interactions. The RI-L model will produce more
variance estimates equal to \(0\) with maximal values of the likelihood
at the boundary of the parameter space. And, when RI-L and gANOVA are
not equal, gANOVA will always have a better likelihood which suggests a
better fit of the model.

\section{Examples of lme4 formulas for CRE-MEM with different correlation structures}\label{ap:formula}

In this appendix, some examples of \texttt{R} formula from simple to
more complex designs (see Table~\ref{tab:model} for the selected
designs). A section that explains the five types of factors is in the
core of the article. We use the notation \texttt{PT} and \texttt{SM} for
the identifier factors of the participants and stimuli respectively.
These two factors as well as the design factor will be columns of the
\texttt{mydata} \texttt{dataframe}, as will be the response variable
\texttt{y}. To correctly interpret the main effects as in the ANOVA
framework, it is extremely important to use contrasts that sum to zero
(like contr.sum or contr.poly) and not the contr.treat default in
\texttt{R}). The fixed part will be assumed to be a full factorial
design in each case. For saturated designs, we drop the highest random
interaction term because we assume no replication of the same
observations (multiple observations associated to the same cell in the
design) and this term is thus not estimable as it is confounded with the
error term.

The way that \texttt{R} interprets formulas including factors is subtle
and not always as some users may expect. In a nutshell, \texttt{R}
assigns to factors the maximum degree of freedom (some sort of
contrasts) available which means that, for the factors \texttt{A} and
\texttt{B} with 2 levels, using the formula
\texttt{ \textasciitilde A*B} (which is internally replaced by
\texttt{ \textasciitilde 1+A+B+A:B}) and \texttt{ \textasciitilde A:B}
(replaced by \texttt{ \textasciitilde 1+A:B}) will assign respectively
\(1\) and \(3\) degrees of freedom for the interaction \texttt{A:B}; as
other example is \texttt{ \textasciitilde A} and
\texttt{ \textasciitilde 0 + A} assign respectively 1 and \(2\) degrees
of freedom for the effect of \texttt{A}. In \texttt{lme4}, this feature
happens when \texttt{R} computes separately the effects; for instance,
\texttt{\textasciitilde (1|PT) + (A|PT)} will assign 2 variances (+ 1
covariance) for the levels of \(A\) and
\texttt{\textasciitilde (A*B||PT)} will assign \(4\) variances (\(+ 6\)
covariances) for the interaction \texttt{A:B}. To estimate some of the
discussed correlation structures, the solution is to convert factors
into numeric variables and we use the \texttt{model.matrix()} function
for that purpose.

\subsection{A simple case, factors $A_P(2)$, $A_S(2)$ and $A_M(2)$}\label{ap:formula-simple}

This design corresponds to model M1 in Table~\ref{tab:param}. In that
simple case, the saturated correlation structure will not have the
highest term of interaction between the factor \(A_M\) and the random
units \texttt{participant:stimulus}. With only 2 levels per factor, the
``sum'' coding will produce similar results to the polynomial coding
variable.

\subsubsection*{RI and RI+}

\begin{Shaded}
\begin{Highlighting}[]
\KeywordTok{lmer}\NormalTok{(y }\OperatorTok{~}\StringTok{ }\NormalTok{Ap}\OperatorTok{*}\NormalTok{As}\OperatorTok{*}\NormalTok{Am }\OperatorTok{+}\StringTok{ }\NormalTok{(}\DecValTok{1}\OperatorTok{|}\NormalTok{PT) }\OperatorTok{+}\StringTok{ }\NormalTok{(}\DecValTok{1}\OperatorTok{|}\NormalTok{SM) }\OperatorTok{+}\StringTok{ }\NormalTok{(}\DecValTok{1}\OperatorTok{|}\NormalTok{PT}\OperatorTok{:}\NormalTok{SM), }\DataTypeTok{data =}\NormalTok{ mydata)}
\end{Highlighting}
\end{Shaded}

\subsubsection*{RI-L and RI-L+}

\begin{Shaded}
\begin{Highlighting}[]
\KeywordTok{lmer}\NormalTok{(y }\OperatorTok{~}\StringTok{ }\NormalTok{Ap}\OperatorTok{*}\NormalTok{As}\OperatorTok{*}\NormalTok{Am }\OperatorTok{+}\StringTok{ }\NormalTok{(}\DecValTok{1}\OperatorTok{|}\NormalTok{PT) }\OperatorTok{+}\StringTok{ }\NormalTok{(}\DecValTok{1}\OperatorTok{|}\NormalTok{PT}\OperatorTok{:}\NormalTok{As) }\OperatorTok{+}\StringTok{ }\NormalTok{(}\DecValTok{1}\OperatorTok{|}\NormalTok{PT}\OperatorTok{:}\NormalTok{Am) }\OperatorTok{+}\StringTok{ }\NormalTok{(}\DecValTok{1}\OperatorTok{|}\NormalTok{PT}\OperatorTok{:}\NormalTok{As}\OperatorTok{:}\NormalTok{Am)}
     \OperatorTok{+}\StringTok{ }\NormalTok{(}\DecValTok{1}\OperatorTok{|}\NormalTok{SM) }\OperatorTok{+}\StringTok{ }\NormalTok{(}\DecValTok{1}\OperatorTok{|}\NormalTok{SM}\OperatorTok{:}\NormalTok{Ap) }\OperatorTok{+}\StringTok{ }\NormalTok{(}\DecValTok{1}\OperatorTok{|}\NormalTok{SM}\OperatorTok{:}\NormalTok{Am)}\OperatorTok{+}\StringTok{ }\NormalTok{(}\DecValTok{1}\OperatorTok{|}\NormalTok{SM}\OperatorTok{:}\NormalTok{Ap}\OperatorTok{:}\NormalTok{Am) }
     \OperatorTok{+}\StringTok{ }\NormalTok{(}\DecValTok{1}\OperatorTok{|}\NormalTok{PT}\OperatorTok{:}\NormalTok{SM), }\DataTypeTok{data =}\NormalTok{ mydata)}
\end{Highlighting}
\end{Shaded}

\subsubsection*{MAX and MAX+}

\begin{Shaded}
\begin{Highlighting}[]
\KeywordTok{lmer}\NormalTok{(y }\OperatorTok{~}\StringTok{ }\NormalTok{Ap}\OperatorTok{*}\NormalTok{As}\OperatorTok{*}\NormalTok{Am }\OperatorTok{+}\StringTok{ }\NormalTok{(As}\OperatorTok{*}\NormalTok{Am}\OperatorTok{|}\NormalTok{PT) }\OperatorTok{+}\StringTok{ }\NormalTok{(Ap}\OperatorTok{*}\NormalTok{Am}\OperatorTok{|}\NormalTok{SM) }\OperatorTok{+}\StringTok{ }\NormalTok{(}\DecValTok{1}\OperatorTok{|}\NormalTok{PT}\OperatorTok{:}\NormalTok{SM), }\DataTypeTok{data =}\NormalTok{ mydata)}
\end{Highlighting}
\end{Shaded}

\subsubsection*{ZCP and ZCP+}

\begin{Shaded}
\begin{Highlighting}[]
\NormalTok{mydata}\OperatorTok{$}\NormalTok{Xs <-}\StringTok{ }\KeywordTok{model.matrix}\NormalTok{(}\OperatorTok{~}\StringTok{ }\NormalTok{As, }\DataTypeTok{data =}\NormalTok{ mydata)[, }\DecValTok{-1}\NormalTok{]}
\NormalTok{mydata}\OperatorTok{$}\NormalTok{Xp <-}\StringTok{ }\KeywordTok{model.matrix}\NormalTok{(}\OperatorTok{~}\StringTok{ }\NormalTok{Ap, }\DataTypeTok{data =}\NormalTok{ mydata)[, }\DecValTok{-1}\NormalTok{]}
\NormalTok{mydata}\OperatorTok{$}\NormalTok{Xm <-}\StringTok{ }\KeywordTok{model.matrix}\NormalTok{(}\OperatorTok{~}\StringTok{ }\NormalTok{Am, }\DataTypeTok{data =}\NormalTok{ mydata)[, }\DecValTok{-1}\NormalTok{]}
\KeywordTok{lmer}\NormalTok{(y }\OperatorTok{~}\StringTok{ }\NormalTok{Ap}\OperatorTok{*}\NormalTok{As}\OperatorTok{*}\NormalTok{Am }\OperatorTok{+}\StringTok{ }\NormalTok{(Xs}\OperatorTok{*}\NormalTok{Xm}\OperatorTok{||}\NormalTok{PT) }\OperatorTok{+}\StringTok{ }\NormalTok{(Xp}\OperatorTok{*}\NormalTok{Xm}\OperatorTok{||}\NormalTok{SM) }\OperatorTok{+}\StringTok{ }\NormalTok{(}\DecValTok{1}\OperatorTok{|}\NormalTok{PT}\OperatorTok{:}\NormalTok{SM), }\DataTypeTok{data =}\NormalTok{ mydata)}
\end{Highlighting}
\end{Shaded}

\subsubsection*{gANOVA and gANOVA+}

The gANOVA package accept two equivalent formula. The first one use the
same notation as the RI-L formula
(\texttt{ + (1|PT) + (1|PT:f1) + (1|PT:f2) + (1|PT:f1:f2)}) and puts
orthonormal coding from the second factors of the right part of the
random effect. Doing so, the order of the factors matters and the first
factor to the right of the bar must be the random unit. Note that it
implies that the interaction participant-stimulus should be written as
only one factor. The second formula is more recommended as it simplifies
the notation quite importantly to the formula \texttt{+ (1|PT|f1*f2)}.
Then gANOVA is specified using:

\begin{Shaded}
\begin{Highlighting}[]
\NormalTok{mydata}\OperatorTok{$}\NormalTok{PTSM <-}\StringTok{ }\KeywordTok{interaction}\NormalTok{(mydata}\OperatorTok{$}\NormalTok{PT, mydata}\OperatorTok{$}\NormalTok{SM)}

\KeywordTok{gANOVA}\NormalTok{(y }\OperatorTok{~}\StringTok{ }\NormalTok{Ap}\OperatorTok{*}\NormalTok{As}\OperatorTok{*}\NormalTok{Am }\OperatorTok{+}\StringTok{ }\NormalTok{(}\DecValTok{1}\OperatorTok{|}\NormalTok{PT) }\OperatorTok{+}\StringTok{ }\NormalTok{(}\DecValTok{1}\OperatorTok{|}\NormalTok{PT}\OperatorTok{:}\NormalTok{As) }\OperatorTok{+}\StringTok{ }\NormalTok{(}\DecValTok{1}\OperatorTok{|}\NormalTok{PT}\OperatorTok{:}\NormalTok{Am) }\OperatorTok{+}\StringTok{ }\NormalTok{(}\DecValTok{1}\OperatorTok{|}\NormalTok{PT}\OperatorTok{:}\NormalTok{As}\OperatorTok{:}\NormalTok{Am)}
          \OperatorTok{+}\StringTok{ }\NormalTok{(}\DecValTok{1}\OperatorTok{|}\NormalTok{SM) }\OperatorTok{+}\StringTok{ }\NormalTok{(}\DecValTok{1}\OperatorTok{|}\NormalTok{SM}\OperatorTok{:}\NormalTok{Ap) }\OperatorTok{+}\StringTok{ }\NormalTok{(}\DecValTok{1}\OperatorTok{|}\NormalTok{SM}\OperatorTok{:}\NormalTok{Am)}\OperatorTok{+}\StringTok{ }\NormalTok{(}\DecValTok{1}\OperatorTok{|}\NormalTok{SM}\OperatorTok{:}\NormalTok{Ap}\OperatorTok{:}\NormalTok{Am)}
          \OperatorTok{+}\StringTok{ }\NormalTok{(}\DecValTok{1}\OperatorTok{|}\NormalTok{PTSM), }\DataTypeTok{data =}\NormalTok{ mydata)}
\end{Highlighting}
\end{Shaded}

or equivalently:

\begin{Shaded}
\begin{Highlighting}[]
\NormalTok{mydata}\OperatorTok{$}\NormalTok{PTSM <-}\StringTok{ }\KeywordTok{interaction}\NormalTok{(mydata}\OperatorTok{$}\NormalTok{PT, mydata}\OperatorTok{$}\NormalTok{SM)}

\KeywordTok{gANOVA}\NormalTok{(y }\OperatorTok{~}\StringTok{ }\NormalTok{Ap}\OperatorTok{*}\NormalTok{As}\OperatorTok{*}\NormalTok{Am }\OperatorTok{+}\StringTok{ }\NormalTok{(}\DecValTok{1}\OperatorTok{|}\NormalTok{PT}\OperatorTok{|}\NormalTok{As}\OperatorTok{*}\NormalTok{Am)}\OperatorTok{+}\StringTok{ }\NormalTok{(}\DecValTok{1}\OperatorTok{|}\NormalTok{SM}\OperatorTok{|}\NormalTok{Ap}\OperatorTok{*}\NormalTok{Am) }\OperatorTok{+}\StringTok{ }\NormalTok{(}\DecValTok{1}\OperatorTok{|}\NormalTok{PTSM), }
       \DataTypeTok{data =}\NormalTok{ mydata)}
\end{Highlighting}
\end{Shaded}

\subsection{A common case, factors $A_P(3)$, $A_{S}(3)$, $A_M(3)$ and $A_M(2)$}\label{ap:formula-common}

This design corresponds to model M3 in Table~\ref{tab:param}. The
highest interaction terms which is confounded with the error term is the
interaction between the 2 factors \(A_{M1}\), \(A_{M2}\) and the
\texttt{participant:stimulus} random unit.

\subsubsection*{RI and RI+}

\begin{Shaded}
\begin{Highlighting}[]
\KeywordTok{lmer}\NormalTok{(y }\OperatorTok{~}\StringTok{ }\NormalTok{Ap}\OperatorTok{*}\NormalTok{As}\OperatorTok{*}\NormalTok{Am1}\OperatorTok{*}\NormalTok{Am2 }\OperatorTok{+}\StringTok{ }\NormalTok{(}\DecValTok{1}\OperatorTok{|}\NormalTok{PT) }\OperatorTok{+}\StringTok{ }\NormalTok{(}\DecValTok{1}\OperatorTok{|}\NormalTok{SM) }\OperatorTok{+}\StringTok{ }\NormalTok{(}\DecValTok{1}\OperatorTok{|}\NormalTok{PT}\OperatorTok{:}\NormalTok{SM), }\DataTypeTok{data =}\NormalTok{ mydata)}
\end{Highlighting}
\end{Shaded}

\subsubsection*{RI-L and RI-L+}

\begin{Shaded}
\begin{Highlighting}[]
\KeywordTok{lmer}\NormalTok{(y }\OperatorTok{~}\StringTok{ }\NormalTok{Ap}\OperatorTok{*}\NormalTok{As}\OperatorTok{*}\NormalTok{Am1}\OperatorTok{*}\NormalTok{Am2 }\OperatorTok{+}\StringTok{ }\NormalTok{(}\DecValTok{1}\OperatorTok{|}\NormalTok{PT) }\OperatorTok{+}\StringTok{ }\NormalTok{(}\DecValTok{1}\OperatorTok{|}\NormalTok{PT}\OperatorTok{:}\NormalTok{As) }\OperatorTok{+}\StringTok{ }\NormalTok{(}\DecValTok{1}\OperatorTok{|}\NormalTok{PT}\OperatorTok{:}\NormalTok{Am1) }\OperatorTok{+}\StringTok{ }\NormalTok{(}\DecValTok{1}\OperatorTok{|}\NormalTok{PT}\OperatorTok{:}\NormalTok{Am2) }
     \OperatorTok{+}\StringTok{ }\NormalTok{(}\DecValTok{1}\OperatorTok{|}\NormalTok{PT}\OperatorTok{:}\NormalTok{As}\OperatorTok{:}\NormalTok{Am1) }\OperatorTok{+}\StringTok{ }\NormalTok{(}\DecValTok{1}\OperatorTok{|}\NormalTok{PT}\OperatorTok{:}\NormalTok{As}\OperatorTok{:}\NormalTok{Am2) }\OperatorTok{+}\StringTok{ }\NormalTok{(}\DecValTok{1}\OperatorTok{|}\NormalTok{PT}\OperatorTok{:}\NormalTok{Am1}\OperatorTok{:}\NormalTok{Am2) }\OperatorTok{+}\StringTok{ }\NormalTok{(}\DecValTok{1}\OperatorTok{|}\NormalTok{PT}\OperatorTok{:}\NormalTok{As}\OperatorTok{:}\NormalTok{Am1}\OperatorTok{:}\NormalTok{Am2)}
     \OperatorTok{+}\StringTok{ }\NormalTok{(}\DecValTok{1}\OperatorTok{|}\NormalTok{SM) }\OperatorTok{+}\StringTok{ }\NormalTok{(}\DecValTok{1}\OperatorTok{|}\NormalTok{SM}\OperatorTok{:}\NormalTok{Ap) }\OperatorTok{+}\StringTok{ }\NormalTok{(}\DecValTok{1}\OperatorTok{|}\NormalTok{SM}\OperatorTok{:}\NormalTok{Am1) }\OperatorTok{+}\StringTok{ }\NormalTok{(}\DecValTok{1}\OperatorTok{|}\NormalTok{SM}\OperatorTok{:}\NormalTok{Am2) }\OperatorTok{+}\StringTok{ }\NormalTok{(}\DecValTok{1}\OperatorTok{|}\NormalTok{SM}\OperatorTok{:}\NormalTok{Ap}\OperatorTok{:}\NormalTok{Am1)}
     \OperatorTok{+}\StringTok{ }\NormalTok{(}\DecValTok{1}\OperatorTok{|}\NormalTok{SM}\OperatorTok{:}\NormalTok{Ap}\OperatorTok{:}\NormalTok{Am2) }\OperatorTok{+}\StringTok{ }\NormalTok{(}\DecValTok{1}\OperatorTok{|}\NormalTok{SM}\OperatorTok{:}\NormalTok{Am1}\OperatorTok{:}\NormalTok{Am2) }\OperatorTok{+}\StringTok{ }\NormalTok{(}\DecValTok{1}\OperatorTok{|}\NormalTok{SM}\OperatorTok{:}\NormalTok{Ap}\OperatorTok{:}\NormalTok{Am1}\OperatorTok{:}\NormalTok{Am2)}
     \OperatorTok{+}\StringTok{ }\NormalTok{(}\DecValTok{1}\OperatorTok{|}\NormalTok{PT}\OperatorTok{:}\NormalTok{SM) }\OperatorTok{+}\StringTok{ }\NormalTok{(}\DecValTok{1}\OperatorTok{|}\NormalTok{PT}\OperatorTok{:}\NormalTok{SM}\OperatorTok{:}\NormalTok{Am1) }\OperatorTok{+}\StringTok{ }\NormalTok{(}\DecValTok{1}\OperatorTok{|}\NormalTok{PT}\OperatorTok{:}\NormalTok{SM}\OperatorTok{:}\NormalTok{Am2), }\DataTypeTok{data =}\NormalTok{ mydata)}
\end{Highlighting}
\end{Shaded}

\subsubsection*{MAX and MAX+}

With more than two levels, it is not advisable to use the \texttt{sum}
coding for the random part and we must create new factors with the
appropriate coding variable. Moreover, \texttt{lmer} does not allow for
different type of coding for the fixed part and the random part. We
suggest creating new variables with the appropriate coding. Here we
choose the orthonormal \texttt{contr.poly} coding:

\begin{Shaded}
\begin{Highlighting}[]
\NormalTok{mydata}\OperatorTok{$}\NormalTok{ApPoly <-}\StringTok{ }\NormalTok{mydata}\OperatorTok{$}\NormalTok{Ap; }\KeywordTok{contrasts}\NormalTok{(mydata}\OperatorTok{$}\NormalTok{ApPoly) <-}\StringTok{ }\NormalTok{contr.poly}
\NormalTok{mydata}\OperatorTok{$}\NormalTok{AsPoly <-}\StringTok{ }\NormalTok{mydata}\OperatorTok{$}\NormalTok{As; }\KeywordTok{contrasts}\NormalTok{(mydata}\OperatorTok{$}\NormalTok{AsPoly) <-}\StringTok{ }\NormalTok{contr.poly}
\NormalTok{mydata}\OperatorTok{$}\NormalTok{Am1Poly <-}\StringTok{ }\NormalTok{mydata}\OperatorTok{$}\NormalTok{Am1; }\KeywordTok{contrasts}\NormalTok{(mydata}\OperatorTok{$}\NormalTok{Am1Poly) <-}\StringTok{ }\NormalTok{contr.poly}
\NormalTok{mydata}\OperatorTok{$}\NormalTok{Am2Poly <-}\StringTok{ }\NormalTok{mydata}\OperatorTok{$}\NormalTok{Am2; }\KeywordTok{contrasts}\NormalTok{(mydata}\OperatorTok{$}\NormalTok{Am2Poly) <-}\StringTok{ }\NormalTok{contr.poly}

\KeywordTok{lmer}\NormalTok{(y }\OperatorTok{~}\StringTok{ }\NormalTok{Ap}\OperatorTok{*}\NormalTok{As}\OperatorTok{*}\NormalTok{Am1}\OperatorTok{*}\NormalTok{Am2 }\OperatorTok{+}\StringTok{ }\NormalTok{(AsPoly}\OperatorTok{*}\NormalTok{Am1Poly}\OperatorTok{*}\NormalTok{Am2Poly}\OperatorTok{|}\NormalTok{PT)}
     \OperatorTok{+}\StringTok{ }\NormalTok{(ApPoly}\OperatorTok{*}\NormalTok{Am1Poly}\OperatorTok{*}\NormalTok{Am2Poly}\OperatorTok{|}\NormalTok{SM) }
     \OperatorTok{+}\StringTok{ }\NormalTok{(Am1Poly }\OperatorTok{+}\StringTok{ }\NormalTok{Am2Poly}\OperatorTok{|}\NormalTok{PT}\OperatorTok{:}\NormalTok{SM), }\DataTypeTok{data =}\NormalTok{ mydata)}
\end{Highlighting}
\end{Shaded}

\subsubsection*{ZCP and ZCP+}

For ZCP, we transform the factors into orthonormal coding variable. We
first need to set the coding using the procedure described in the
previous section. Then, we transform the factors into numeric variables
using the \texttt{model.matrix()} function.

\begin{Shaded}
\begin{Highlighting}[]
\NormalTok{dataAp <-}\StringTok{ }\KeywordTok{data.frame}\NormalTok{(}\KeywordTok{model.matrix}\NormalTok{( }\OperatorTok{~}\StringTok{ }\NormalTok{ApPoly, }\DataTypeTok{data =}\NormalTok{ mydata)[,}\OperatorTok{-}\DecValTok{1}\NormalTok{])}
\KeywordTok{colnames}\NormalTok{(dataAp) <-}\StringTok{ }\KeywordTok{c}\NormalTok{(}\StringTok{"Xpa"}\NormalTok{, }\StringTok{"Xpb"}\NormalTok{)}

\NormalTok{dataAs <-}\StringTok{ }\KeywordTok{data.frame}\NormalTok{(}\KeywordTok{model.matrix}\NormalTok{( }\OperatorTok{~}\StringTok{ }\NormalTok{AsPoly, }\DataTypeTok{data =}\NormalTok{ mydata)[,}\OperatorTok{-}\DecValTok{1}\NormalTok{])}
\KeywordTok{colnames}\NormalTok{(dataAs) <-}\StringTok{ }\KeywordTok{c}\NormalTok{(}\StringTok{"Xsa"}\NormalTok{, }\StringTok{"Xsb"}\NormalTok{)}

\NormalTok{dataAm1 <-}\StringTok{ }\KeywordTok{data.frame}\NormalTok{(}\KeywordTok{model.matrix}\NormalTok{( }\OperatorTok{~}\StringTok{ }\NormalTok{Am1Poly, }\DataTypeTok{data =}\NormalTok{ mydata)[,}\OperatorTok{-}\DecValTok{1}\NormalTok{])}
\KeywordTok{colnames}\NormalTok{(dataAm1) <-}\StringTok{ }\KeywordTok{c}\NormalTok{(}\StringTok{"Xm1a"}\NormalTok{, }\StringTok{"Xm1b"}\NormalTok{)}

\NormalTok{dataAm2 <-}\StringTok{ }\KeywordTok{data.frame}\NormalTok{(}\KeywordTok{model.matrix}\NormalTok{( }\OperatorTok{~}\StringTok{ }\NormalTok{Am2Poly, }\DataTypeTok{data =}\NormalTok{ mydata)[,}\OperatorTok{-}\DecValTok{1}\NormalTok{])}
\KeywordTok{colnames}\NormalTok{(dataAm2) <-}\StringTok{ }\KeywordTok{c}\NormalTok{(}\StringTok{"Xm2a"}\NormalTok{)}

\NormalTok{mydata <-}\StringTok{ }\KeywordTok{cbind}\NormalTok{(mydata, dataAp, dataAs, dataAm1, dataAm2)}

\KeywordTok{lmer}\NormalTok{(y }\OperatorTok{~}\StringTok{ }\NormalTok{Ap}\OperatorTok{*}\NormalTok{As}\OperatorTok{*}\NormalTok{Am1}\OperatorTok{*}\NormalTok{Am2 }\OperatorTok{+}\StringTok{ }\NormalTok{((Xsa }\OperatorTok{+}\StringTok{ }\NormalTok{Xsb)}\OperatorTok{*}\NormalTok{(Xm1a }\OperatorTok{+}\StringTok{ }\NormalTok{Xm1b)}\OperatorTok{*}\NormalTok{Xm2a}\OperatorTok{||}\NormalTok{PT)}
     \OperatorTok{+}\StringTok{ }\NormalTok{((Xpa }\OperatorTok{+}\StringTok{ }\NormalTok{Xpb)}\OperatorTok{*}\NormalTok{(Xm1a }\OperatorTok{+}\StringTok{ }\NormalTok{Xm1b)}\OperatorTok{*}\NormalTok{Xm2a}\OperatorTok{||}\NormalTok{SM) }
     \OperatorTok{+}\StringTok{ }\NormalTok{(Xm1a }\OperatorTok{+}\StringTok{ }\NormalTok{Xm1b }\OperatorTok{+}\StringTok{ }\NormalTok{Xm2a}\OperatorTok{||}\NormalTok{PT}\OperatorTok{:}\NormalTok{SM), }\DataTypeTok{data =}\NormalTok{ mydata)}
\end{Highlighting}
\end{Shaded}

\subsubsection*{gANOVA and gANOVA+}

As explained previously, the interaction participant:stimuli should be
written as only one factor when we can run the gANOVA function:

\begin{Shaded}
\begin{Highlighting}[]
\NormalTok{mydata}\OperatorTok{$}\NormalTok{PTSM <-}\StringTok{ }\KeywordTok{interaction}\NormalTok{(mydata}\OperatorTok{$}\NormalTok{PT, mydata}\OperatorTok{$}\NormalTok{SM)}

\KeywordTok{gANOVA}\NormalTok{(y }\OperatorTok{~}\StringTok{ }\NormalTok{Ap}\OperatorTok{*}\NormalTok{As}\OperatorTok{*}\NormalTok{Am1}\OperatorTok{*}\NormalTok{Am2 }\OperatorTok{+}\StringTok{ }\NormalTok{(}\DecValTok{1}\OperatorTok{|}\NormalTok{PT}\OperatorTok{|}\NormalTok{As}\OperatorTok{*}\NormalTok{Am1}\OperatorTok{*}\NormalTok{Am2) }\OperatorTok{+}\StringTok{ }\NormalTok{(}\DecValTok{1}\OperatorTok{|}\NormalTok{SM}\OperatorTok{|}\NormalTok{Ap}\OperatorTok{*}\NormalTok{Am1}\OperatorTok{*}\NormalTok{Am2) }
       \OperatorTok{+}\StringTok{ }\NormalTok{(}\DecValTok{1}\OperatorTok{|}\NormalTok{PTSM}\OperatorTok{|}\NormalTok{Am1}\OperatorTok{+}\NormalTok{Am2), }\DataTypeTok{data =}\NormalTok{ mydata)}
\end{Highlighting}
\end{Shaded}

\subsection{A complex case, factors $A_P(3)$, $A_{S}(3)$, $A_M(3)$, $A_M(2)$, $A_{PS}(2)$, $A_{O}(2)$}\label{ap:formula-complex}

This design corresponds to model M5 of Table~\ref{tab:param}. The
highest interaction term which is confounded with the error term is the
interaction between the 3 factors \(A_{M1}\), \(A_{M2}\), \(A_O\) and
the \texttt{participant:stimulus} random unit.

\subsubsection*{RI and RI+}

\begin{Shaded}
\begin{Highlighting}[]
\KeywordTok{lmer}\NormalTok{(y }\OperatorTok{~}\StringTok{ }\NormalTok{Ap}\OperatorTok{*}\NormalTok{As}\OperatorTok{*}\NormalTok{Am1}\OperatorTok{*}\NormalTok{Am2}\OperatorTok{*}\NormalTok{Aps}\OperatorTok{*}\NormalTok{Ao }\OperatorTok{+}\StringTok{ }\NormalTok{(}\DecValTok{1}\OperatorTok{|}\NormalTok{PT) }\OperatorTok{+}\StringTok{ }\NormalTok{(}\DecValTok{1}\OperatorTok{|}\NormalTok{SM) }\OperatorTok{+}\StringTok{ }\NormalTok{(}\DecValTok{1}\OperatorTok{|}\NormalTok{PT}\OperatorTok{:}\NormalTok{SM), }\DataTypeTok{data =}\NormalTok{ mydata)}
\end{Highlighting}
\end{Shaded}

\subsubsection*{RI-L and RI-L+}

\begin{Shaded}
\begin{Highlighting}[]
\KeywordTok{lmer}\NormalTok{(y }\OperatorTok{~}\StringTok{ }\NormalTok{Ap}\OperatorTok{*}\NormalTok{As}\OperatorTok{*}\NormalTok{Am1}\OperatorTok{*}\NormalTok{Am2}\OperatorTok{*}\NormalTok{Aps}\OperatorTok{*}\NormalTok{Ao }\OperatorTok{+}\StringTok{ }\NormalTok{(}\DecValTok{1}\OperatorTok{|}\NormalTok{PT) }\OperatorTok{+}\StringTok{ }\NormalTok{(}\DecValTok{1}\OperatorTok{|}\NormalTok{PT}\OperatorTok{:}\NormalTok{As) }\OperatorTok{+}\StringTok{ }\NormalTok{(}\DecValTok{1}\OperatorTok{|}\NormalTok{PT}\OperatorTok{:}\NormalTok{Am1) }\OperatorTok{+}\StringTok{ }\NormalTok{(}\DecValTok{1}\OperatorTok{|}\NormalTok{PT}\OperatorTok{:}\NormalTok{Am2) }
     \OperatorTok{+}\StringTok{ }\NormalTok{(}\DecValTok{1}\OperatorTok{|}\NormalTok{PT}\OperatorTok{:}\NormalTok{Aps) }\OperatorTok{+}\StringTok{ }\NormalTok{(}\DecValTok{1}\OperatorTok{|}\NormalTok{PT}\OperatorTok{:}\NormalTok{Ao) }\OperatorTok{+}\StringTok{ }\NormalTok{(}\DecValTok{1}\OperatorTok{|}\NormalTok{PT}\OperatorTok{:}\NormalTok{As}\OperatorTok{:}\NormalTok{Am1) }\OperatorTok{+}\StringTok{ }\NormalTok{(}\DecValTok{1}\OperatorTok{|}\NormalTok{PT}\OperatorTok{:}\NormalTok{As}\OperatorTok{:}\NormalTok{Am2) }\OperatorTok{+}\StringTok{ }\NormalTok{(}\DecValTok{1}\OperatorTok{|}\NormalTok{PT}\OperatorTok{:}\NormalTok{As}\OperatorTok{:}\NormalTok{Amsi) }
     \OperatorTok{+}\StringTok{ }\NormalTok{(}\DecValTok{1}\OperatorTok{|}\NormalTok{PT}\OperatorTok{:}\NormalTok{As}\OperatorTok{:}\NormalTok{Ao) }\OperatorTok{+}\StringTok{ }\NormalTok{(}\DecValTok{1}\OperatorTok{|}\NormalTok{PT}\OperatorTok{:}\NormalTok{Am1}\OperatorTok{:}\NormalTok{Am2) }\OperatorTok{+}\StringTok{ }\NormalTok{(}\DecValTok{1}\OperatorTok{|}\NormalTok{PT}\OperatorTok{:}\NormalTok{Am1}\OperatorTok{:}\NormalTok{Aps) }\OperatorTok{+}\StringTok{ }\NormalTok{(}\DecValTok{1}\OperatorTok{|}\NormalTok{PT}\OperatorTok{:}\NormalTok{Am1}\OperatorTok{:}\NormalTok{Ao) }
     \OperatorTok{+}\StringTok{ }\NormalTok{(}\DecValTok{1}\OperatorTok{|}\NormalTok{PT}\OperatorTok{:}\NormalTok{Am2}\OperatorTok{:}\NormalTok{Aps) }\OperatorTok{+}\StringTok{ }\NormalTok{(}\DecValTok{1}\OperatorTok{|}\NormalTok{PT}\OperatorTok{:}\NormalTok{Am2}\OperatorTok{:}\NormalTok{Ao) }\OperatorTok{+}\StringTok{ }\NormalTok{(}\DecValTok{1}\OperatorTok{|}\NormalTok{PT}\OperatorTok{:}\NormalTok{Aps}\OperatorTok{:}\NormalTok{Ao) }\OperatorTok{+}\StringTok{ }\NormalTok{(}\DecValTok{1}\OperatorTok{|}\NormalTok{PT}\OperatorTok{:}\NormalTok{As}\OperatorTok{:}\NormalTok{Am1}\OperatorTok{:}\NormalTok{Am2)}
     \OperatorTok{+}\StringTok{ }\NormalTok{(}\DecValTok{1}\OperatorTok{|}\NormalTok{PT}\OperatorTok{:}\NormalTok{As}\OperatorTok{:}\NormalTok{Am1}\OperatorTok{:}\NormalTok{Aps) }\OperatorTok{+}\StringTok{ }\NormalTok{(}\DecValTok{1}\OperatorTok{|}\NormalTok{PT}\OperatorTok{:}\NormalTok{As}\OperatorTok{:}\NormalTok{Am1}\OperatorTok{:}\NormalTok{Ao) }\OperatorTok{+}\StringTok{ }\NormalTok{(}\DecValTok{1}\OperatorTok{|}\NormalTok{PT}\OperatorTok{:}\NormalTok{As}\OperatorTok{:}\NormalTok{Am2}\OperatorTok{:}\NormalTok{Amsi) }\OperatorTok{+}\StringTok{ }\NormalTok{(}\DecValTok{1}\OperatorTok{|}\NormalTok{PT}\OperatorTok{:}\NormalTok{As}\OperatorTok{:}\NormalTok{Aps}\OperatorTok{:}\NormalTok{Ao)}
     \OperatorTok{+}\StringTok{ }\NormalTok{(}\DecValTok{1}\OperatorTok{|}\NormalTok{PT}\OperatorTok{:}\NormalTok{Am1}\OperatorTok{:}\NormalTok{Am2}\OperatorTok{:}\NormalTok{Aps) }\OperatorTok{+}\StringTok{ }\NormalTok{(}\DecValTok{1}\OperatorTok{|}\NormalTok{PT}\OperatorTok{:}\NormalTok{Am2}\OperatorTok{:}\NormalTok{Aps}\OperatorTok{:}\NormalTok{Ao) }\OperatorTok{+}\StringTok{ }\NormalTok{(}\DecValTok{1}\OperatorTok{|}\NormalTok{PT}\OperatorTok{:}\NormalTok{As}\OperatorTok{:}\NormalTok{Am1}\OperatorTok{:}\NormalTok{Am2}\OperatorTok{:}\NormalTok{Aps)}
     \OperatorTok{+}\StringTok{ }\NormalTok{(}\DecValTok{1}\OperatorTok{|}\NormalTok{PT}\OperatorTok{:}\NormalTok{As}\OperatorTok{:}\NormalTok{Am1}\OperatorTok{:}\NormalTok{Am2}\OperatorTok{:}\NormalTok{Ao) }\OperatorTok{+}\StringTok{ }\NormalTok{(}\DecValTok{1}\OperatorTok{|}\NormalTok{PT}\OperatorTok{:}\NormalTok{As}\OperatorTok{:}\NormalTok{Am1}\OperatorTok{:}\NormalTok{Aps}\OperatorTok{:}\NormalTok{Ao) }\OperatorTok{+}\StringTok{ }\NormalTok{(}\DecValTok{1}\OperatorTok{|}\NormalTok{PT}\OperatorTok{:}\NormalTok{As}\OperatorTok{:}\NormalTok{Am2}\OperatorTok{:}\NormalTok{Aps}\OperatorTok{:}\NormalTok{Ao) }
     \OperatorTok{+}\StringTok{ }\NormalTok{(}\DecValTok{1}\OperatorTok{|}\NormalTok{PT}\OperatorTok{:}\NormalTok{Am1}\OperatorTok{:}\NormalTok{Am2}\OperatorTok{:}\NormalTok{Aps}\OperatorTok{:}\NormalTok{Ao) }\OperatorTok{+}\StringTok{ }\NormalTok{(}\DecValTok{1}\OperatorTok{|}\NormalTok{PT}\OperatorTok{:}\NormalTok{As}\OperatorTok{:}\NormalTok{Am1}\OperatorTok{:}\NormalTok{Am2}\OperatorTok{:}\NormalTok{Aps}\OperatorTok{:}\NormalTok{Ao)}
     \OperatorTok{+}\StringTok{ }\NormalTok{(}\DecValTok{1}\OperatorTok{|}\NormalTok{SM) }\OperatorTok{+}\StringTok{ }\NormalTok{(}\DecValTok{1}\OperatorTok{|}\NormalTok{SM}\OperatorTok{:}\NormalTok{Ap) }\OperatorTok{+}\StringTok{ }\NormalTok{(}\DecValTok{1}\OperatorTok{|}\NormalTok{SM}\OperatorTok{:}\NormalTok{Am1) }\OperatorTok{+}\StringTok{ }\NormalTok{(}\DecValTok{1}\OperatorTok{|}\NormalTok{SM}\OperatorTok{:}\NormalTok{Am2) }
     \OperatorTok{+}\StringTok{ }\NormalTok{(}\DecValTok{1}\OperatorTok{|}\NormalTok{SM}\OperatorTok{:}\NormalTok{Aps) }\OperatorTok{+}\StringTok{ }\NormalTok{(}\DecValTok{1}\OperatorTok{|}\NormalTok{SM}\OperatorTok{:}\NormalTok{Ao) }\OperatorTok{+}\StringTok{ }\NormalTok{(}\DecValTok{1}\OperatorTok{|}\NormalTok{SM}\OperatorTok{:}\NormalTok{Ap}\OperatorTok{:}\NormalTok{Am1) }\OperatorTok{+}\StringTok{ }\NormalTok{(}\DecValTok{1}\OperatorTok{|}\NormalTok{SM}\OperatorTok{:}\NormalTok{Ap}\OperatorTok{:}\NormalTok{Am2) }\OperatorTok{+}\StringTok{ }\NormalTok{(}\DecValTok{1}\OperatorTok{|}\NormalTok{SM}\OperatorTok{:}\NormalTok{Ap}\OperatorTok{:}\NormalTok{Amsi) }
     \OperatorTok{+}\StringTok{ }\NormalTok{(}\DecValTok{1}\OperatorTok{|}\NormalTok{SM}\OperatorTok{:}\NormalTok{Ap}\OperatorTok{:}\NormalTok{Ao) }\OperatorTok{+}\StringTok{ }\NormalTok{(}\DecValTok{1}\OperatorTok{|}\NormalTok{SM}\OperatorTok{:}\NormalTok{Am1}\OperatorTok{:}\NormalTok{Am2) }\OperatorTok{+}\StringTok{ }\NormalTok{(}\DecValTok{1}\OperatorTok{|}\NormalTok{SM}\OperatorTok{:}\NormalTok{Am1}\OperatorTok{:}\NormalTok{Aps) }\OperatorTok{+}\StringTok{ }\NormalTok{(}\DecValTok{1}\OperatorTok{|}\NormalTok{SM}\OperatorTok{:}\NormalTok{Am1}\OperatorTok{:}\NormalTok{Ao) }
     \OperatorTok{+}\StringTok{ }\NormalTok{(}\DecValTok{1}\OperatorTok{|}\NormalTok{SM}\OperatorTok{:}\NormalTok{Am2}\OperatorTok{:}\NormalTok{Aps) }\OperatorTok{+}\StringTok{ }\NormalTok{(}\DecValTok{1}\OperatorTok{|}\NormalTok{SM}\OperatorTok{:}\NormalTok{Am2}\OperatorTok{:}\NormalTok{Ao) }\OperatorTok{+}\StringTok{ }\NormalTok{(}\DecValTok{1}\OperatorTok{|}\NormalTok{SM}\OperatorTok{:}\NormalTok{Aps}\OperatorTok{:}\NormalTok{Ao) }\OperatorTok{+}\StringTok{ }\NormalTok{(}\DecValTok{1}\OperatorTok{|}\NormalTok{SM}\OperatorTok{:}\NormalTok{Ap}\OperatorTok{:}\NormalTok{Am1}\OperatorTok{:}\NormalTok{Am2)}
     \OperatorTok{+}\StringTok{ }\NormalTok{(}\DecValTok{1}\OperatorTok{|}\NormalTok{SM}\OperatorTok{:}\NormalTok{Ap}\OperatorTok{:}\NormalTok{Am1}\OperatorTok{:}\NormalTok{Aps) }\OperatorTok{+}\StringTok{ }\NormalTok{(}\DecValTok{1}\OperatorTok{|}\NormalTok{SM}\OperatorTok{:}\NormalTok{Ap}\OperatorTok{:}\NormalTok{Am1}\OperatorTok{:}\NormalTok{Ao) }\OperatorTok{+}\StringTok{ }\NormalTok{(}\DecValTok{1}\OperatorTok{|}\NormalTok{SM}\OperatorTok{:}\NormalTok{Ap}\OperatorTok{:}\NormalTok{Am2}\OperatorTok{:}\NormalTok{Amsi) }\OperatorTok{+}\StringTok{ }\NormalTok{(}\DecValTok{1}\OperatorTok{|}\NormalTok{SM}\OperatorTok{:}\NormalTok{Ap}\OperatorTok{:}\NormalTok{Aps}\OperatorTok{:}\NormalTok{Ao)}
     \OperatorTok{+}\StringTok{ }\NormalTok{(}\DecValTok{1}\OperatorTok{|}\NormalTok{SM}\OperatorTok{:}\NormalTok{Am1}\OperatorTok{:}\NormalTok{Am2}\OperatorTok{:}\NormalTok{Aps) }\OperatorTok{+}\StringTok{ }\NormalTok{(}\DecValTok{1}\OperatorTok{|}\NormalTok{SM}\OperatorTok{:}\NormalTok{Am2}\OperatorTok{:}\NormalTok{Aps}\OperatorTok{:}\NormalTok{Ao) }\OperatorTok{+}\StringTok{ }\NormalTok{(}\DecValTok{1}\OperatorTok{|}\NormalTok{SM}\OperatorTok{:}\NormalTok{Ap}\OperatorTok{:}\NormalTok{Am1}\OperatorTok{:}\NormalTok{Am2}\OperatorTok{:}\NormalTok{Aps)}
     \OperatorTok{+}\StringTok{ }\NormalTok{(}\DecValTok{1}\OperatorTok{|}\NormalTok{SM}\OperatorTok{:}\NormalTok{Ap}\OperatorTok{:}\NormalTok{Am1}\OperatorTok{:}\NormalTok{Am2}\OperatorTok{:}\NormalTok{Ao) }\OperatorTok{+}\StringTok{ }\NormalTok{(}\DecValTok{1}\OperatorTok{|}\NormalTok{SM}\OperatorTok{:}\NormalTok{Ap}\OperatorTok{:}\NormalTok{Am1}\OperatorTok{:}\NormalTok{Aps}\OperatorTok{:}\NormalTok{Ao)}\OperatorTok{+}\StringTok{ }\NormalTok{(}\DecValTok{1}\OperatorTok{|}\NormalTok{SM}\OperatorTok{:}\NormalTok{Ap}\OperatorTok{:}\NormalTok{Am2}\OperatorTok{:}\NormalTok{Aps}\OperatorTok{:}\NormalTok{Ao) }
     \OperatorTok{+}\StringTok{ }\NormalTok{(}\DecValTok{1}\OperatorTok{|}\NormalTok{SM}\OperatorTok{:}\NormalTok{Am1}\OperatorTok{:}\NormalTok{Am2}\OperatorTok{:}\NormalTok{Aps}\OperatorTok{:}\NormalTok{Ao) }\OperatorTok{+}\StringTok{ }\NormalTok{(}\DecValTok{1}\OperatorTok{|}\NormalTok{SM}\OperatorTok{:}\NormalTok{Ap}\OperatorTok{:}\NormalTok{Am1}\OperatorTok{:}\NormalTok{Am2}\OperatorTok{:}\NormalTok{Aps}\OperatorTok{:}\NormalTok{Ao)}
     \OperatorTok{+}\StringTok{ }\NormalTok{(}\DecValTok{1}\OperatorTok{|}\NormalTok{PT}\OperatorTok{:}\NormalTok{SM) }\OperatorTok{+}\StringTok{ }\NormalTok{(}\DecValTok{1}\OperatorTok{|}\NormalTok{PT}\OperatorTok{:}\NormalTok{SM}\OperatorTok{:}\NormalTok{Am1) }\OperatorTok{+}\StringTok{ }\NormalTok{(}\DecValTok{1}\OperatorTok{|}\NormalTok{PT}\OperatorTok{:}\NormalTok{SM}\OperatorTok{:}\NormalTok{Am2) }\OperatorTok{+}\StringTok{ }\NormalTok{(}\DecValTok{1}\OperatorTok{|}\NormalTok{PT}\OperatorTok{:}\NormalTok{SM}\OperatorTok{:}\NormalTok{Ao) }
     \OperatorTok{+}\StringTok{ }\NormalTok{(}\DecValTok{1}\OperatorTok{|}\NormalTok{PT}\OperatorTok{:}\NormalTok{SM}\OperatorTok{:}\NormalTok{Am1}\OperatorTok{:}\NormalTok{Am2) }\OperatorTok{+}\StringTok{ }\NormalTok{(}\DecValTok{1}\OperatorTok{|}\NormalTok{PT}\OperatorTok{:}\NormalTok{SM}\OperatorTok{:}\NormalTok{Am1}\OperatorTok{:}\NormalTok{Ao) }\OperatorTok{+}\StringTok{ }\NormalTok{(}\DecValTok{1}\OperatorTok{|}\NormalTok{PT}\OperatorTok{:}\NormalTok{SM}\OperatorTok{:}\NormalTok{Am2}\OperatorTok{:}\NormalTok{Ao), }\DataTypeTok{data =}\NormalTok{ mydata)}
\end{Highlighting}
\end{Shaded}

\subsection*{MAX and MAX+}

We set the orthonormal coding using the following functions. Note that
the factors with 2 levels do not need to change from coding of type
``sum'' to coding of type ``polynomial''.

\begin{Shaded}
\begin{Highlighting}[]
\NormalTok{mydata}\OperatorTok{$}\NormalTok{ApPoly <-}\StringTok{ }\NormalTok{mydata}\OperatorTok{$}\NormalTok{Ap; }\KeywordTok{contrasts}\NormalTok{(mydata}\OperatorTok{$}\NormalTok{ApPoly) <-}\StringTok{ }\NormalTok{contr.poly}
\NormalTok{mydata}\OperatorTok{$}\NormalTok{AsPoly <-}\StringTok{ }\NormalTok{mydata}\OperatorTok{$}\NormalTok{As; }\KeywordTok{contrasts}\NormalTok{(mydata}\OperatorTok{$}\NormalTok{AsPoly) <-}\StringTok{ }\NormalTok{contr.poly}
\NormalTok{mydata}\OperatorTok{$}\NormalTok{Am1Poly <-}\StringTok{ }\NormalTok{mydata}\OperatorTok{$}\NormalTok{Am1; }\KeywordTok{contrasts}\NormalTok{(mydata}\OperatorTok{$}\NormalTok{Am1Poly) <-}\StringTok{ }\NormalTok{contr.poly}
\NormalTok{mydata}\OperatorTok{$}\NormalTok{Am2Poly <-}\StringTok{ }\NormalTok{mydata}\OperatorTok{$}\NormalTok{Am2; }\KeywordTok{contrasts}\NormalTok{(mydata}\OperatorTok{$}\NormalTok{Am2Poly) <-}\StringTok{ }\NormalTok{contr.poly}
\NormalTok{mydata}\OperatorTok{$}\NormalTok{ApsPoly <-}\StringTok{ }\NormalTok{mydata}\OperatorTok{$}\NormalTok{Aps; }\KeywordTok{contrasts}\NormalTok{(mydata}\OperatorTok{$}\NormalTok{ApsPoly) <-}\StringTok{ }\NormalTok{contr.poly}
\NormalTok{mydata}\OperatorTok{$}\NormalTok{AoPoly <-}\StringTok{ }\NormalTok{mydata}\OperatorTok{$}\NormalTok{Ao; }\KeywordTok{contrasts}\NormalTok{(mydata}\OperatorTok{$}\NormalTok{AoPoly) <-}\StringTok{ }\NormalTok{contr.poly}
\end{Highlighting}
\end{Shaded}

\begin{Shaded}
\begin{Highlighting}[]
\KeywordTok{lmer}\NormalTok{(y }\OperatorTok{~}\StringTok{ }\NormalTok{Ap}\OperatorTok{*}\NormalTok{As}\OperatorTok{*}\NormalTok{Am1}\OperatorTok{*}\NormalTok{Am2}\OperatorTok{*}\NormalTok{Aps}\OperatorTok{*}\NormalTok{Ao }\OperatorTok{+}\StringTok{ }\NormalTok{(AsPoly}\OperatorTok{*}\NormalTok{Am1Poly}\OperatorTok{*}\NormalTok{Am2Poly}\OperatorTok{*}\NormalTok{ApsPoly}\OperatorTok{*}\NormalTok{AoPoly}\OperatorTok{|}\NormalTok{PT) }
     \OperatorTok{+}\StringTok{ }\NormalTok{(ApPoly}\OperatorTok{*}\NormalTok{Am1Poly}\OperatorTok{*}\NormalTok{Am2Poly}\OperatorTok{*}\NormalTok{ApsPoly}\OperatorTok{*}\NormalTok{AoPoly}\OperatorTok{|}\NormalTok{SM) }
     \OperatorTok{+}\StringTok{ }\NormalTok{(Am1Poly }\OperatorTok{+}\StringTok{ }\NormalTok{Am2Poly }\OperatorTok{+}\StringTok{ }\NormalTok{AoPoly }\OperatorTok{+}\StringTok{ }\NormalTok{Am1Poly}\OperatorTok{:}\NormalTok{PolyAm2 }\OperatorTok{+}\StringTok{ }\NormalTok{Am1Poly}\OperatorTok{:}\NormalTok{AoPoly }\OperatorTok{+}\StringTok{ }\NormalTok{Am2Poly}\OperatorTok{:}\NormalTok{AoPoly}\OperatorTok{|}\NormalTok{PT}\OperatorTok{:}\NormalTok{SM), }
     \DataTypeTok{data =}\NormalTok{ mydata)}
\end{Highlighting}
\end{Shaded}

\subsection*{ZCP and ZCP+}

See the MAX model to change the coding of the factors.

\begin{Shaded}
\begin{Highlighting}[]
\NormalTok{dataAp <-}\StringTok{ }\KeywordTok{data.frame}\NormalTok{(}\KeywordTok{model.matrix}\NormalTok{( }\OperatorTok{~}\StringTok{ }\NormalTok{ApPoly, }\DataTypeTok{data =}\NormalTok{ mydata)[,}\OperatorTok{-}\DecValTok{1}\NormalTok{])}
\KeywordTok{colnames}\NormalTok{(dataAp) <-}\StringTok{ }\KeywordTok{c}\NormalTok{(}\StringTok{"Xpa"}\NormalTok{, }\StringTok{"Xpb"}\NormalTok{)}

\NormalTok{dataAs <-}\StringTok{ }\KeywordTok{data.frame}\NormalTok{(}\KeywordTok{model.matrix}\NormalTok{( }\OperatorTok{~}\StringTok{ }\NormalTok{AsPoly, }\DataTypeTok{data =}\NormalTok{ mydata)[,}\OperatorTok{-}\DecValTok{1}\NormalTok{])}
\KeywordTok{colnames}\NormalTok{(dataAs) <-}\StringTok{ }\KeywordTok{c}\NormalTok{(}\StringTok{"Xsa"}\NormalTok{, }\StringTok{"Xsb"}\NormalTok{)}

\NormalTok{dataAm1 <-}\StringTok{ }\KeywordTok{data.frame}\NormalTok{(}\KeywordTok{model.matrix}\NormalTok{( }\OperatorTok{~}\StringTok{ }\NormalTok{Am1Poly, }\DataTypeTok{data =}\NormalTok{ mydata)[,}\OperatorTok{-}\DecValTok{1}\NormalTok{])}
\KeywordTok{colnames}\NormalTok{(dataAm1) <-}\StringTok{ }\KeywordTok{c}\NormalTok{(}\StringTok{"Xm1a"}\NormalTok{, }\StringTok{"Xm1b"}\NormalTok{)}

\NormalTok{mydata}\OperatorTok{$}\NormalTok{Xm2 <-}\StringTok{ }\KeywordTok{model.matrix}\NormalTok{( }\OperatorTok{~}\StringTok{ }\NormalTok{Am2Poly, }\DataTypeTok{data =}\NormalTok{ mydata)[, }\DecValTok{-1}\NormalTok{]}
\NormalTok{mydata}\OperatorTok{$}\NormalTok{Xps <-}\StringTok{ }\KeywordTok{model.matrix}\NormalTok{( }\OperatorTok{~}\StringTok{ }\NormalTok{ApsPoly, }\DataTypeTok{data =}\NormalTok{ mydata)[, }\DecValTok{-1}\NormalTok{]}
\NormalTok{mydata}\OperatorTok{$}\NormalTok{Xo <-}\StringTok{ }\KeywordTok{model.matrix}\NormalTok{( }\OperatorTok{~}\StringTok{ }\NormalTok{AoPoly, }\DataTypeTok{data =}\NormalTok{ mydata)[, }\DecValTok{-1}\NormalTok{]}
  
\NormalTok{mydata <-}\StringTok{ }\KeywordTok{cbind}\NormalTok{(mydata,dataAp,dataAs,dataAm1)}
\end{Highlighting}
\end{Shaded}

\begin{Shaded}
\begin{Highlighting}[]
\KeywordTok{lmer}\NormalTok{(y }\OperatorTok{~}\StringTok{ }\NormalTok{Ap}\OperatorTok{*}\NormalTok{As}\OperatorTok{*}\NormalTok{Am1}\OperatorTok{*}\NormalTok{Am2}\OperatorTok{*}\NormalTok{Aps}\OperatorTok{*}\NormalTok{Ao }\OperatorTok{+}\StringTok{ }\NormalTok{((Xsa}\OperatorTok{+}\NormalTok{Xsb)}\OperatorTok{*}\NormalTok{(Xm1a}\OperatorTok{+}\NormalTok{Xm1b)}\OperatorTok{*}\NormalTok{Xm2}\OperatorTok{*}\NormalTok{Xps}\OperatorTok{*}\NormalTok{Xo}\OperatorTok{||}\NormalTok{PT) }
     \OperatorTok{+}\StringTok{ }\NormalTok{((Xpa}\OperatorTok{+}\NormalTok{Xpb)}\OperatorTok{*}\NormalTok{(Xm1a}\OperatorTok{+}\NormalTok{Xm1b)}\OperatorTok{*}\NormalTok{Xm2}\OperatorTok{*}\NormalTok{Xps}\OperatorTok{*}\NormalTok{Xo}\OperatorTok{||}\NormalTok{SM) }
     \OperatorTok{+}\StringTok{ }\NormalTok{((Xm1a}\OperatorTok{+}\NormalTok{Xm1b) }\OperatorTok{+}\StringTok{ }\NormalTok{Xm2 }\OperatorTok{+}\StringTok{ }\NormalTok{Xo }\OperatorTok{+}\StringTok{ }\NormalTok{(Xm1a}\OperatorTok{+}\NormalTok{Xm1b)}\OperatorTok{:}\NormalTok{Xm2 }\OperatorTok{+}\StringTok{ }\NormalTok{(Xm1a}\OperatorTok{+}\NormalTok{Xm1b)}\OperatorTok{:}\NormalTok{Xo }\OperatorTok{+}\StringTok{ }\NormalTok{Xm2}\OperatorTok{:}\NormalTok{Xo}\OperatorTok{||}\NormalTok{PT}\OperatorTok{:}\NormalTok{SM), }
     \DataTypeTok{data =}\NormalTok{ mydata)}
\end{Highlighting}
\end{Shaded}

\subsubsection*{gANOVA and gANOVA+}

\begin{Shaded}
\begin{Highlighting}[]
\NormalTok{mydata}\OperatorTok{$}\NormalTok{PTSM <-}\StringTok{ }\KeywordTok{interaction}\NormalTok{(mydata}\OperatorTok{$}\NormalTok{PT, mydata}\OperatorTok{$}\NormalTok{SM)}

\KeywordTok{gANOVA}\NormalTok{(y }\OperatorTok{~}\StringTok{ }\NormalTok{Ap}\OperatorTok{*}\NormalTok{As}\OperatorTok{*}\NormalTok{Am1}\OperatorTok{*}\NormalTok{Am2}\OperatorTok{*}\NormalTok{Aps}\OperatorTok{*}\NormalTok{Ao }\OperatorTok{+}\StringTok{ }\NormalTok{(}\DecValTok{1}\OperatorTok{|}\NormalTok{PT}\OperatorTok{|}\NormalTok{As}\OperatorTok{*}\NormalTok{Am1}\OperatorTok{*}\NormalTok{Am2}\OperatorTok{*}\NormalTok{Aps}\OperatorTok{*}\NormalTok{Ao)}
       \OperatorTok{+}\StringTok{ }\NormalTok{(}\DecValTok{1}\OperatorTok{|}\NormalTok{SM}\OperatorTok{|}\NormalTok{Ap}\OperatorTok{*}\NormalTok{Am1}\OperatorTok{*}\NormalTok{Am2}\OperatorTok{*}\NormalTok{Aps}\OperatorTok{*}\NormalTok{Ao) }
       \OperatorTok{+}\StringTok{ }\NormalTok{(}\DecValTok{1}\OperatorTok{|}\NormalTok{PTSM}\OperatorTok{|}\NormalTok{Am1 }\OperatorTok{+}\StringTok{ }\NormalTok{Am2 }\OperatorTok{+}\StringTok{ }\NormalTok{Ao }\OperatorTok{+}\StringTok{ }\NormalTok{Am1}\OperatorTok{:}\NormalTok{Am2 }\OperatorTok{+}\StringTok{ }\NormalTok{Am1}\OperatorTok{:}\NormalTok{Ao }\OperatorTok{+}\StringTok{ }\NormalTok{Am2}\OperatorTok{:}\NormalTok{Ao), }
       \DataTypeTok{data =}\NormalTok{ mydata)}
\end{Highlighting}
\end{Shaded}

\renewcommand\refname{References}
  \bibliography{ms}

\end{document}